\documentclass[12pt]{article}

\usepackage{amstext,amsmath,amssymb,amsfonts,bbm,slashed}
\usepackage[latin1]{inputenc}
\usepackage{epsfig}
\usepackage{hyperref}
\usepackage{amsthm}
\usepackage{subfigure}
\usepackage{color}
\usepackage{multirow}
\usepackage{calc}

\usepackage{stmaryrd}

\usepackage{cancel}
\usepackage{graphicx}

\usepackage{float}

\usepackage{soul}

\usepackage[normalem]{ulem}
\usepackage{amsmath}
\newcommand{\stkout}[1]{\ifmmode\text{\sout{\ensuremath{#1}}}\else\sout{#1}\fi}

\usepackage[T1]{fontenc}
\usepackage{array}
\usepackage{makecell}
\newcolumntype{x}[1]{>{\centering\arraybackslash}p{#1}}
\usepackage{tikz}

\usepackage{pbox}

\newcommand{\kin}{{\rm{kin\,}}}
\newcommand{\ext}{{\rm{ext\,}}}
\newcommand{\inter}{{\rm{int\,}}}

\newcommand{\pet}{\eta_{+}}
\newcommand{\bet}{\eta_{\bullet}}
\newcommand{\tet}{\eta_{\times}}
\newcommand{\rhop}{\rho_{+}}
\newcommand{\rhot}{\rho_{\times}}

\newcommand{\omp}{\omega_{{\rm d};+}}
\newcommand{\omt}{\omega_{{\rm d};\times}}
\newcommand{\omb}{\omega_{{\rm d};\bullet}}

\newcommand{\lex}{ {l_{\rm ext}} }

\def\C{{\mathbbm C}}

\def\Z{{\mathbbm Z}}

\newcommand{\cG}{{\mathcal G}}
\newcommand{\bG}{{\partial\mathcal G}}
\newcommand{\tJ}{{\widetilde{J}}}

\newcommand{\cexG}{\mathcal G_{\text{color}}}

\newcommand{\cL}{{\mathcal L}}
\newcommand{\cV}{{\mathcal V}}
\newcommand{\cF}{{\mathcal F}}

\newcommand{\bdel}{ {\boldsymbol{\delta}} }

\newcommand{\bmu}{\boldsymbol{\mu}}
\newcommand{\col}{ \rm{color} }

\newcommand{\Tr}{{\rm Tr}}

\newcommand{\Sym}{ {\rm Sym} }

\newtheorem{lemma}{Lemma}

\newtheorem{theorem}{Theorem}
\newtheorem{proposition}{Proposition}

\newcommand{\bea}{\begin{eqnarray}}
\newcommand{\eea}{\end{eqnarray}}
\newcommand{\beq}{\begin{equation}}
\newcommand{\eeq}{\end{equation}}

\newcommand{\be}{\begin{equation}}
\newcommand{\ee}{\end{equation}}

\makeatletter

\begin{document}

\begin{titlepage}
\begin{flushright}
ICMPA-MPA/2017/20
\end{flushright}

\vspace{20pt}

\begin{center}

{\Large\bf Renormalizable Enhanced Tensor Field Theory: \\

\medskip

The quartic melonic case
\medskip 
}
\vspace{15pt}

{\large Joseph Ben Geloun$^{a,c,\dag}$ and Reiko Toriumi$^{b,\ddag} $}

\vspace{15pt}

$^{a}${\sl Laboratoire d'Informatique de Paris Nord UMR CNRS 7030}\\
{\sl Universit\'e Paris 13, 99, avenue J.-B. Clement, 93430 Villetaneuse, France} \\

\vspace{5pt}

$^{b}${\sl Institute for Mathematics, Astrophysics, and Particle Physics, Radboud University, }\\
{\sl Heyendaalseweg 135, 6525 AJ, Nijmegen, the Netherlands }

\vspace{5pt}

$^{c}${\sl International Chair in Mathematical Physics 
and Applications\\ (ICMPA-UNESCO Chair), University of Abomey-Calavi,\\
072B.P.50, Cotonou, Rep. of Benin}\\

\vspace{5pt}

E-mails:  {\sl $^{\dag}$bengeloun@lipn.univ-paris13.fr,
$^\ddag$reiko.toriumi@science.ru.nl}

\vspace{10pt}

\begin{abstract}

Amplitudes of ordinary tensor models are dominated at large $N$  by the so-called melonic graph amplitudes. Enhanced tensor models extend tensor models with special scalings of their interactions which allow, in the same limit, that the sub-dominant amplitudes to be ``enhanced'', that is to be as dominant as the melonic ones. These models were introduced to explore new large $N$ limits and to probe different phases for tensor models.  Tensor field theory is the quantum field theoretic counterpart of tensor models and enhanced tensor field theory enlarges this theory space to accommodate enhanced tensor interactions.  We undertake  the multi-scale renormalization analysis for two types of enhanced quartic melonic theories  with rank $d$ tensor fields  $\phi: (U(1)^{D})^{d} \to \mathbb{C}$ and  with interactions of the form  $p^{2a}\phi^4$  reminiscent of derivative couplings expressed in momentum space.  Scrutinizing the degree of divergence  of both theories,  we identify generic conditions for their renormalizability  at all orders of perturbation.  For a first type of theory, we identify a 2-parameter space of  just-renormalizable models for generic $(d,D)$. These models have dominant non-melonic four-point functions.  Finally, by specifying the parameters, we detail the renormalization analysis of a second type   of model.  Lying in between just- and super-renormalizability, that model is more exotic: all four-point amplitudes are convergent, however it exhibits an infinite family of divergent two-point amplitudes.  
  
\end{abstract}

\today

\end{center}

\noindent  Pacs numbers:  11.10.Gh, 04.60.-m, 02.10.Ox
\\

\bigskip

\setcounter{footnote}{0}

\end{titlepage}


\tableofcontents

\section{Introduction}
\label{intro}

Tensor models 
\cite{ambj3dqg,mmgravity,sasa1,boul,oog},
and their field theory version, tensor field theories are approaches 
to Quantum Gravity (QG) 
which propose a background-independent  quantization and, in the field theory case, 
an ultraviolet-consistent completion of General Relativity.
They study a discrete-to-continuum transition for discretized path integrals summing over not only metrics of a discretized Einstein-Hilbert action, but also over topologies.
The partition function of tensor models spans weighted
triangulations for every piecewise-linear manifold in any dimensions, hence they are naturally a random-geometric approach to QG.
In this regard, they can be considered to fall under the umbrella of discretization approaches to QG, such as quantum Regge calculus 
\cite{regge, hamber} and (causal) dynamical triangulations 
\cite{Ambjorn:2005jj,Ambjorn:2012jv,Ambjorn:2013hma}. 

Historically, tensor models were introduced as  higher dimensional generalizations 
of matrix models which saw their celebrated
success in describing 2-dimensional QG \cite{Di Francesco:1993nw}.
It was, however, not so straightforward to generalize matrix models' achievements to higher dimensions mainly because the organizing principle of and computational tools for the partition function 
of tensor models were lacking; diagonalization of tensors is not obvious and 
techniques on which matrix model calculations relied on did not  find extensions
to tensors. In particular,  matrix models generate maps sorted by their genus. Their partition function then admits a genus expansion and, at large size $N$ of the matrix \cite{thooft}, 
calculations can be made exact and matrix models become solvable. 
The large $N$ limit is crucial to achieve the continuum limit of matrix models, 
as a 2D theory of gravity coupled with a Liouville conformal 
field \cite{Kazakov:1985ds,KKM, David:1985nj, KPZ,david,kawai}.
This is one of the most acclaimed results pertaining to 2D QG.

The large $N$ limit for tensor models \cite{Gur3,GurRiv,Gur4} was finally unveiled after the advent of colored tensor models generating triangulations shown to be pseudo-manifolds \cite{color,Gurau:2009tz,Gurau:2010nd,Gurau:2011xp}.
The partition function of colored tensor models can be catalogued in terms of a new quantity 
called the degree of the tensor graph which plays the role of the genus 
in higher dimensions. Such a discovery, as anticipated, led  to a wealth of developments in random tensors   in  areas as diverse as statistical mechanics, quantum field theory, constructive field theory, combinatorics, probability theory, 
 geometry and topology \cite{Bonzom:2011zz}--\cite{BenGeloun:2017vwn}.
The following references provide 
comprehensive 
reviews on random tensors and tensor field theories \cite{sigma,razvanbook,Rivasseau:2011hm,Carrozza:2013mna}.  
Furthermore, more recently, tensor models gathered an attention in a new direction:
they turn out to be desirable toy models for holographic duality  
\cite{Witten:2016iux, Gurau:2016lzk,Klebanov:2016xxf,Krishnan:2016bvg,Ferrari:2017ryl,
Gurau:2017xhf,Bonzom:2017pqs}. The 
large $N$ limit in range of the disorder of the famous Sachdev-Ye-Kitaev (SYK) model 
\cite{sachdevye,kitaev,Maldacena:2016hyu,Gross:2016kjj},
corresponds to the large $N$ limit of colored tensor models thought of as quantum mechanical models without disorder \cite{Witten:2016iux}. 

Despite all its remarkable achievements, colored tensor models have
not  yet succeeded to define a ``nice'' continuum limit in which an emergent 3D
or 4D space 
could be identified. In colored tensor models,
some graphs which are particular triangulations of a sphere, called melons,
are found to be dominant at large $N$ \cite{Bonzom:2011zz}.
In the world of melons, colored tensor models undergo a phase transition towards
the so-called branched polymer phase which is not of the characteristics 
({\it e.g.,} Hausdorff and spectral dimensions) that our large and smooth space-time manifold holds \cite{Gurau:2013cbh}.
In order to improve the critical behavior of tensor models, 
it was then put forward to go beyond the melonic sector,  by 
modifying the weights of interactions in order 
to  include a wider class of graphs that could be resummed at large $N$. 
Such a proposal has been called ``enhancing'' tensor models and has  first been investigated 
in the work by Bonzom et al. \cite{Bonzom:2012wa,Bonzom:2015axa}. 
The upshot of this analysis is somehow encouraging: some enhanced tensor models undergo a phase transition
from branched polymers to a 2D QG phase (with positive entropy exponents). 
Let us be more specific at this point: the previous studies on enhanced tensor models 
focused on increasing the statistical weights of non-melonic tensor interactions called necklaces 
(which are only present in the tensor rank $d \ge 4$).
In a different perspective along with its very own set of questions, our proposal is to use the framework of field theories, therefore working in tensor field theories rather than in tensor models, and to explore new ways of building enhanced models 
in which non-melonic graphs could contribute to the analysis at large $N$. 

Once one promotes tensor models to field theories, which possess now with infinite degrees of freedom, we call them tensor field theories.  Note that, from the 90's, 
Boulatov introduced a gauge invariant version of tensor models by embedding
them  in lattice gauge  field theory over $SU(2)$ \cite{boul}. This approach was 
considerably appealing  to make contact with other QG approaches and was at the inception of Group Field Theory (GFT)
 \cite{Freidel:2005qe, oriti, Krajewski:2012aw}. Hence, chronologically, the first field theoretic approach of tensor models was GFT.  GFT implements a constraint
 (referred to as the gauge invariance constraint) on the fields to achieve a geometrical interpretation of the combinatorial simplices associated with 
 the tensor field and their interactions, 
 along with a flatness condition for the gluing of simplices. On the other hand, in GFT, as the name refers to it, the group as a manifold where the fields live is a central concept: the group law is used in the underlying lattice gauge field theory. 
Tensor field theory distinguishes itself with GFT as it might not have these constraints.

There are other motivations for introducing fields in the search of an emergent
spacetime. For instance, one makes another progress by regarding the simplicial complex associated with the tensor (in a tensor model) as a true (combinatorial) quantum of space. 
These fields live in an abstract internal space and are endowed with a given dynamics
and consequently a flow.  The goal is then to provide a phase portrait of 
that theory space and, in particular, to detect the presence of interesting (fixed) points.
Such fixed points would be associated with interesting physics. 
Thus, rather than tuning  a given tensor model at criticality and seeing  a new phase
for geometry emerging, we might  give initial conditions of a model in a 
field theory space and let it flow towards the corresponding fixed point. 
To define a  flow, a parameter or a scale is needed. This makes  the presence of  propagators or regulators of paramount importance  in usual field theory. Hence embedding tensor models into a field theoretic context, that
is giving them a propagator, provides them with a  flow. 
This naturally steer us towards other interesting questions.

Quantum field theories have many well-established  tools 
in order to reveal the properties of high energy physics and condensed matter systems. 
However, tensor field theory as  a quantum field theory also inherits several of its 
drawbacks like divergent amplitudes due to the existence of infinitely many 
degrees of freedom. The treatment of divergences, hence 
the renormalization program for tensor field theories becomes even more intricate because they are non-local field theories, {\it i.e.,} their interactions 
occur in a region of the configuration space. 
As a result, to import  the quantum field theoretic methods to tensor field theories was an important axis of investigations in the recent years. 

The  Renormalization Group (RG) program has been successfully applied  to tensor
field theory and also GFT leading to the discovery of entirely new families of 
renormalizable non-local quantum field theories
\cite{BenGeloun:2011rc,BenGeloun:2012yk,Geloun:2012bz,Carrozza:2012uv,Samary:2012bw,Geloun:2013saa,Carrozza:2013wda}. 
These
models  can be regarded as a rightful extension of matrix field theories 
like the Grosse and Wulkenhaar model \cite{Grosse:2004yu,Grosse:2012uv,Grosse:2016qmk}, an asymptotically safe  non-local quantum field theory stemming
from noncommutative geometry. 
The parametric representation and the ensuing dimensional regularization 
have been extended to tensor field theory with the emergence of 
new Symanzik polynomial invariants  for tensor graphs
\cite{Geloun:2014ema}.  Moreover,  the computations of 
the perturbative  $\beta$-functions for $\phi^4$-
and $\phi^6$-like models were achieved in the UV
  \cite{BenGeloun:2012yk, BenGeloun:2012pu,Carrozza:2012uv,
Carrozza:2013wda,Carrozza:2014rya,Carrozza:2014rba,Rivasseau:2015ova}.
The perturbative results \cite{BenGeloun:2012pu, BenGeloun:2012yk, Rivasseau:2015ova} suggested a generic asymptotic freedom for tensor field theories.  
This result was somehow surprising at first as they are not gauge theories, however, the (combinatorially) non-local nature of the tensor interactions drives the presence 
of a non-trivial wave-function renormalization, which then eventually dominates the renormalization of coupling constants. 
Afterwards, careful (perturbative) studies on $\phi^6$-like 
models hinted that the asymptotic safety may be possible in GFT \cite{Carrozza:2014rba,Carrozza:2014rya}. As a consequence, this last result strongly 
prompted that $\phi^6$ theories could have a more complicated behavior in the UV 
even for tensor field theories and the fact that asymptotic freedom might not hold
for these particular models.

The perturbative renormalization reveals  interesting UV properties
for tensor field theories which were encouraging to proceed to the next level. 
The exact renormalization group equations via Polchinski \cite{Krajewski:2015clk,Krajewski:2016svb} and via Wetterich (Functional Renormalization Group (FRG))
equations were fruitfully applied in all rank $d\ge 2$ matrix and tensor models
with compelling corroborations on the existence of Gaussian and non-Gaussian 
UV and IR points 
 \cite{Eichhorn:2014xaa,Benedetti:2014qsa,Benedetti:2015yaa,Geloun:2016xep,Carrozza:2016tih,Eichhorn:2017xhy,Geloun:2016qyb,Carrozza:2017vkz}. 
 Within the ordinary consistency checks on the FRG methods ({\it i.e.,} extensions
 of the truncation at higher orders and a change of the theory regulator),
non-perturbative calculations show that several $\phi^4$ models
 are asymptotically free and a $\phi^6$ model is asymptotically safe
 \cite{Benedetti:2014qsa,Geloun:2015qfa,Geloun:2016qyb}. 
In the GFT setting, similar conclusions were  reached 
 using the same tools with an extension of the truncation \cite{Carrozza:2016tih,Carrozza:2017vkz}. 
Hence, we conclude with a certain degree of confidence that, generically 
 in tensor field theory, renormalizable $\phi^4$ models are
 UV asymptotically free, and renormalizable $\phi^6$ models are
UV asymptotically safe.  
 
 The notable UV behavior of renormalizable tensor field theories is only one interesting
 aspect among other results brought by the FRG analysis. Another
 result concerns strong evidences for the existence of infrared (IR) fixed points.  
 For  tensor field theories,
the existence of a IR fixed point could play an important role. 
 Indeed, one aim of the FRG program 
is to identify the phase portrait of field theories. Stable IR  and UV fixed 
points define complete trajectories which allow to distinguish different regimes
of the theory, in other words, the existence of such trajectories 
could provide evidences for phase transitions in the models. 
A known mechanism characterizing a phase transition in ordinary field theory 
is spontaneous symmetry breaking. 
In fact, from preliminary calculations  in \cite{Benedetti:2014qsa,Geloun:2016qyb}, the phase diagrams of some 
tensor field theories show a IR fixed point which is similar to 
the Wilson-Fisher fixed point of a scalar field theory 
 (it  however occurs in different dimensions). This would likely  imply 
 that there is a phase transition in tensor field theory.
 If one shows that this phase transition results from a spontaneous symmetry breaking
 in these models, this transition will be described 
 in terms of a symmetric phase and a broken or a condensed phase. 
The broken phase would correspond to a new vacuum state 
corresponding to some geometry, characterized by a non-zero expectation value of the field.  This may validate the scenario in which homogeneous and isotropic geometries emerge as a condensate in GFT \cite{gielen}.

In this paper, 
we undertake the study of the theory space of enhanced tensor field theories by
addressing the perturbative renormalization of classes of enhanced models. 
We study tensor field models with quartic melonic interactions
with a momentum weight mimicking derivative couplings. 
The effect of the new couplings is to make the non-melonic graph amplitudes larger 
than or as large as the melonic graph amplitudes. 
Note that derivative couplings are well established in ordinary 
 renormalizable quantum field theory  {\it e.g.} appearing as in non-Abelian Yang-Mills  theories. The issue addressed in this work is to find a class of renormalizable theories 
endowed with non-local and weighted interactions. 
As one can expect, the presence of these interactions  bears  
additional subtleties as it naturally tends to increase the divergence degree of a graph. 

The  enhanced models that we study radically differ from that of \cite{Bonzom:2015axa} and \cite{Carrozza:2017vkz}, as we do not enhance non-melonic interactions of the necklace type but  melonic interactions. 
We could apply the same idea of derivative-type couplings over necklaces and expect that the resulting kind of enhanced tensor field theories to be closely related to the one of the above references. 

Specifically, we focus on  $\phi^4$-melonic couplings
 which are endowed with extra powers of momenta 
$|p|^{2 a}$; we call the resulting models $p^{2a}\phi^4$-models, where $a\geq 0$ is a parameter. 
The study is put on a very general ground, at any rank $d$
of the tensor field defined on a Abelian group of dimension $d\times D$. 
The propagator of the model is of the form 
$(\sum |p|^{2b} + \mu)^{-1}$, 
where $b>0$. 
Hence our model is parametrized by $(d,D, a, b)$. The case $a=0$ stands for the standard tensor field theory. Initially proposed by \cite{Geloun:2015lta}, these  models 
were found tractable at fixed ranks $d=3,4$, $D=1$, and $b=1$, and there were indications of their super-renormalizability
without a full-fledge proof of this statement. 
We carry on detailed analyses for these models, extending them at any rank
and any dimension. The method that we use is the so-called multi-scale renormalization
\cite{Rivasseau:1991ub}. It proves to be efficient enough to address non-local 
field theories (like tensor field theories) by achieving a perturbative 
power counting theorem and then the renormalization at all orders. 
Using the multi-scale analysis, we then find conditions 
on the tuple $(d,D,a,b)$ for potentially renormalizable enhanced models 
of two different types. 

- For the first type of theory, 
quite remarkably, we show that for generic $(d,D)$ parameters, there exists a 
just-renormalizable model at all orders.  Theorem \ref{theoren+} summarizes this result.  

- For the second type of theory, 
we prove the renormalizability  at all orders of a  specific model for a  
choice of parameters. Theorem \ref{theoremx} is another main result of our analysis. 

The plan of the paper is as follows:
In section \ref{sect:actio}, we introduce two models:  the model $+$ and the model $\times$ with different enhancements  in the $\phi^4$-tensor interactions.
In section \ref{sect:perturb}, preparing for the power counting analyses, we give an explicit expression for the amplitudes of a given graph $\cG$.
Section \ref{sect:pow} addresses the multi-scale analysis: we optimally bound 
a generic graph amplitude in terms of  combinatorial quantities of the graph.
In section \ref{sect:enhancedmelon}, we determine the parameter spaces of $(d,D, a, b)$ which could potentially give rise to renormalizable models $+$ and $\times$. 
Concretely, we investigate further instances of renormalizable  models:
 (1) section \ref{sect:renmo+} presents a generic model $+$ with arbitrary $D$, $d$, $a= D(d-2)/2$, and $b = D(2d-3)/4$;  (2) section \ref{sect:renmox} addresses a model $\times$ with $D=1$, $d=3$, $a=1/2$ and $b=1$. 
We prove that these models determined by such parameters are indeed renormalizable at all orders of perturbation theory. 
We give a summary of our results and future prospectives in section \ref{concl}. 
Closing the manuscript, in appendix \ref{app:sums}, the reader will find the detail of the spectral sums to be used for bounding the amplitudes, and 
appendices   \ref{app:mod+} and \ref{app:modx} respectively 
illustrate some representative and divergent graphs appearing in specific models $+$
and $\times$.

\section{Enhanced $p^{2a}\phi^4$ tensor field theories}
\label{sect:actio}

We consider a field theory defined by a rank $d$ complex  tensor $\phi_{\bf P}$, with ${\bf P}=(p_1,p_2,\dots,p_d)$ a multi-index, and $\bar\phi_{\bf P}$ denotes its complex conjugate. 
From a field theory standpoint, introducing a complex function $\phi: (U(1)^{D})^{\times d} \to \C$, where $D$ will be called dimension of the group $U(1)^D$, $\phi_{\bf P}$ is the Fourier component of the field and the indices $p_s$ are by themselves multi-indices: 
\be
p_{s} =(p_{s,1},p_{s,2},\dots,p_{s,D}) \,,\; \, p_{s,i} \in \Z\,. 
\ee
Let us make a few remarks. First, considering $\phi_{\bf P}$
as a rank $d$ tensor is a slight abuse because the modes $p_{k,s}$ range
up to infinity.  
Cutting sharply off all modes to $N$, then the resulting multi-index
object $\phi_{{\bf P};N}$
transforms under the fundamental representation of $U(N)^{D\times d}$
and hence is a tensor.  For convenience, we keep the name of tensor 
 for the field $\phi_{{\bf P}}$. 
Second, $\phi_{\bf P}$ could be considered as 
a $d\times D$ multi-index tensor, we shall call it a rank $d$ tensor, 
because  $d$ and $D$ will play different roles in the following. 
Third, several of the results derived hereafter could be extended to any 
compact Lie group $G_D$ of dimension $D$ 
admitting a Peter-Weyl decomposition  (see, for instance, how
a treatment for $SU(2)^{D'}$, $D=3D'$, can be achieved using 
tools in \cite{Geloun:2013saa}). 
 The treatment of the corresponding models
could have been achieved with some extra work.   
Finally, a last remark is that the dimension $D$
has nothing to see with the space dimension associated with 
the discrete geometry encoded by the tensor contractions
as we will discuss soon. Thus referring in the following
to UV and IR should be related with small and large 
distances on the group $U(1)^D$.

A general action  $S$ built by a sum of convolutions of the tensors $\phi_{\bf P}$ and $\bar\phi_{\bf P}$ can be written as:  
\bea\label{eq:actiond}
&&
S[\bar\phi,\phi]=\Tr_2 (\bar\phi \cdot 
{\bf K}
\cdot \phi) 
+ \mu
\, \Tr_2 (\phi^2) + S^{\inter}[\bar\phi,\phi]\,, 
\cr\cr
&&
\Tr_2 (\bar\phi \cdot 
{\bf K}
\cdot \phi) =
\sum_{{\bf P}, \, {\bf P}'} \bar\phi_{{\bf P}} \, 
{\bf K}({\bf P};{\bf P}')
\, \phi_{{\bf P}' } \,, 
\qquad 
\Tr_{2}(\phi^2) = \sum_{{\bf P}} \bar\phi_{{\bf P}}\phi_{{\bf P}}\,, 
 \cr\cr
&& S^{\inter}[\bar\phi,\phi]=  \sum_{n_b} 
\lambda_{n_b}  \Tr_{n_b}(\bar\phi^{n_b}\cdot 
{\bf V}_{n_b}
\cdot \phi^{n_b})\,,
\eea
where $\Tr_{n_b}$ are sums over all indices $p_{k,s}$ of ${\bf P}$
of $n_b$ tensors $\phi$ and $\bar\phi$. Then $\Tr_{n_b}$ are considered as traces over indices of the tensors. 
In \eqref{eq:actiond},  the kernels 
${\bf K}$
and   
${\bf V}_{n_b}$
are to be specified,
 $\mu$ is a mass coupling and $\lambda_{n_b}$
is a  coupling constant. If 
${\bf V}_{n_b}$
corresponds to a simple pairing between 
tensor indices (by delta functions identifying indices), then 
$  \Tr_{n_b}(\bar\phi^{n_b}\cdot  {\bf V}_{n_b} \cdot \phi^{n_b})$ spans the space of unitary invariants \cite{Gurau:2011tj, Gurau:2012ix,Bonzom:2012hw}. 

There is a geometrical interpretation of the interaction  
$\Tr_{n_b}(\bar\phi^{n_b}\cdot 
{\bf V}_{n_b}
\cdot \phi^{n_b})$. 
If each tensor field is regarded as a $d$-simplex, 
the generalized trace $\Tr_{n_b}$ corresponds to a pairing or an identification 
of the $(d-1)$-simplices on the boundary of the $d$-simplexes to form a $d+1$ dimensional discrete geometry. If the kernel 
${\bf V}_{n_b}$ 
 is not a simple pairing, it 
 then assigns a weight to each of those discrete geometries.

A model is specified after giving the data of the kernels 
${\bf K}$
and   
${\bf V}_{n_b}$.
Let us  introduce some convenient notations: 
\bea\label{delmom}
&&
\bdel_{ { \bf P}; {\bf P'} } =    \prod_{s=1}^d\prod_{i=1}^D  \delta_{ p_{s,i}, p'_{s,i}}  \,,\qquad
{\bf P}^{2b} = \sum_{s=1}^d |p_s|^{2b}\,,  \qquad  |p_{s}|^{2b} = \sum_{i=1}^D |p_{s,i}|^{2b} \,,\cr\cr
&&
\phi_{12\dots d} = \phi_{p_1,p_2,\dots, p_d} = \phi_{\bf P} \,.
\eea
for a real parameter $b\geq 0$, and where $ \delta_{ p,q}$ is the usual Kronecker symbol on $\Z$. 

We introduce the following class of kernels for the kinetic term 
\beq
\label{eq:3dkin}
{\bf {K}}_b({ \bf P}; {\bf P'} ) 
=\bdel_{ { \bf P}; {\bf P'} } {\bf P}^{2b}   \,.
\eeq
${\bf K}_b$ 
therefore represents 
a sum of the power of eigenvalues of  $d$ Laplacian operators over the $d$ copies of $U(1)^D$.
The case $b=1$  corresponds precisely to Laplacian 
eigenvalues on the torus. Seeking renormalizable theories, from the fact that we are dealing with a nonlocal model, we might be led to choose values of $b$ different from integers.  In usual quantum field theory (QFT)
$b$ should have an upper bound $b\leq 1$ to ensure the Osterwalder-Schrader (OS)
positivity axiom \cite{Rivasseau:1991ub}. Whether or not such 
a condition (or any OS axioms) might be kept for  tensor field theories is still in debate \cite{Rivasseau:2011hm}.   
Thus, for the moment, to avoid putting strong constraints on the models, 
we let $b$ as a free strictly positive real parameter.

 We will be interested in 2 models distinguished  by their interactions. 
Introduce a parameter $a\in (0,\infty)$ and write:
\bea
&& 
\Tr_{4;1}(\phi^4) = \sum_{p_{s},  p'_{s} \in \Z^D} 
\phi_{12\dots d} \,\bar\phi_{1'23\dots d} \,\phi_{1'2'3'\dots d'} \,\bar\phi_{12'3'\dots d'} \,, 
\label{phi4sim}\\
&&
\Tr_{4;1}([p^{2a}+p'^{2a}]\,\phi^4) = \sum_{p_{s}, p'_{s} \in \Z^D} 
\Big( |p_{1}|^{2a} + |{p'}_{1}|^{2a}\Big)\phi_{12\dots d} \,\bar\phi_{1'23\dots d} \,\phi_{1'2'3'\dots d'} \,\bar\phi_{12'3'\dots d'}\,, \cr\cr
&& = 
 2\sum_{p_{s}, p'_{s} \in \Z^D} 
 |p_{1}|^{2a}\,\phi_{12\dots d} \,\bar\phi_{1'23\dots d} \,\phi_{1'2'3'\dots d'} \,\bar\phi_{12'3'\dots d'} = 2\,\Tr_{4;1}(p^{2a}\,\phi^4)  \,,
\label{intplus}  \\ 
&&
\Tr_{4;1}([p^{2a}p'^{2a}]\,\phi^4) = \sum_{p_{s}, p'_{s} \in \Z^D} 
\Big( |p_{1}|^{2a}  |{p'}_{1}|^{2a}\Big)\phi_{12\dots d} \,\bar\phi_{1'23\dots d} \,\phi_{1'2'3'\dots d'} \,\bar\phi_{12'3'\dots d'} \,.
\label{intprod} 
 \eea
Note that in \eqref{phi4sim}, \eqref{intplus} and \eqref{intprod}, the color index 1 plays
a special role. We  sum over all possible
color  indices  and obtain colored symmetric  interactions: 
\bea
\label{eq:3dinter}
&& 
 \Tr_{4}(\phi^4)
 :=  \Tr_{4;1} (\phi^4) + \Sym (1 \to 2 \to \dots \to d) 
\,, \cr\cr
&& 
\Tr_{4}(p^{2a}\,\phi^4)
 :=  \Tr_{4;1} (p^{2a}\,\phi^4)+ \Sym (1 \to 2 \to \dots \to d) \,, \cr\cr
&& 
\Tr_{4}([p^{2a}p'^{2a}]\,\phi^4)
 :=  \Tr_{4;1} ([p^{2a}p'^{2a}]\,\phi^4)+ \Sym (1 \to 2 \to \dots \to d) \,. 
\eea
The momentum weights 
in the interactions $\Tr_{4}(p^{2a}\,\phi^4)$ and $\Tr_{4}([p^{2a}p'^{2a}]\,\phi^4)$
can be viewed as derivative couplings for particular choices
of $a$. 
This is why, at times, we will call them coupling derivatives. 
Written in the momentum space, the interactions  are however put in a more general setting using $|p|^{2a}$, for positive values of $a$. Once again,  achieving renormalizability will be our sole constraint for fixing $a$.  
These interactions are called  enhanced compared to $\Tr_{4}(\phi^4)$ (the usual quartic melonic graph studied for instance in \cite{BenGeloun:2012pu}) because they can generate amplitudes which are
more divergent, and so enhanced, compared to those generated by 
$\Tr_{4}(\phi^4)$ alone.  
As a second property, we discussed that enhanced interactions represent
weighted discrete geometries. The contraction pattern of the four tensors
shows us that the weight here has a subtle sense:
we are weighting a particular $(d-1)$-simplex in the $(d+1)$-simplex
representing the interaction.

 It turns out that the renormalization analysis performed in sections 
\ref{sect:renmo+} and \ref{sect:renmox}
leads  us to new 2-point diverging graphs. Then we must add to the kinetic term
the new terms: 
\bea
 \Tr_2 (p^{2\xi}\phi^2) = 
\Tr_2 (\bar\phi \cdot 
{\bf K}_{\xi}
\cdot \phi)\,, \qquad \xi = a,2a\,,
\eea
in addition to the kinetic term $\Tr_2 (p^{2\xi}\phi^2)$, where $\xi = b$.

We will need counter-terms for each term in 
the action. In particular, the counter-term $CT_{2}$ of the form of the mass, 
 $CT_{2;b}$ for the wave function, and new 2-point interactions $CT_{2;a}$ and $CT_{2;2a}$, that will be important for renormalizing two-point functions. We define
\be\label{counter}
CT_{2}[\bar\phi,\phi]  =\delta_{\mu}\Tr_2(\phi^2) \,, \;\,
CT_{2;\xi}[\bar\phi,\phi]  
= Z_{\xi}\Tr_2 (p^{2\xi}\phi^2)
 \,, \;\,  \xi=a,2a,b
\,, \ee 
where $\delta_\mu$ and  $Z_\xi$ are counter-term couplings. 
Note that, in the following, $Z_b$ is called wave function renormalization. 

The models that we will study have the following kinetic terms and  interactions: 
\bea
\text{model }+: 
&&
S^{\inter}_+[\bar\phi,\phi] =  \frac{ \lambda}{2}\,\Tr_{4}(\phi^4)  
 +\frac{\pet}{2}\, \Tr_{4}(p^{2a}\,\phi^4)
  +CT_{2}[\bar\phi,\phi] + \sum_{\xi=a,b}CT_{2;\xi}[\bar\phi,\phi] 
 \cr\cr
&& 
S^{\kin}_+[\bar\phi,\phi] =  
\sum_{\xi=a,b}  \Tr_2 (p^{2 \xi} \phi^2) 
 + \mu \Tr_2 (\phi^2) \,,
\label{model1}\\
\text{model }\times: &&
S^{\inter}_\times[\bar\phi,\phi] =   \frac{ \lambda}{2}\,\Tr_{4}(\phi^4)  
 +\frac{\tet}{2}\, \Tr_{4}([p^{2a}p'^{2a}]\,\phi^4)
 + CT_{2}[\bar\phi,\phi] +\sum_{\xi=a,2a,b}CT_{2;\xi}[\bar\phi,\phi]  \,,
 \cr\cr
&& 
S^{\kin}_\times[\bar\phi,\phi] =  
\sum_{\xi=a,2a,b}  \Tr_2 (p^{2 \xi} \phi^2) 
 + \mu \Tr_2 (\phi^2)
\label{model2}
\eea
where $\lambda$, $\pet$ and $\tet$ 
are coupling constants.  

It is an interesting question to list the classical symmetries of the models
$+$ and $\times$ given by the generalized Noether theorem for such non-local theories 
 \cite{Kegeles:2016wfg,Kegeles:2015oua}. To apply the  Lie symmetry algorithm 
 as worked out in these references could be an interesting exercise for 
derivative coupling theories and could bear important 
consequences for the Ward identities.

The present theory space is clearly much more involved than the usual unitary invariant theory space where the vertices of the model do not have any momentum weight. 
It will result from our analysis that a new combinatorics provides our models  
with a genuinely different renormalization procedure. 
Then, the comparison could be made with the models in Table 8 in \cite{Geloun:2013saa} 
which are unitary invariant models. We seize
this opportunity to correct that table: the just-renormalizable $\phi^6$-models should be UV asymptotically safe (rather than free) under the light of many recent results  
\cite{Carrozza:2014rba,Carrozza:2014rya,Geloun:2016xep,Eichhorn:2017xhy, Carrozza:2016tih,Carrozza:2017vkz}.

In \cite{Geloun:2015lta}, a power counting theorem was proved for
the model $+$ restricted at rank $d=3$ and $d=4$ and $D=1$. Nevertheless, the optimization procedure to reach a power counting was quite complicated. 
There  were indications of  potentially super-renormalizable 
enhanced models without finalizing the proof of such a renormalizability.
In this work, we will improve that analysis by noting 
that the relevant interaction is rather 
$\Tr_{4}(p^{2a}\,\phi^4)$ \eqref{eq:3dinter}. 
Before reaching this point, our next task is 
to  express generic amplitudes in the enhanced models.

\section{Amplitudes}
\label{sect:perturb}

Models $+$ and $\times$ associated with actions given 
by  \eqref{model1} and \eqref{model2}, respectively, give
the quantum models determined by the partition function
\be
Z_\bullet  = \int d\nu_{C_\bullet}(\bar\phi,\phi) \; e^{-S^{\inter}_{\bullet}[\bar\phi,\phi]}\,, 
\ee
where $\bullet =+,\times$, and $d\nu_{C_\bullet}(\bar\phi,\phi)$ is a field Gaussian measure with covariance $C_\bullet$
given by the inverse of the kinetic term:
 \be
C_\bullet({\bf P};{\bf P'}) =\tilde{C}_\bullet({\bf P})\, \bdel_{{\bf P},{\bf P}'}\,,\qquad
\tilde{C}_\bullet({\bf P})\,= \frac{1}{\sum_\xi {\bf P}^{2 \xi}+ \mu }\,. 
\ee 
where, if $\bullet =+$, $\xi=a,b$ and if $\bullet =\times$,  $\xi=a,2a,b$.
Dealing with the interactions, we have the vertex kernels 
${\bf V}_{4;s}$ 
and 
${\bf V}_{+;4;s}$ 
associated with \eqref{model1} and 
${\bf V}_{4;s}$ 
and 
${\bf V}_{\times;4;s}$ 
associated with \eqref{model2}. These kernels are given by 
\bea\label{vertexkernel}
&&
{\bf V}_{4;s}({\bf P };{\bf P}';{\bf P }'';{\bf P }''') 
= \frac\lambda2\delta_{4;s}({\bf P };{\bf P}';{\bf P }'';{\bf P }''')\,,\cr\cr
&&
{\bf V}_{+;4;s}({\bf P };{\bf P}';{\bf P }'';{\bf P }''') 
=  \frac{\eta_+}{2} |p_s|^{2a} \, \delta_{4;s}({\bf P };{\bf P}';{\bf P }'';{\bf P }''')\,,\cr\cr
&&
{\bf V}_{\times;4;s}({\bf P };{\bf P}';{\bf P }'';{\bf P }''')
= \frac{\eta_\times}{2} |p_s|^{2a}|{p'}_s|^{2a} \, \delta_{4;s}({\bf P };{\bf P}';{\bf P }'';{\bf P }''')\,,
\eea
$s=1,2,\dots, d$,  
where the operator $\delta_{4;s}(-)$ is a product of Kronecker deltas identifying 
the different momenta according to the pattern dictated by 
the interaction $\Tr_{4;s}(\phi^4)$. Note that 
${\bf V}_{\bullet; 4;s}$ 
has a color index.  The vertex operator  
${\bf V}_{2}$
associated with 
the mass counter-term is a delta function $\bdel_{ { \bf P}; {\bf P'} } $; 
the vertex operators 
${\bf V}_{2;\xi;s}$
$\xi=a,2a,b$,
 associated with the counter-terms  $CT_{2;\xi}[\bar\phi,\phi]$ 
are delta functions  weighted by momenta $|p_s|^{2\xi}$. 

\

\noindent 
{\bf Feynman tensor graphs.}
There are two equivalent graphical representations of Feynman 
graphs in tensor models. The first one is called ``stranded graph'' and it incorporates more details of the structure
of the Feynman graph (used and explained in \cite{color}
and \cite{Geloun:2013saa}). The other representation of a Feynman
graph in this theory is  a bipartite colored graph \cite{color,Gurau:2013cbh,Bonzom:2012hw,Gurau:2011xp}.
We  mostly use the latter  because
it is convenient and economic. The first representation will be used in this section
to make explicit the notion of faces associated with momentum loops. 
 
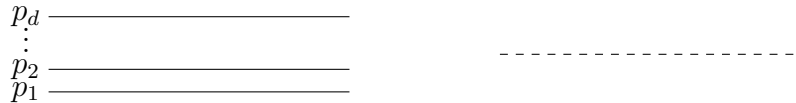
\begin{figure}[H]
\centering
     \begin{minipage}[t]{0.7\textwidth}
\begin{tikzpicture}
\node (r1) at (-0.3,0.0)  {$p_{1}$};  
\node (r2) at (-0.3,0.3)  {$p_{2}$};
\node (r1) at (-0.3,0.8)  {$\vdots$};  
\node (r3) at (-0.3,1)  {$p_{d}$};  
\draw (0,0.3) -- (4,0.3) ;
\draw (0,1) -- (4,1) ;
\draw (0,0) -- (4,0) ;
\draw[dashed] (6,0.5) -- (10,0.5);
\end{tikzpicture}
\caption{\small  The propagator of the theory:
the stranded representation (left) made with 
$d$ segments representing $d$ momenta; 
the colored representation (right) denoted by a dotted line. }
\label{fig:propacol}
\end{minipage}
\end{figure}

\begin{figure}[H]
 \centering
     \begin{minipage}{1\textwidth}
     \centering
\includegraphics[angle=0, width=14cm, height=6cm]{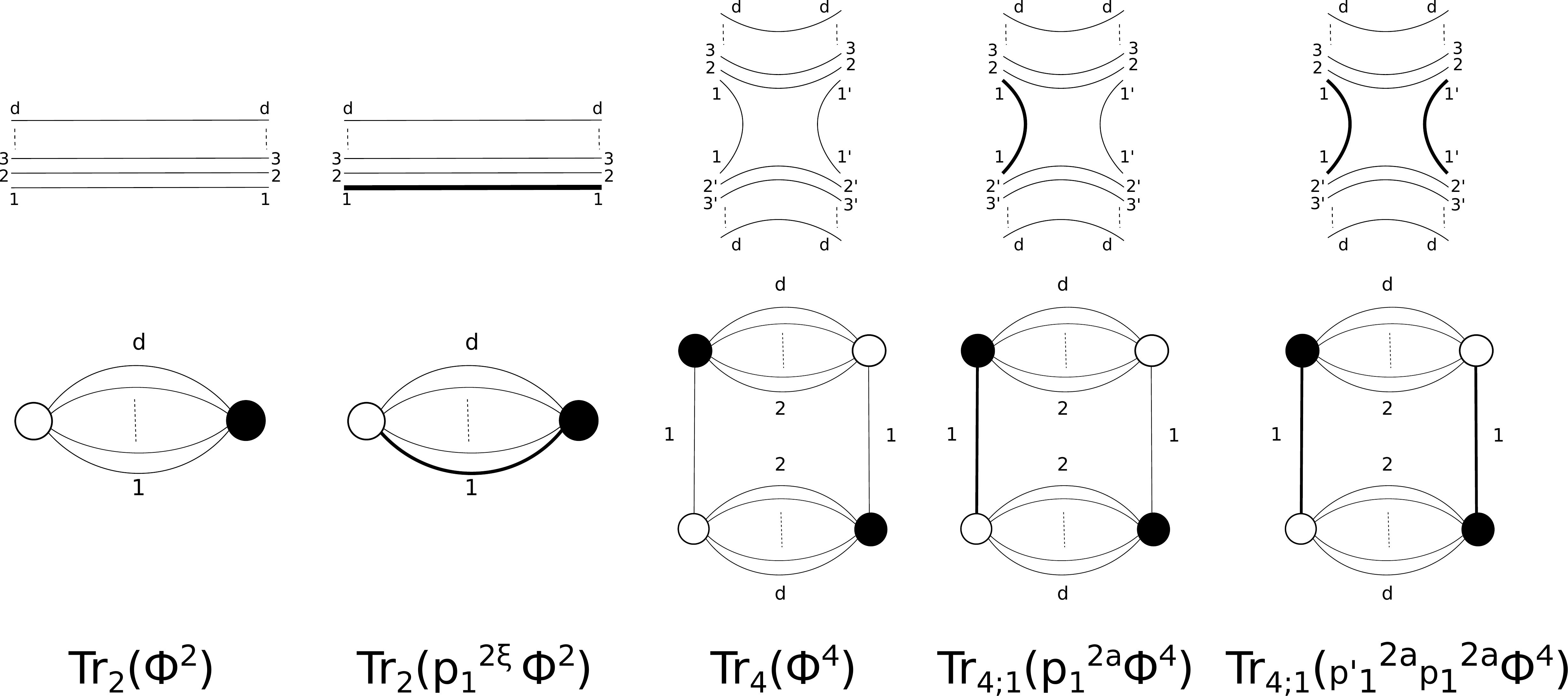}
\caption{ {\small Rank $d$  vertices of the mass, $\phi^2$- and $\phi^4$-terms.   
}} 
\label{fig:4vertex}
\end{minipage}
\end{figure}

At the graphical level the propagator is drawn as a collection of $d$ segments
called strands (see Figure \ref{fig:propacol}).
Each interaction is sketched as a stranded vertex 
or by a $d$-regular colored bipartite graph called a ``bubble.'' 
The bipartiteness of the graph comes from the representation
of each field $\phi$ as a white vertex and each field $\bar\phi$ by a black one. 
For instance, see bubbles corresponding to $\phi^2$ vertices (mass and wave functions vertices),
 and $\phi^4$-interactions in Figure \ref{fig:4vertex}. Note that, 
 the bubbles representing the vertex kernel 
${\bf V}_{\bullet;4;s}$, $\bullet=+,\times$, 
 appear with one or two bold edges, respectively. The color of
a bold edge corresponds to the color index of the
enhanced momentum.  
The bubbles that describe the vertices are particular contractions of tensors and
are called melons. 

\begin{figure}[H]
 \centering
     \begin{minipage}{1\textwidth}
     \centering
\includegraphics[angle=0, width=12cm, height=5cm]{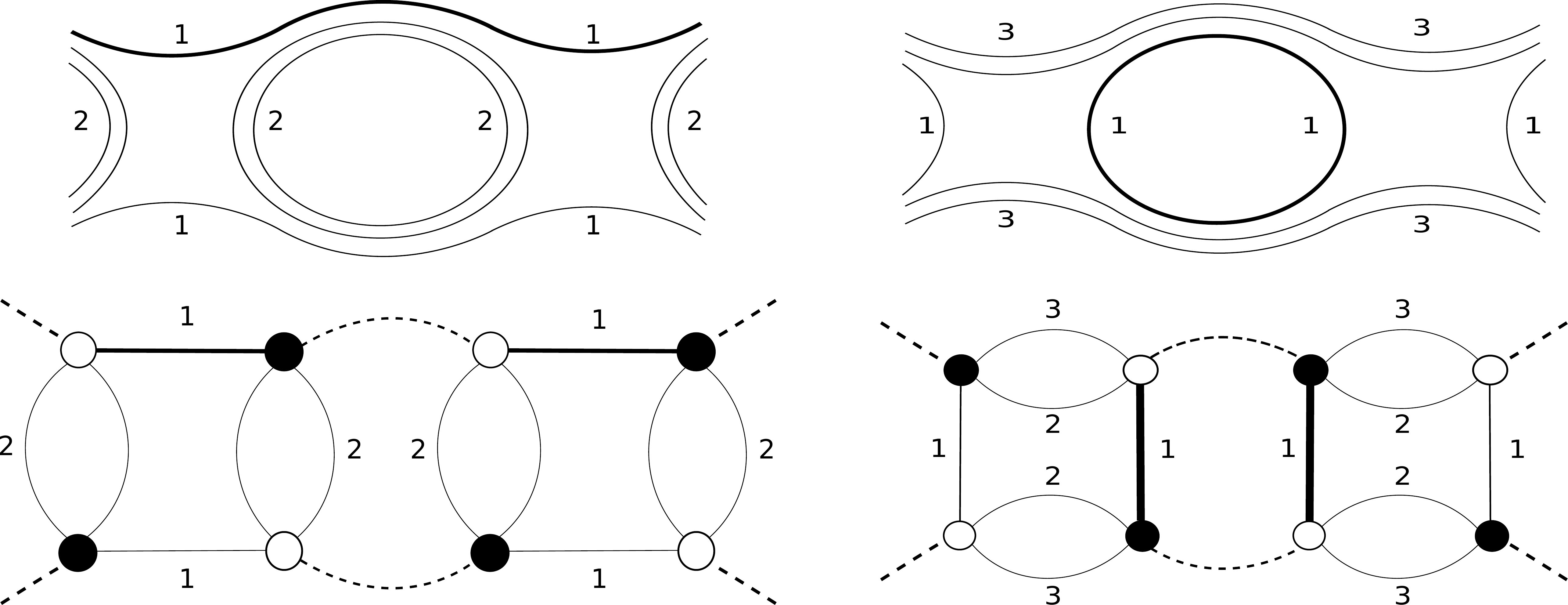} 
\caption{ {\small Rank $d=3$  Feynman graphs.   }} 
\label{fig:graphs}
\end{minipage}
\end{figure}

Perturbation theory 
tells us that, via  the Wick theorem, we should glue
vertices by propagator lines to produce a Feynman graph. 
Some examples of Feynman tensor graphs
by the above rule are depicted in Figure  \ref{fig:graphs}.
We put half-lines or external legs on vertices to reflect the presence
of external fields. In the following, a Feynman tensor graph is simply called a graph
and is denoted by $\cG$. 
In the stranded picture, closed cycles (homeomorphic to circles) in the 
graphs are called  closed or internal faces and  strands homeomorphic to segments are called open or external faces. The set of closed faces is denoted by $\cF_{\inter}$ and the set of open faces $\cF_{\ext}$.
As expected, the presence of an internal face is associated
with a sum over infinite values of momenta which can make the amplitude
divergent, hence the need of regularization and renormalization
for the model. 
In the colored graph representation, note that an extra color $0$
could be attributed to all dotted propagator lines.
The cycles in that $(d+1)$ colored graph 
have two colors. The internal faces of $\cG$, elements of $\cF_{\inter}$, 
are associated with bicolored cycles of colors $0s$, with  $s=1,2,\dots,d$.   
To obtain the subset of $\cF_{\inter}$ (or of $\cF_{\ext}$) of faces of colors $0s$ 
 from the $d+1$ colored graph, we remove all edges except those of colors $0$ and $s$
   and observe the remaining cycles (or open strands, respectively).  In the end, for simplicity, we omit the color
   0 in the couple $0s$ and claim that a (internal or external) face is of color $s$. 

\

\noindent{\bf Amplitudes.}
Given a connected graph $\cG$ with  vertex set $\cV$ (with $V=|\cV|$) and  line  or  propagator set  $\cL$  (with $L=|\cL|$), we formally write
the amplitude of $\cG$
\beq\label{ampli}
A_{\cG} = \sum_{{\bf P}_v} 
\prod_{l \in \cL } C_{\bullet;l} ( {\bf P}_{v(l)} ; {\bf P}'_{v'(l)}) 
\prod_{v \in \cV} (
- 
{\bf V}_{v}
(\{{\bf P}_{v}\})\,.
\eeq
The above formula shows that propagators $C_l$ have a line index $l$ and momentum arguments  ${\bf P}_{v(l)}$, with $v(l)$ the source or target of the
line $l$. The vertex constraints 
${\bf V}_{v}$
convolute the set 
of momenta and can be of the form 
${\bf V}_{4;s}$, ${\bf V}_{\bullet; 4;s}$, ${\bf V}_{2}$, ${\bf V}_{2;\xi;s}$, 
$\xi=a,2a,b$.  
 The presence of these weights makes the amplitude quite different from those of unitary invariant
 theories. For instance, as opposed to the ordinary situation, the amplitudes do not directly factorize in terms of internal faces.

To derive a power counting theorem we need to study graph amplitudes
 $A_{\cG}$ coming from the perturbative expansion of correlators
 of the form 
 \bea
 &&
\langle  \phi_{\bf P} \bar \phi_{\bf P'}\phi_{\bf P'} \bar\phi_{\bf P'''} \rangle \,, 
\label{phi4}\\ 
&&
\langle |p_{1}|^{2a}\, \phi_{\bf P} \bar\phi_{\bf P'}\phi_{\bf P'}\bar \phi_{\bf P'''} \delta_{4;s}({\bf P };{\bf P}';{\bf P }'';{\bf P }''')\rangle
\label{pPhi4+}\\
&&
\langle |p_{1}|^{2a}|p_{1'}|^{2a} \,\phi_{\bf P} \bar \phi_{\bf P'}\phi_{\bf P'} \bar\phi_{\bf P'''} \delta_{4;s}({\bf P };{\bf P}';{\bf P }'';{\bf P }''')\rangle \,.
\label{pPhi4x}
\eea

In  tensor graphs,
consider the faces as  previously introduced.  A face $f_s$ with color $s$ has an  $s$-colored $\Z^D$ conserved momentum 
$p_{f_s}$, and passes through some vertices $v_{s}$, with vertex kernel of the form 
${\bf V}_{4;s}$,  ${\bf V}_{\bullet;4;s}$, ${\bf V}_{2}$, or ${\bf V}_{2;\xi;s}$, 
$\xi=a,2a,b$. This face
 may also pass through some other vertices with color $s'\ne s$. 
More generally, a face $f$ can pass through a vertex $v$ a number of times, say $\alpha$. Denote this statement by $v^{\alpha} \in f$. 
Because of the coloring,  $\alpha$ can only be $0,1,2$
($v_s \in f_s $ will mean $v_s^{1}\in f_s$).  
We therefore define the incidence matrix between faces and
vertices by 
\beq\label{incivf}
\epsilon_{v_sf_{s'}} = \left\{ \begin{array}{cl} \alpha ,&  (s=s') \wedge (v_s^{\alpha} \in f_{s}) ,\\
0,& \text{otherwise.} \end{array}
\right. 
\eeq 
Given two faces $f_{1;s_1}$ and $f_{2;s_2}$ and 
a vertex $v_s$,  we introduce another multi-index object that we denote by $\epsilon_{v_sf_{1;s_1}f_{2;s_2}}$ defined as
\beq\label{inctens}
\epsilon_{v_sf_{1;s_1}f_{2;s_2}}= \left\{ \begin{array}{cl} 1,&  (s=s_1=s_2) \wedge  (v_s  \in f_{1;s_1}) \wedge \; (v_s  \in f_{2;s_2}),\\
0,& \text{otherwise.} \end{array}
\right.
\eeq
The case $f_{1;s_1} = f_{2;s_2}$ could also occur. 
A first observation is that $\epsilon_{v_sf_{1;s_1}f_{2;s_2}} = 
\epsilon_{v_sf_{s_1}}\epsilon_{v_sf_{s_2}}$ in the case when $v_s \in f_{1;s_1}$
and $v_s \in f_{2;s_2}$.  Looking 
at the diagonal, {\it i.e.} $f_{1;s_1}=f_{2;s_2}$, $\epsilon_{v_sf_{1;s}f_{1;s}} =1$  if and only
if  $v^2_s \in f_{1;s}$. 

We are in position to re-express the interaction weights 
${\bf V}_{\bullet;4;s}$ 
in \eqref{vertexkernel}. 
Fix a color $s$, the weight of a vertex kernel of $v_{s}$ of the kind 
${\bf V}_{\bullet;4;s}$ 
can be written as 
\bea
\text{model }+: &&\frac{\eta_+}{2}  \sum_{f_{s'}} \epsilon_{v_s,f_{s'}}\, |p_{f_{s'}}|^{2a} \, , \cr\cr
\text{model }\times: &&
\frac{\eta_\times}{2}  \sum_{f_{s'},f_{s''}} \epsilon_{v_s,f_{s'},f_{s''}}\, |p_{f_{s'}}|^{2a}\, 
|{p}_{f_{s''}}|^{2a} .
\eea
We stress, at this point, that the two models $+$ and $\times$ will be studied  
separately, then there is no confusion to adopt a single notation as:
\bea
\text{model}\; \bullet:\qquad \frac{\eta }{2} 
 (\epsilon \, p )_{v_s} \,. 
\eea
 The weight of degree 2 vertices (in both models) which
 belong to  $\cV_{2;\xi;s}$ is of the form 
$Z_{\xi} \sum_{f_s}\epsilon_{v_s,f_{s'}}|p_{f_{s'}}|^{2\xi }=Z_\xi  (\epsilon \, p )_{v_s}$, 
where $\xi=a,2a,b$.

 Let us introduce:
 
 -  the set $\cV_{4;s}$ of vertices 
with kernel  
${\bf V}_{4;s}$, 
$\cV_4 = \sqcup_{s=1}^d \cV_{4;s}$ (disjoint union notation), 

-  the set $\cV_{\bullet; 4;s}$ of vertices with vertex 
kernel 
${\bf V}_{\bullet;4;s}$,
$\bullet = +, \times$, 
$\cV_{\bullet;4}= \sqcup_{s =1}^d\cV_{\bullet; 4;s}$,

- the set $\cV_2$  of mass vertices with kernel 
${\bf V}_{2}$,
 the set  $\cV_{2;\xi;s}$ of vertices with kernels 
${\bf V}_{2;\xi;s}$,
 $\cV_{2;s} = \cup_{\xi} \cV_{2; \xi;s}$. 
  
We denote
the cardinalities
$|\cV_{4;s}|=V_{4;s}$,
$|\cV_{4}|=V_4$,
$|\cV_{\bullet;4;s}|=V_{\bullet;4;s}$, 
$|\cV_{\bullet;4}|=V_{\bullet;4}$, 
$\bullet=+,\times$; 
$|\cV_{2;\xi;s}|=V_{2;\xi;s}$,  $V_{2;\xi}= \sum_s V_{2;\xi;s}$. 
Then, 
$\cV = \sqcup_{s =1}^d (\cV_{4;s} \cup \cV_{\bullet;4; s} \cup \cV_{2;s})$,
$|\cV|=V$. 

Using the Schwinger parametric form of the propagator kernel as
\beq
\tilde C_\bullet({\bf P}) = \int_0^{\infty} d\alpha \; e^{-\alpha(\sum_\xi{\bf P}^{2 \xi} + \mu
)}\,, 
\eeq
integrating all deltas from  propagators and vertex
operators, we put  the amplitude \eqref{ampli} in the form 
\bea\label{amplf}
A_{\cG} &=& 
\kappa(\lambda, \bet, Z_{\xi})
\sum_{p_{f_s}} 
\int \Big[\prod_{l\in \cL} d\alpha_l\, e^{-\alpha_l \mu
} \Big]\; 
\Big[\prod_{f_s\in \cF_{\ext}} e^{-(\sum_{l\in f_s} \alpha_l)
\sum_\xi |p^{\ext}_{f_s}|^{2 \xi}}\Big] \cr\cr
&& 
\times \Big[
\prod_{f_s\in \cF_{\inter}} e^{-(\sum_{l\in f_s} \alpha_l)
\sum_\xi |p_{f_s}|^{2 \xi}}\Big]
 \Big[\prod_{s=1}^{d} \prod_{v_s \in \cV_{\bullet;4; s} \cup \cV_{2;s}} (\epsilon\,\tilde{p})_{v_s} \Big] \,, 
\eea
where $\kappa(\lambda,  \bet,Z_{\xi})$ includes symmetry factors
and coupling constants, $p^{\ext}_{f_s}$ are external momenta which are not summed, whereas $p_{f_s}$ are  internal momenta  and are
 summed. In the last line, 
$\tilde{p}_{f_s}$ refers to an internal or an external momentum. 
The sum over infinite values of momenta produces divergent amplitudes \eqref{amplf}. 
In the next section, we will address the nature of these divergences
through a power counting theorem.

\section{Power counting theorems for $p^{2a}\phi^4$-models}
\label{sect:pow}

For simplicity, we will study a connected graph amplitude without $\cV_{2;\xi;s}$ vertices.
To add these vertices towards the end can be easily done. 

\

\noindent{\bf Multiscale analysis.}
We slice the propagator in a geometric progression with 
the parameter $M>1$, and then bound each slice of the propagator:
\bea\label{bounds}
&&
\tilde C_\bullet ({\bf P}) = \int_0^{\infty} d\alpha\; e^{-\alpha( \sum_\xi {\bf P}^{2\xi} + \mu)}
 = \sum_{i=0}^\infty C_{\bullet;i}  ({\bf P})\,, \cr\cr
&&
C_{\bullet; 0(}{\bf P})  = \int_{1}^{\infty} d\alpha\;
e^{-\alpha(\sum_\xi {\bf P}^{2\xi}  + \mu)} \leq K\,,\cr\cr
&&
C_{\bullet; i}  ({\bf P})= \int_{M^{-2(i+1)}}^{M^{-2i}} d\alpha\;
e^{-\alpha( \sum_\xi {\bf P}^{2\xi} + \mu)} 
\leq K' M^{-2i}  
e^{- M^{-2  i}( \sum_\xi {\bf P}^{2 \xi}  + \mu)}\cr\cr
&&
\leq 
K M^{-2i}  \;
e^{- \delta M^{-i}(\sum_\xi \sum_{s=1}^d \sum_{l=1}^D|p_{s;l}|^{\xi} + \mu)} 
\leq K M^{-2i}  \;
e^{- \delta M^{-i}(\sum_\xi \sum_s |p_s|^{\xi} + \mu)} \,,
\eea
$|p_{s}|^\xi = \sum_{l=1}^D|p_{s;l}|^{\xi}$, 
for some constants $K$, $K'$ and $\delta$. 

The slice decomposition  yields the standard interpretation that  high values of $i$ select high 
momenta of order $\sim M^{i}$ and this refers to the UV
(this coincides with small distances on the group $U(1)^D$). Meanwhile, low momenta are picked around the slice $i=0$, and correspond to the IR. Note that, since we are dealing
with a compact group, the latter limit is harmless. 
Introduce a cut-off $\Lambda$ on the slices $i$, and 
then cut off the propagators as $C_\bullet ^{\Lambda}=\sum_{i=0}^{\Lambda} C_{\bullet; i}$. 
We will not display $\Lambda$ in the following expression.  

Cutting off all propagators in \eqref{ampli},  the amplitude $A_{\cG}$  becomes $\sum_{\bmu} A_{\cG;\bmu}$ where $\bmu=\{i_l\}_{l\in \cL}$ is a multi-index called momentum assignment  which  collects the  propagator indices $i_l \in [0,\Lambda]$, and 
\be
A_{\cG;\bmu} =\kappa(\lambda,\eta_{\bullet})\sum_{p_{v;s}} 
\Big[\prod_{l \in \cL }C _{\bullet; i_l} ({\bf P}_{v(l)}; {\bf P}'_{v'(l)}) \Big]
\Big[\prod_{s=1}^{d} \prod_{v_s \in \cV_{\bullet;4; s}} (\epsilon\,\tilde{p})_{v_s} \Big]  .
\ee
Using \eqref{bounds}, the above expression finds the form
\bea
&&
|A_{\cG;\bmu}|\leq  \kappa(\lambda,\eta_{\bullet})
K^L
K_1^V  \; K_2^{F_{\ext}} \Big[ \prod_{l \in \cL} M^{-2 i_l} \Big]\cr\cr
&& \times 
\sum_{p_{f_s}} 
\Big[
\prod_{f_s\in \cF_{\inter}} e^{-\delta(\sum_{l\in f_s} M^{-i_l})\sum_\xi |p_{f_s}|^{\xi}} \Big]
\Big[\prod_{s=1}^{d} \prod_{v_s \in \cV_{\bullet;4; s}} (\epsilon\,\tilde{p})_{v_s} \Big],
\label{amplinit}
\eea
where  $K_{1,2}$ are constants. 

$A_{\cG;\bmu }$ is the focus of our attention and
is the quantity that  must be bounded by an optimization procedure.  
A standard procedure detailed in \cite{Rivasseau:1991ub}
will allow one to sum over the assignments $\bmu$
 after renormalization.

The next definition can be found in  \cite{Rivasseau:1991ub}.
It paves the way to the notion of locality of the theory
through the definition of quasi-local subgraphs.  
Let $\cG$ be a graph,  with  line set $\cL$.
Fix $i$   a  slice index and define $\cG^i$ to be the subgraph of
$\cG$ built with propagator lines with indices obeying  $\forall \ell \in \cL(\cG^i)\cap\cL$, $i_\ell \geq i$. It might
happen that $\cG^i$ disconnects in several components; we denote these
connected components $G^i_{k}$ and call them  quasi-local subgraphs. 
It is important to give a characterization of the quasi-local
subgraphs. Given $g$, a subgraph of $\cG$ 
with   internal line set $\cL(g)$ and   external line set $\cL_{\ext}(g)$. 
Consider a momentum assignment $\bmu$ of $\cG$,
and define $i_{g}(\bmu)=\inf_{\ell \in \cL(g)} i_\ell$ and $e_{g}(\bmu)=\sup_{\ell \in \cL_{\ext}(g)}i_\ell$. 
We can identify $g$ with a  quasi-local subgraph of $\cG$ if and only if 
$i_{g}(\bmu)>e_{g}(\bmu)$. 

The set  $\{G^i_k\}$ of quasi-local subgraphs of $\cG$ 
 is partially ordered under inclusion. The inclusion can be put
in a form of an abstract tree   (with vertices the $G^i_k$'s)
 called the Gallavotti-Nicol\`o
(GN) tree \cite{Galla}.  
We 
perform the internal momenta sums 
in \eqref{amplinit} in an optimal way, and show that
the result can be expressed in terms of the quasi-local subgraphs. 
This condition, called the compatibility condition with the GN tree,
turns out to be crucial when performing the sum over the 
momentum attribution.

All external momenta  must be at a lower scale than internal momenta,
thus for any external faces $f_{s}$ and internal face $f_{s'}$, $p^{\ext}_{f_s} \ll p_{f_{s'}}$.  
We  bound all factors or terms with $p^{\ext}_{f_s}$ and obtain:
\beq
|A_{\cG;\bmu}|\leq  K_3
\Big[\prod_{l \in \cL} M^{-2 i_l}\Big] \, 
\sum_{p_{f_s}} 
\big[\prod_{f_s\in \cF_{\inter}} e^{-\delta(\sum_{l\in f_s} M^{-i_l})\sum_\xi |p_{f_s}|^\xi} 
\big]\big[\prod_{s=1}^{d}\prod_{v_s \in \cV_{\bullet;4; s}} (\epsilon \,p^{\,2a})_{v_s}\big],
\label{ineq}
\eeq
where $K_3=\kappa(\lambda,\eta_{\bullet})
K^L K_1^V  \; K_2^{F_{\ext}} K'$, and $K'$ is a constant obtained from
the  bound over the external momenta present in the vertex kernel 
 $\prod_{s=1}^{d} \prod_{v_s \in \cV_{\bullet; s}} (\epsilon\,\tilde{p})_{v_s} $;
 note that  
 $\epsilon $ in \eqref{amplinit} is now restricted  
 to  internal faces in \eqref{ineq}. 

Performing the sum over internal momenta 
$p_{f_s}$  must be done in a way to get the 
lowest possible divergence in \eqref{ineq}. This is an optimization procedure that we  detail
now.

We  determine the behavior of some momentum sums. The following results
have been detailed in appendix \ref{app:sums}. 
For constants, $B>0$, $c>0$, $b>0$, $a>0$ and $a'>0$, 
and an integer $n\ge 0$, we have   
\bea
&&
\sum_{p_1, \dots, p_{D}=1}^{\infty} (\sum_{l=1}^D p_l^{c})^{n} 
e^{-B \sum_{l=1}^D(p_l^b+p_l^a)} 
 = k B^{-\frac{(cn+D)}{b}}
e^{-B^{1 - \frac{a}{b}}} (1+ O(B^{\frac{1}{b}}))
\,,
\label{sumsABC2}
\\
&&
\sum_{p_1, \dots, p_{D}=1}^{\infty} (\sum_{l=1}^D p_l^{c})^{n} 
e^{-B (\sum_{l=1}^D(p_l^b+p_l^a+p_l^{a'}))} 
=
 k B^{-\frac{(cn+D)}{b}} e^{-B^{2 - \frac{(a+a')}{b}}}  (1+ O(B^{\frac{1}{b}}))
\,,
\label{sumsABDC3}
\eea
for a constant $k$. 
At this point,
we make two assumptions on the parameters 
$a,a',b$: 

- for the model $+$, $a \leq b$, 
\be
\sum_{p_1, \dots, p_{D}=1}^{\infty} (\sum_{l=1}^D p_l^{c})^{n} 
e^{-B \sum_{l=1}^D(p_l^b+p_l^a)} 
 = k B^{-\frac{(cn+D)}{b}}
(1+ O(B^{\frac{1}{b}}) + O(B^{1-\frac{a}{b}}) )
\,;
\label{sumsfinABC}
\ee

- for the model $\times$, $a +a'\leq 2b$, 
\be
\sum_{p_1, \dots, p_{D}=1}^{\infty} (\sum_{l=1}^D p_l^{c})^{n} 
e^{-B (\sum_{l=1}^D(p_l^b+p_l^a+p_l^{a'}))} 
=
k B^{-\frac{(cn+D)}{b}}  (1+ O(B^{\frac{1}{b}})  + O(B^{2 - \frac{(a+a')}{b}}))
\,.
\label{sumsfinABDC}
\ee 
Finally, the integration of internal momenta can be performed in the amplitudes. 

\

\noindent{\bf Model + -} Given a face  $f$ (the subscript $s$ is not useful at this stage), we target the line $l_f$ such that $i_{l_{f}}=\min_{l\in f} i_l=i_{f}$. After the integration,
it will generate the lowest factor $M^{i_{f} \times m}$, where $m$
is yet to be determined.  

 In the product 
$\prod_{s=1}^{d}\prod_{v_s \in \cV_{+; 4;s}} (\epsilon \,p^{\,2a})_{v_s}$, 
we choose the factor of a given $p_{f}$  and perform the sum  
$\sum_{p_f}(|p_{f}|^{2a})^{\rho_{f}} e^{-\delta M^{-i_f} |p_{f}|^b}$, 
with  $\rho_f$ an integer, such that the bound \eqref{ineq} still holds. 
Performing this sum using \eqref{sumsfinABC}, we get a product of $M^{\frac{i_f}{b}(2a \rho_{f} +D)}$ with the lowest possible power.  
Take a closed face $f_s$ of color $s$,  the integer $\rho_{f_s}$ counts how many times  $f_s$ passes through vertices of $\cV_{+;4;s}$.  We have
\beq\label{rho}
 \rho_{f_{s}} = \sum_{v_s \in \cV_{+; 4;s}} \epsilon_{v_s,f_{s}}\,, \qquad 
\rhop(\cG) =  \sum_s\sum_{f_{s}} \rho_{f_{s}}\,.
\eeq
We then write a new bound
\beq
|A_{\cG;\bmu}|\leq  \kappa_1
\Big[\prod_{l \in \cL} M^{-2 i_l} \Big]\, 
\sum_{p_{f_s}} \Big[ 
\prod_{f_s\in \cF_{\inter}} e^{-\delta M^{-i_{f_s}} \sum_\xi |p_{f_s}|^\xi } \Big] 
\Big[ \prod_{s'=1}^{d}\prod_{f_{s'} } |p_{f_{s'}}|^{\,2a \rho_{f_{s'}}}\Big]\,, 
\eeq
where $\kappa_1$ is a new constant incorporating the previous constant $K_3$.  
Performing the sum over internal momenta, one
gets using \eqref{sumsfinABC} with $a\leq b$,
\beq\label{ad}
|A_{\cG;\bmu}|\leq  \kappa_2
 \prod_{l \in \cL} M^{-2  \, i_l} \, 
\prod_{f_s\in \cF_{\inter}} M^{\frac{i_{f_s}}{b}(2a\rho_{f_{s}}+D)} \,, 
\eeq
where $\kappa_2$ is another constant depending on the graph 
that includes $\kappa_1$ and new constants
coming from the summation over internal momenta.

We re-express the above bound in terms of the quasi-local subgraphs $G^{i}_k$.
The product over lines can be written \cite{Rivasseau:1991ub} as
\be\label{lines}
 \prod_{l \in \cL} M^{-2  \, i_l}  = 
 \prod_{l \in \cL}\prod_{(i,k)/\, l\in \cL(G^i_k)} M^{-2} 
 = \prod_{(i,k)} M^{-2 L(G^i_k)}\,.
\ee
The second product over faces splits in two factors. The first
term can be treated as: 
\be
\prod_{f_s\in \cF_{\inter}} M^{\frac{D}{b}  i_{f_s}}
 = \prod_{ f_s\in \cF_{\inter} } \prod_{(i,k)/\, l_{f_s} \in \cL(G^{i}_k)}M^{\frac{D}{b}} 
= \prod_{(i,k)} \prod_{ f_s\in \cF_{\inter} \cap G^{i}_k} M^{\frac{D}{b}} 
 =  \prod_{(i,k)} M^{\frac{D}{b} \, F_{\inter}(G^i_k)}
\label{faces}
  \,. 
\ee
 The last product involving  $\rho_{f_s}$ can be treated as
\bea
\prod_{ f_s\in \cF_{\inter} } 
\prod_{ (i,k)/\, l_{f_s}\in G^{i}_{k}  }  M^{  \frac{2a}{b} i_{f_s} \rho_{f_s} } 
= \prod_{(i,k)} \prod_{  f_s\in \cF_{\inter}\cap G^i_k }
M^{\frac{2a}{b} \rho_{f_s} } = \prod_{(i,k)} M^{\frac{2a}{b} \rhop(G^i_k)}\,,
\label{enhan}
\eea
where $\rhop(\cdot)$ has been defined in \eqref{rho}.

Now, if we introduce the counter-term and the wave function vertices $V_{2;a;s}$ and $V_{2;b;s}$,
they might bring an additional momentum enhancement to faces. We want to keep the definition of $\rho_{f_s}$ as in \eqref{rho} and we must now add to it the 
contributions of the $2$-point vertices of any types.  
Hence $\rho_{f_s} \to \rho_{f_s} + \rho_{2;a;f_s}+ \rho_{2;b;f_s}$, where $\rho_{2;\xi;f_s}=\sum_{v_s \in \cV_{2;\xi;s} }\epsilon_{v_s,f_s}$
is the number of times that $f_s$ visits $\cV_{2;\xi;s}$ vertices, $\xi=a,b$.
 To the above power counting, we should therefore add the following factor
\bea
\prod_{ f_s\in \cF_{\inter} } 
\prod_{ (i,k)/\, l_{f_s}\in G^{i}_{k}  }M^{ i_{f_s} [\frac{2a}{b}\rho_{2;a;f_s} +\frac{2b}{b}\rho_{2;b;f_s}] }  
 = 
\prod_{(i,k)} \prod_{  f_s\in \cF_{\inter}\cap G^i_k }
M^{[\frac{2a}{b} \rho_{2;a;f_s} + 2 \rho_{2;b;f_s}] } \,. 
\eea
Note that a vertex of  $\cV_{2;\xi;s}$  has a single strand with enhanced 
momentum  $p_s^{2\xi}$, $\xi=a,b$. When a face uses that
strand, the corresponding vertex contributes exactly once to the power counting. 
Then, 
\be\label{rho2xi}
\rho_{2;\xi} = \sum_{f_s \in F_{\inter}(G^i_k)}\rho_{2;\xi;f_s}
\ee
counts the number of vertices of   $\cV_{2;\xi;s}$ in $G^i_k$. In the end, we have 
\be
\prod_{(i,k)} \prod_{  f_s\in \cF_{\inter}\cap G^i_k }
M^{[\frac{2a}{b} \rho_{2;a;f_s} + 2 \rho_{2;b;f_s}] }  
 = \prod_{(i,k)} M^{ \frac{2a}{b}\rho_{2;a}(G^i_k) + 2 \rho_{2;b}(G^i_k) } \,. 
\ee

Changing $M \to M^{b}$, we obtain a power counting of the amplitude 
\eqref{ad} for the model $+$, under the condition $a\leq b$, as
\beq\label{theo+}
|A_{\cG;\bmu}| \le \kappa \prod_{(i,k) \subset N^2} M^{\omp(G^i_k)}\,,
\eeq
where $\kappa$ is a constant and the degree of divergence of $G^i_k$ is given by
\beq\label{deg+}
\omp(G^i_k) = - 2 b L (G^i_k) + DF_{\rm int} (G^i_k)+ 2 a \rhop (G^i_k)
 +\sum_{\xi=a,b}2\xi \rho_{2;\xi}(G^i_k)  \,. 
\eeq
Putting $a$ to 0 leads to the ordinary power counting theorem
of usual tensor field theories. 

\ 

\noindent{\bf Model $\times$ -}
The analysis is very similar to the above. 
We count how many times a face $f_s$ passes through all vertices of the type $\cV_{\times;s}$ and this defines the following quantities
\be \label{varrho}
\varrho_{f_s} =\sum_{v_{s'},f_{s''}}\epsilon_{v_{s'}f_{s}f_{s''}}\,, 
\qquad \rhot(\cG)=\sum_{s}\sum_{f_{s}}\varrho_{f_{s}}\,. 
\ee 
With a similar calculation as above, 
using \eqref{sumsfinABDC} with $3a \leq 2b$, 
introducing also vertices of $\cV_{2;\xi;s}$,
$\xi=a,2a,b$, and $\rho_{2;\xi;f_s}$ as the number of times
that a closed face $f_s$ runs through vertices of $\cV_{2;\xi;s}$
and $\rho_{2;\xi}$ still obeys \eqref{rho2xi}, 
we obtain the power counting of the model $\times$ as 
\be\label{theox}
|A_{\cG;\bmu}| \le \kappa \prod_{(i,k) \subset N^2} M^{\omt(G^i_k)}\,,
\ee
where $\kappa$ is a constant and the degree of divergence of $G^i_k$ is given by
\beq\label{degx}
\omt(G^i_k) = - 2 b \, L (G^i_k) + D  F_{\inter} (G^i_k)+ 2 a \rhot (G^i_k)
+ \sum_{\xi=a,2a,b}2\xi \rho_{2;\xi}(G^i_k)  \,. 
\eeq
From \eqref{deg+} and \eqref{degx} and  for convenience, we can use unified notations
 $\omb$ with $\bullet =+,\times$, with the sum of $\xi$ being  appropriately chosen.

\section{Analyses of the potentially renormalizable models}
\label{sect:enhancedmelon}

In this section, we explore the parameter spaces of potentially renormalizable models  $+$ and $\times$.

In the analyses below, we need the number of internal faces of a connected graph $\cG$, in any rank $d \ge 3$ tensorial model, which is given in \cite{Samary:2012bw}:
\beq
F_{\rm int} = - {2 \over (d^-)!} ( \omega(\cG_{\rm color}) - \omega (\partial \cG))  - (C_{\partial \cG} - 1) - {d^- \over 2} N_{\rm ext} + d^- - {d^- \over 4} (4 - 2 n) \cdot V,
\label{eq:face}
\eeq
where $d^- = d-1$ with $d$ being the  rank of the tensor field,
$\cG_{\rm color}$ is the colored extension of $\cG$, $\bG$ denotes the boundary of $\cG$ \cite{BenGeloun:2011rc}, with $C_{\bG}$  the number of connected components of $\bG$,
 $N_{\ext}$ is the number of external legs of the graph, $V_k$ is the number of vertices of coordination number $k$, $ V = \sum_k V_k$ is the total number of vertices in $\cG$, and $n \cdot V = \sum_k k V_k$ is the number of half lines emanating  from vertices.
$\omega(\cG_{\rm color}) = \sum_{J_{\cG_{\rm color}}} g_{\tJ_{\cexG}}$, $\omega (\partial \cG) = \sum_{J_{\partial \cG}} g_{J_{\partial \cG}}$ with
 genus $g_{J}$, the genus of a ribbon graph $J$ called jacket \cite{Gur4}. 
 A jacket is nothing but a particular embedding of the bipartite colored graph $\cG$. 
 The jackets of  $\cexG$ are denoted $J_{\cexG}$ and they must be ``closed'' to
 define a closed surface $\tJ_{\cexG}$ on which a genus 
 $g_{\tJ_{\cexG}}$ could be identified. The boundary graph $\bG$ itself maps to 
 a rank $d-1$ colored tensor graph.  $\bG$  therefore has jackets denoted $J_{\bG}$. 
 
 The quantity $\omega(\cexG)$ is called the 
 degree of the colored tensor graph $\cG_{\rm color}$. It replaces the genus and
 allows one to define a large $N$ expansion for  colored tensor models \cite{Gur4}. 
 A graph $\cG$ is called a melon if and only if its colored extension $\cexG$
 is a melon and that is if $\omega(\cexG)=0$ (all jackets  $\tJ_{\cexG}$ are planar). 
We shall need a few properties of the quantity $\omega(\cG_{\rm color}) - \omega (\partial \cG)$ withdrawn from \cite{Geloun:2012fq} that we will recall
at some point. 

Let us recall the following terminology:
a ``bridge''  in a graph is a line such that cutting that line adds another connected component  to this graph.  The ``cut of a bridge'' means the removal of the bridge 
from the graph  and letting two external legs where its extremities were incident. 
A graph  is called one-particle reducible (1PR) graph if it has bridges, otherwise
it is called one-particle irreducible (1PI).

\begin{lemma}[$\rho_{\bullet}$ and $\rho_{2;\xi}$ for 1PR graph]\label{lem:bridg}
Let $\cG$ be a graph with bridges (or a 1PR graph) such that cutting the bridges gives the family $\{\cG_j\}$ of subgraphs. 
then 
\be
\rho_{\bullet}(\cG) = \sum_{j} \rho_{\bullet}(\cG_j) \,,
\qquad 
\rho_{2;\xi}(\cG) = \sum_{j} \rho_{2;\xi}(\cG_j)\,, 
\ee
where ${\bullet}= +$, and $\xi=a,b$, or ${\bullet}=\times$, and $\xi=a,2a,b$. 
\end{lemma}
\proof  This  follows from the fact that through a bridge no closed face passes.
The quantities $\rho_{\bullet}(\cG)$ and $\rho_{2;\xi}(\cG)$ 
can be computed with the block diagonal matrix $\epsilon_{vf}$ using vertices and 
closed faces in each connected component $\cG_j$. 

\qed 

The following proposition is easy to prove. 

\begin{lemma}[Bounds on $\rho_{2;\xi}$]\label{lem:boundxi}
Let $\cG$ be a graph of the model $\bullet$. 
Then $\rho_{2;\xi}(\cG) \leq V_{2;\xi}$.  If $\cG$ is 1PI then  
\be
\rho_{2;\xi}(\cG) = V_{2;\xi}(\cG)\,.  
\ee
\end{lemma}

\subsection{Models $+$}
\label{sect:modelsplus}

Consider the ``contraction'' operation of a
degree-2 vertex $v$ (belonging to $\cV_2$ or to $\cV_{2;s}$) on the graph $\cG$  which removes $v$  and replaces it by a propagator line with the same external momenta. 
Consider the graph $\tilde\cG$ resulting from the contractions
of all degree-2 vertices in $\cG$. 
Note that if $\cG$ is 1PR or 1PI then so is $\tilde\cG$
and the number of degree-4 vertices  and external legs 
coincide in both graphs.  
We define the number $Br$ of c-bridges (chain-bridges) of $\cG$ to be the number
of bridges in $\tilde\cG$. Note a c-bridge of $\cG$ can be very well 
associated with a bridge $\cG$. 
We also introduce $V_4 + V_{+;4} = V_{(4)}$. 

\begin{lemma}[Bound of $\rho_{+}$]\label{lem:rhob}
Let $\cG$ be a graph with $N_{\ext} > 0$ external legs. Then,  
$\rhop(\cG) \le V_{+;4}$. If $\cG$ is melonic

- $V_{(4)} =1$,  then $\rhop(\cG)=0$. 

- $V_{(4)}  >1$, then   $\rhop(\cG) \le V_{(4)}- {N_{\rm ext} \over 2}- Br$, 

where $Br$ is the number of c-bridges in the graph $\cG$. 
\end{lemma}

\proof   The first statement is clear from the combinatorial procedure
counting at most $V_{+;4}$ for $\rhop(\cG) $ for an arbitrary graph. 
Now this bound can be refined for a melonic $N_{\ext}$-point 
graph. 
If $V_{(4)}=1$, then 
either $N_{\ext} =4 $,  and then  $\rho_+(\cG)=0$, or $N_{\ext}=2$, and 
we have a melonic tadpole or a melonic graph with one c-bridge  
which gives again $\rho_+(\cG)=0$.

A 1PI graph $\cG$ with 4 valent vertices can have at most 2 external legs per vertex.
 Consider a melonic graph $\cG$ and its colored extension $\cG_{\col}$:
then each vertex in $\cG_{\col}$ comes with a partner (see for instance Figure 1 in \cite{Geloun:2012fq}).
Note that the two partner vertices belong to the same vertex in $\cG$. 
If one vertex $v$ has a propagator $l$ and 
its partner $\tilde v$
has no propagator (hence has an external leg) then  $l$ must be a bridge.
Focusing on 1PI bipartite melons, then either  $v$ and $\tilde v$ have both propagators or have both external legs. The presence of $N_{\ext}$ external legs in 1PI bipartite melons implies
that these external legs must be hooked to $N_{\ext}/2$ vertices. 
Take any vertex $v_s$ with color $s$ where an  external leg is incident, then  an external leg is also incident to $\tilde v_s$. 
None of the open faces with color $0s$,
 which can be enhanced, could bring any contribution to $\rhop(\cG)$.  Repeating the argument for $N_{\ext}/2$ vertices, we see that these vertices could not
be part of the optimization procedure computing $\rhop(\cG)$ and
so $\rhop(\cG) \le V_{(4)} - N_{\ext}/2$.
 
Now we treat the case  of a 1PR graph $\cG$.
Consider its resulting $\tilde\cG$ after the contraction of all of its degree-2 vertices. 
Cut all bridges in $\tilde\cG$ to obtain a family of 1PI subgraphs. On each component $\tilde\cG_{j}$
the bound $\rhop(\tilde\cG_j) \le V_{(4)}(\tilde\cG_j)- {N_{\rm ext}(\tilde\cG_j) \over 2}$ holds. 
Summing this relation over 1PI subgraphs and using Lemma \ref{lem:bridg}, we get 
\be\label{rhop1pr}
\rhop(\tilde\cG) =
\sum_{j}\rhop(\tilde\cG_j) 
\le \sum_{j}[V_{(4)}(\tilde\cG_j)- {N_{\rm ext}(\tilde\cG_j) \over 2} ]
= V_{(4)} - \frac{1}{2} N_{\ext} - \sharp \text{bridges }\,, 
\ee
where we used that each bridge cut brings two additional external legs compared
to $N_{\ext}$.  Finally, we can use the relation $\rhop(\cG)=\rhop(\tilde\cG)$
because degree-2 vertices are not involved in the counting
of $\rhop$ and $\sharp \text{bridges }=Br$.
In summary, we can also use   \eqref{rhop1pr} 
for 1PI graph with $Br =0$. 

\qed 

\begin{figure}[H]\
\begin{center}
     \begin{minipage}{.7\textwidth}
     \centering
   \includegraphics[angle=0, width=4cm, height=3.5cm]{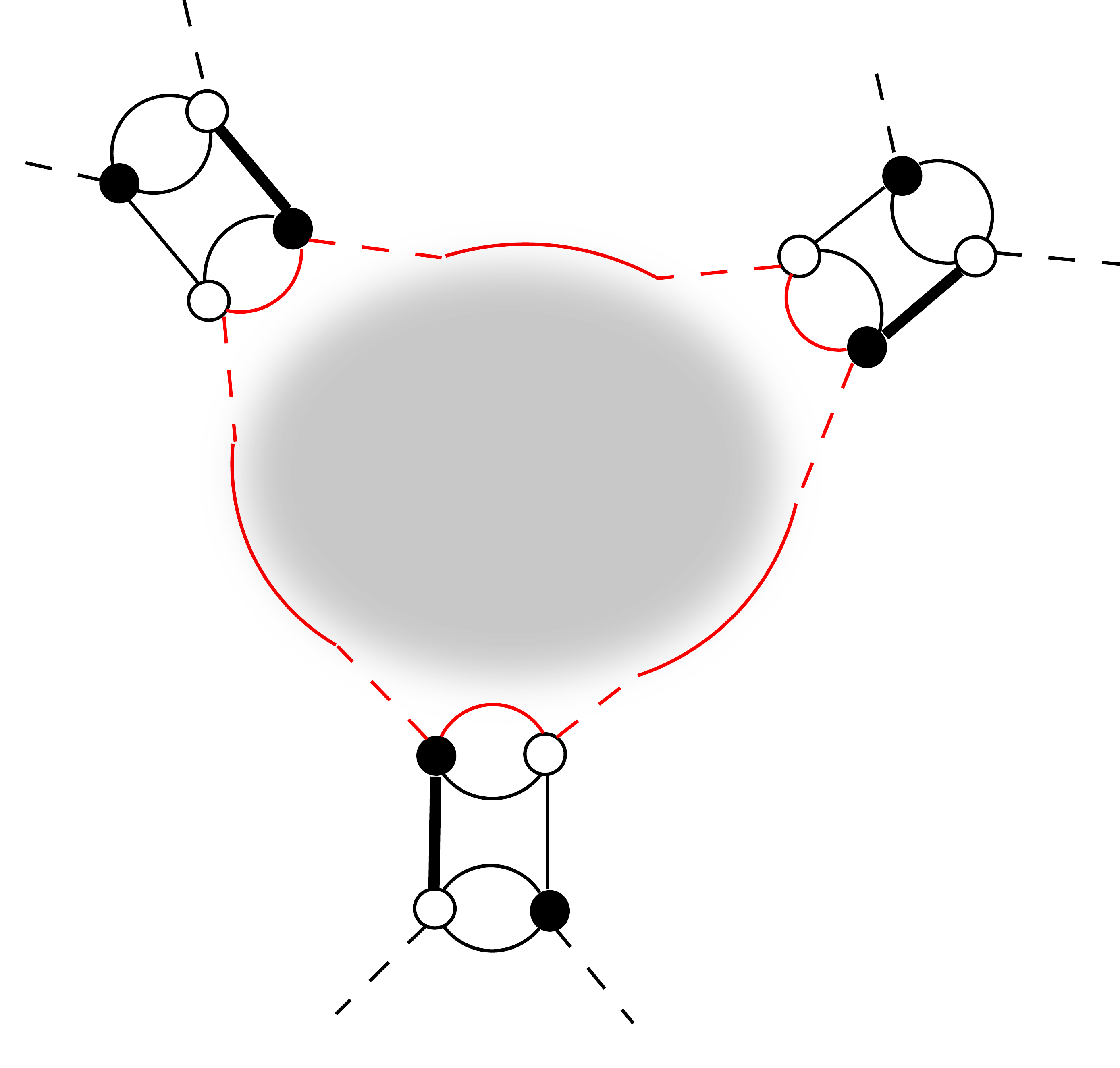}
\includegraphics[angle=0, width=4cm, height=3.5cm]{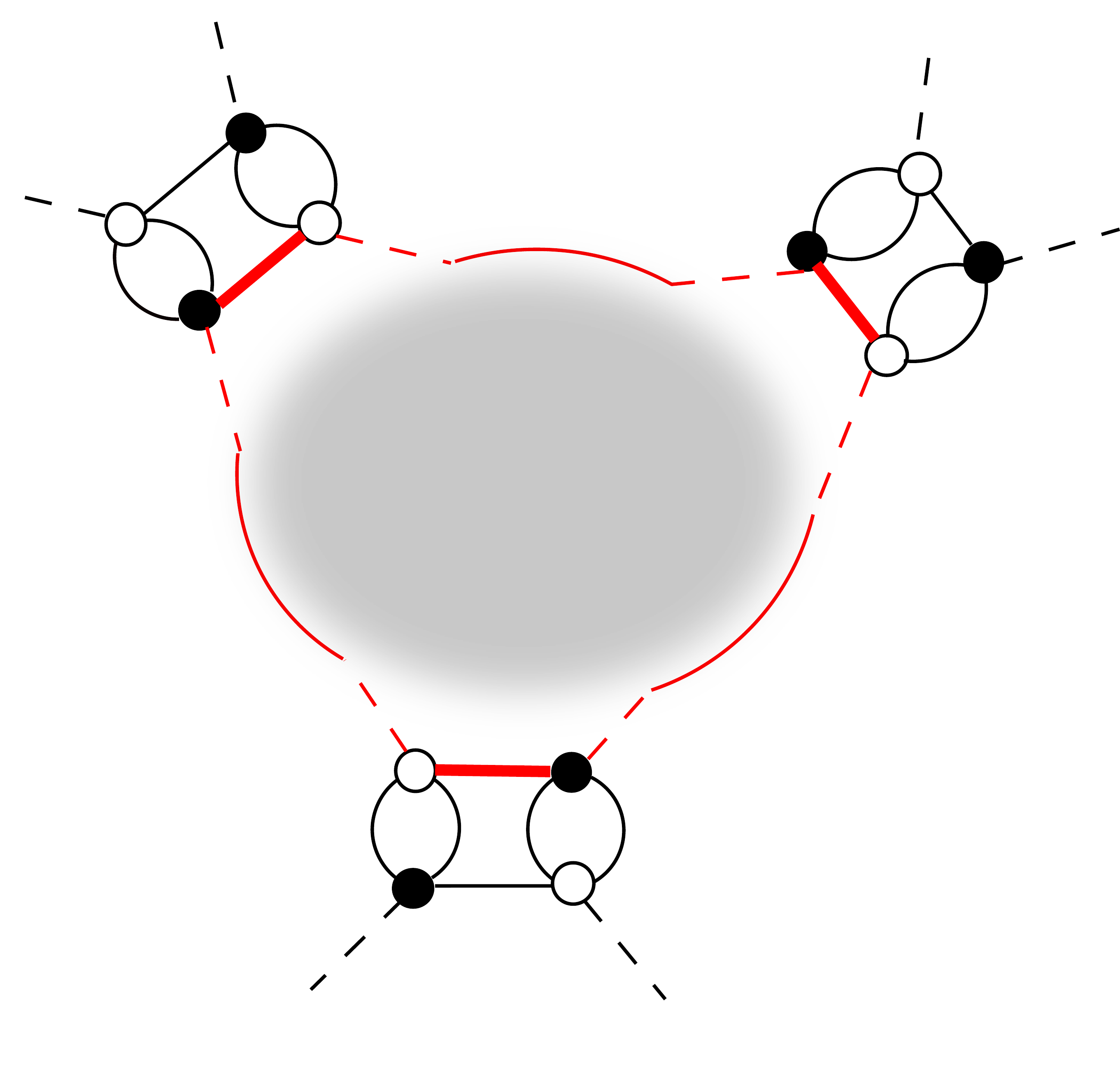}
\caption{ {\small Examples of $N_{\ext} = 6$-point functions in rank $d=3$ of a melonic and a non-melonic type. }} 
\label{fig:rhomaxmelons0}
\end{minipage}
\end{center}
\end{figure}

As an illustration of Lemma \ref{lem:rhob}, consider the graphs of Figure \ref{fig:rhomaxmelons0}. 
Consider the melonic graph at the left hand side. ${N_{\ext} \over 2} = 3 $ vertices which have external legs will not contribute to $\rhop (\cG) $.
Hence, $\rhop (\cG) \le V_{(4)}  - {N_{\ext} \over 2}$.
On the other hand, consider the non-melonic graph on the right hand side. $ 3$ vertices which have external legs contribute to $\rhop (\cG) $.

For a melonic graph, Lemma \ref{lem:rhob} gives in fact two bounds. The bound $\rhop(\cG)\leq V_{+;4}$ is sharper than the other, if and only if 
\be
V_4 \ge \frac{N_\ext}{2} + Br \,. 
\ee

\

\noindent{\bf Potentially renormalizable models.} 
We restrict now to primitively divergent graphs which can be considered
connected and with $Br=0$, in other words to 1PI graphs. 
The degree of divergence of this model is, by
combining \eqref{deg+} and \eqref{eq:face} and using $2 L = n \cdot V - N_{\ext}$,
$a\leq b$,
\bea
\omp(\cG)
 &=&  - {2 D \over (d^-)!} ( \omega(\cG_{\rm color}) - \omega (\partial \cG))  - D (C_{\partial \cG} - 1) 
-{1\over 2} \left[ ( D \, d^- - 2 b) N_{\ext} - 2 D \, d^- \right] 
\crcr
&&+ {1 \over 2} \left[ -2 D\, d^- + (D \, d^- - 2 b) n \right] \cdot V  + 2 a \rhop
+2a\rho_{2;a}+ 2b \rho_{2;b}  \,.
\eea
From Lemma \ref{lem:rhob}, we have 
\bea
\triangle^{\rm melon}_{+} &=& 
\left\{
\begin{array}{lc}
0,  & V_{(4)} = 1 \cr
 V_{(4)} - {N_{\ext} \over 2}- \rhop(\cG^{\rm melon}) \ge 0, & V_{(4)}>1 \cr
 \end{array}\right. 
\crcr
\triangle_{+}  &=&   V_{+;4} -  \rhop(\cG)  \ge 0\,.
\eea
The case $V_{(4)}>1$ is the most important one
when we study all orders of perturbation and we will focus on that. 
Using the Lemma \ref{lem:boundxi}, and 
further inserting that $\omega(\cG_{\rm color})=0$ and $\omega (\partial \cG) = 0$ for melonic graphs,
\bea
&&
\omp(\cG^{\rm melon}) 
\le 
-  D (C_{\partial \cG} - 1) 
- {1 \over 2} \left[ (D\, d^- - 2 b + 2 a)N_{\rm ext} - 2D \,d^-\right] 
\crcr
&&
- 2 bV_2 -  2 (b-a) V_{2;a}
+ ( D \, d^- - 4 b + 2 a) V_{(4)}
- 2 a \triangle^{\rm melon}_{+}   \cr\cr
&&
\le
-  D (C_{\partial \cG} - 1) 
- {1 \over 2} \left[ (D\, d^- - 2 b + 2 a)N_{\rm ext} - 2D \,d^-\right] 
\crcr
&&
- 2 bV_2 -  2 (b-a) V_{2;a}
+ ( D \, d^- - 4 b + 2 a) V_{(4)}\,. 
\label{eq:1omegameldelta0}
\eea
There is another bound for melonic graphs: 
\bea
&&
\omp(\cG^{\rm melon}) 
\le
-  D (C_{\partial \cG} - 1) - {1 \over 2} \left[ (
D d^- - 2b)N_{\rm ext} - 2 D d^-\right]
\crcr
&&
- 2 b V_2 - 2 (b-a) V_{2;a}
+ (Dd^- -4b)V_{4} 
+ ( D d^- - 4 b + 2a) V_{+;4} \cr\cr
&&
\leq 
-  D (C_{\partial \cG} - 1) -  \left[  bN_{\rm ext} - D d^-\right]
\crcr
&&
- 2 b V_2 - 2 (b-a) V_{2;a}
+ (Dd^- -4b)\Big(V_{4} - \frac{N_\ext}{2}\Big)  
+ ( D d^- - 4 b + 2a) V_{+;4}
\,.
\label{eq:2omegamel}
\eea
Either choosing \eqref{eq:2omegamel} or 
\eqref{eq:1omegameldelta0} as a sharper bound,
leads to the same  result. 

 Meanwhile, for non-melonic graphs, using  
 $\omega(\cG_{\rm color}) - \omega (\partial \cG)\ge 
{1 \over 2} (d^- - 1) d^- !$ \cite{Geloun:2012fq},   we get 
\bea
&&
\omp(\cG^{\rm non-melon}) 
\le
- D (d^- - 1) 
-  D (C_{\partial \cG} - 1) - {1 \over 2} \left[ (
D d^- - 2b)N_{\rm ext} - 2 D d^-\right]
\crcr
&&
- 2 b V_2 - 2 (b-a) V_{2;a}
+ (Dd^- -4b)V_{4} 
+ ( D d^- - 4 b + 2a) V_{+;4}
- 2 a \triangle_{+} \cr\cr
&&
\le 
- D (d^- - 1) 
-  D (C_{\partial \cG} - 1) - {1 \over 2} \left[ (
D d^- - 2b)N_{\rm ext} - 2 D d^-\right]
\crcr
&&
- 2 b V_2 - 2 (b-a) V_{2;a}
+ (Dd^- -4b)V_{4} 
+ ( D d^- - 4 b + 2a) V_{+;4}
\,.
\label{eq:1omeganonmeldelta0}
\eea
 For  renormalizable models, we require the coefficients of  vertices to be negative. 
 This is demanding that, since $a>0$, $b>0$, 
\beq 
D \, d^- - 4  b + 2 a \le 0 \,, \qquad b\ge a \,,
\eeq
which give for $a$,
\beq
a\; \le  \;2 \, b - {1 \over 2} D \, d^-\,.
\label{eq:validity2}
\eeq
We see that the condition $a\leq b$ coming from the sum over internal
momenta has been naturally incorporated in the analysis. 
Then, to achieve just-renormalizability, we use 
$a = 2 b - {1 \over 2} D d^- \geq 0$ (and $a\le b$ implies that $b\leq  {1 \over 2} D d^- $) given in
\eqref{eq:validity2} into \eqref{eq:1omegameldelta0} and \eqref{eq:1omeganonmeldelta0},  and see if the conditions 
\be
\quad \omp(\cG^{\rm melon})|_{N_{\rm ext} \ge 6} < 0  \,, 
\qquad 
\quad \omp(\cG^{\rm non-melon})|_{N_{\rm ext} \ge 6} < 0
\label{eq:range}
\ee
can be accomplished.
These conditions translate into
\bea
&&
\omp(\cG^{\rm melon}) |_{N_{\rm ext} \ge 6} 
\le \cr
&&
\Big[
- D (C_{\partial \cG} - 1) 
+ D d^-
-b N_{\rm ext} 
- 2b V_2 -2 (b-a)V_{2;a} 
\Big]
\Big|_{N_{\rm ext} \ge 6} 
< 0 
\,,
\label{eq:1omegameljust} 
\\\cr
&&
\omp(\cG^{\rm non-melon}) |_{N_{\rm ext} \ge 6} 
\le \cr
&&
\Big[
D
- D (C_{\partial \cG} - 1) 
- ({1 \over 2} D \, d^- - b)N_{\rm ext} 
- 2 b  V_2  - 2(b-a)V_{2;a} 
- 2a V_{4}
\Big]\Big|_{N_{\rm ext} \ge 6} 
< 0 
\,.
\label{eq:1omeganonmeljust}
\eea
As $N_{\ext}$ increases, $\omp$ decreases, so $\omp$ is maximum at $N_{\ext} =6$ for
melonic graphs;  $\omp$ is maximum at $N_{\ext}=6$
 as long as  $b < { D d^- \over 2}$,  for non-melonic graphs. Thus, the conditions
 for having convergent $N_{\ext}=6$-pt functions are:
\bea
\omp(\cG^{\rm melon}) |_{N_{\ext} = 6}
&\le& 
-  D (C_{\partial \cG} - 1) 
+ D \,d^-
-6 b 
- 2  b V_2 - 2(b-a)V_{2;a}
\crcr
&\le&
 D \,d^-
-6 b 
< 0,
\label{eq:1omegameljust6}
\\
\omp(\cG^{\rm non-melon}) |_{N_{\ext}=6}
&\le& 
D
- D (C_{\partial \cG} - 1)  
-3 D d^-
+6 b 
- 2 b  V_2   - 2(b-a)V_{2;a} -2aV_{4}
\crcr
&\le& 
D -3 D d^- +6 b
< 0 
\,.
\label{eq:1omeganonmeljust6}
\eea
The above inequalities further reduce to 
\beq
{D d^- \over 6} < b < { D ( 3 d^- -1 ) \over 6}\,. 
\label{eq:1bjust6}
\eeq
Note here that ${ D ( 3 d^- - 1) \over 6} < {D d^- \over 2}$ is always true for $D >0$, thus 
we have improved the bound on $b$. Under  \eqref{eq:1bjust6}, the  degree of divergence for $N_{\ext} \ge 6$ is maximum at $N_{\ext} = 6$ and strictly negative. Furthermore, we demand that $a>0$ and so that $b > \frac{Dd^-}{4}$. We finally 
get the bound
\beq
\boxed{
{D d^- \over 4} < b < { D ( 3 d^- -1 ) \over 6}
}
\,.
\label{eq:1bjust6final}
\eeq
Now we use \eqref{eq:1bjust6final} to find a bound on $a = 2 b - {1 \over 2} D d^-$ \eqref{eq:validity2} as
\beq
{\boxed{
0 < a < { D ( 3 d^- -2) \over 6}
}}
\label{eq:1ajust6final}
\eeq
Combining \eqref{eq:1ajust6final} and \eqref{eq:1bjust6final} for given $D$ and $d^-\geq 2$,
we obtain the ranges of values of $a$ and $b$ in Table \ref{table:table1} which could
lead to just-renormalizable models: 
\begin{table}[H]
\setlength{\extrarowheight}{0.2cm}
\centering
\begin{tabular}{lcccc}  \hline  \hline
&$d^-=2$&$d^-=3$&$d^-=4$&$d^-=5$ \\
\hline\hline
$D = 1$ 
&\pbox{2.5cm} {$ {0} < a < {2 \over 3} $ \\ ${1 \over 2} < b  <{5 \over 6}$ } 
& \pbox{2.5cm}{${0} < a  < { 7 \over6} $ \\ ${3 \over 4} < b < { 4 \over 3}$} 
&\pbox{2.5cm}{${0} < a <{ 5 \over 3}$ \\ ${1} < b <{11 \over 6}$}
&\pbox{2.5cm}{${0} < a <{13 \over 6} $ \\ ${5 \over 4} < b <{7 \over 3}$}
 \\
 \hline
$D = 2$ 
&\pbox{2.5cm}{${0} < a <{ 4 \over 3} $ \\ ${ 1} < b <{5 \over 3} $} 
&\pbox{2.5cm}{$0 < a < { 7 \over 3}$ \\ ${3 \over 2} < b <{8 \over 3} $} 
&\pbox{2.5cm}{$0 < a <{10 \over 3}$ \\ ${2} < b <{11 \over 3}$} 
&\pbox{2.5cm}{$0 < a < {13 \over 3}$ \\ ${ 5 \over 2} < b <{14 \over 3}$} 
  \\
 \hline
$D = 3$ 
&\pbox{2.5cm}{${0} < a <2 $ \\ ${ 3 \over 2} < b <{5 \over 2}$} 
&\pbox{2.5cm}{${0} < a <{ 7 \over 2}$ \\ ${ 9 \over 4} < b < 4$}
&\pbox{2.5cm}{${ 0} < a <5$ \\ ${3} < b <{11 \over 2}$}
&\pbox{2.5cm}{$0 < a <{13 \over 2} $ \\ ${ 15 \over 4} < b <7$}
  \\
\hline
$D = 4$ 
&\pbox{2.5cm}{${0} < a <{8 \over 3} $ \\ $2 < b <{10 \over 3}$} 
&\pbox{2.5cm}{${0} < a  <{14 \over 3}$ \\ $3 < b <{16 \over 3}$}
&\pbox{2.5cm}{$0 < a <{20 \over 3}$ \\ $4 < b <{22 \over 3}$}
&\pbox{2.5cm}{$0 < a  <{26 \over 3}$ \\ $5 < b <{28 \over 3}$}
\\
 \hline \hline
\end{tabular}
\caption{Allowed region of the values of $a$ and $b$  for potentially just-renormalizable  models $+$
with $d^- \le  5$ and $D \le 4$.
}
\label{table:table1} 
\end{table}
\vskip 10pt

This table shows that there might be uncountable models which could be just renormalizable. We note that the limit cases $a=0$ lead to the renormalizable
invariant tensor field theories studied in \cite{BenGeloun:2012pu} ($d=3, D=1,b=\frac12$) and \cite{Geloun:2013saa}  [$(d=4,D=1 ,b=\frac34); 
(d=5,D=1,  b=1); (d=3, D=2, b=1)$].

Let us seek further conditions leading to interesting models with $a>0$. 
One of these conditions is to achieve logarithmic divergence for non-melonic graphs at $N_{\ext} = 4$. For this, achieving 
\beq
 \omp(\cG^{\rm non-melon})|_{N_{\rm ext} = 4} = 0
\label{eq:justrenlog4}
\eeq
 entails
\beq
\boxed{
b = {1 \over 2} D ( d^- - {1 \over 2})\,, \qquad 
a= {1 \over 2} D ( d^- - 1) }
\label{eq:validityb}
\eeq
which  is consistent with \eqref{eq:1bjust6final}, since
${1 \over 2} D ( d^- - {1 \over 2}) < { D ( 3 d^- -1 ) \over 6}$ for $D>0$  and 
${D d^- \over 4} < {1 \over 2} D ( d^- - {1 \over 2})$ for $ d^->1$.
In Table \ref{table:table2}, we explicitly show the valid values of $a$ and $b$ given in \eqref{eq:validityb}.
\begin{table}[H]
\setlength{\extrarowheight}{0.2cm}
\centering
\begin{tabular}{x{2cm}x{2cm}x{2cm}x{2cm}x{2cm}}  \hline\hline
&$d^-=2$&$d^-=3$&$d^-=4$&$d^-=5$ \\
\hline\hline
$D=1$
&\pbox{2cm}{$a = {1 \over 2}$ \\ $b= {3 \over 4}$} 
& \pbox{2cm}{$a = {1}$ \\ $b= {5 \over 4}$} 
&\pbox{2cm}{$a = {3 \over 2}$ \\ $b= {7 \over 4}$}
&\pbox{2cm}{$a = { 2}$ \\ $b= {9\over 4}$}
\\ \hline
$D=2$
&\pbox{2cm}{$a = {1}$ \\ $b= {3 \over 2}$} 
&\pbox{2cm}{$a = {2}$ \\ $b= {5 \over 2}$} 
&\pbox{2cm}{$a = {3}$ \\ $b= {7\over 2}$} 
&\pbox{2cm}{$a = {4}$ \\ $b= {9 \over 2}$} 
\\ \hline
$D=3$
&\pbox{2cm}{$a = {3 \over 2}$ \\ $b= {9 \over 4}$} 
&\pbox{2cm}{$a = {3}$ \\ $b= {15 \over 4}$}
&\pbox{2cm}{$a = {9 \over 2}$ \\ $b= {21 \over 4}$}
&\pbox{2cm}{$a = {6}$ \\ $b= {27 \over 4}$}
\\ \hline
$D=4$
&\pbox{2cm}{$a = {2}$ \\ $b= {3}$} 
&\pbox{2cm}{$a = {4}$ \\ $b= {5}$}
&\pbox{2cm}{$a = {6}$ \\ $b= {7}$}
&\pbox{2cm}{$a = {8}$ \\ $b= {9}$}
\\ \hline \hline
\end{tabular}
\caption{Values of $a$ and $b$ for potentially just-renormalizable theories with $\omp(\cG^{\rm non-melon})|_{N_{\rm ext} = 4} = 0$
with $d^- \le 5$ and $D  \le 4$.
}
\label{table:table2} 
\end{table}
Table \ref{table:table1} and Table \ref{table:table2} are consistent
 for just-renormalizable models with the superficial degree of divergence which does not depend on $V_4$, with   logarithmic divergence for graphs with $N_{\ext} =4$, and with 
 convergent graphs with $N_{\ext} \ge 6$.
Let us discuss the behavior of  melonic graphs.  Concentrating on $N_{\ext}=4$, 
we evaluate $\omp(\cG^{\rm melon})|_{N_{\ext} = 4}$ 
keeping in mind \eqref{eq:1bjust6final} and obtain 
\beq
\omega_d(\cG^{\rm melon}) |_{N_{\ext} = 4}
\le 
D d^- - 4 b  
\,,
\eeq
which gives
\beq
\omega_d(\cG^{\rm melon}) |_{N_{\ext} = 4}
<
0
\,.
\eeq
Therefore, we have convergent melonic graphs at $N_{\ext} = 4$.
Divergent non-melonic graphs at $N_{\ext} = 4$ dominate all 
melonic graphs. 

Insisting on having a derivative coupling  in the direct space, that is on $(U(1)^D)^{d}$, we  impose that $a$ and $b$ are integers. In that situation,  
we have the obvious solutions to make $D$ a multiple  of $4$. 
Having covered the parameter space for finding interesting models, we will 
prove that,  in section \ref{sect:renmo+}, all models for generic $(d,D)$
(including those of Table \ref{table:table2}) are in fact  just-renormalizable.

\subsection{Models $\times$}
\label{sect:modelstimes}

We work under the same definition and conditions as in section \ref{sect:modelsplus},
where $V_{(4)}$ presently denotes $V_4 + V_{\times ;4}$.

\begin{lemma}[Bound of $\rho_{\times}$]\label{lem:rhobx}
Let $\cG$ be a graph with $N_{\ext}$ external legs and $Br$ c-bridges.
We have $\rhot(\cG) \le 2 V_{\times;4}$.  

If $\cG$ is such that

-a-  $V_{(4)} =1$ and $N_{\ext}=4$, then $\rhot(\cG)=0$ 

-b- $V_{(4)}=1$, $N_{\ext}=2$, $\rhot(\cG)\leq 1$

-c- $V_{(4)}>1$, then $\rhot(\cG) \le 2 V_{(4)} - {N_{\ext} \over 2} - Br$.

If $\cG$ is melonic and

-d- $V_{(4)} =1$, $N_{\ext}=2$  then $\rhot(\cG)=0$. 

-e- $V_{(4)}>1$, then  $\rhot(\cG) \le 2 V_{(4)}- N_{\rm ext}- 2 Br$.

\end{lemma}

\proof  The first statement should not bring any difficulties. 
  Now let us consider a general 1PI  graph. 
Assume  $V_{(4)}=1$ and  $N_{\ext}=4$, then $\rhot(\cG)  =0$,
there are no closed faces and so nothing to count. 
 Now if $V_{(4)}=1$, $N_{\ext}=2$, 
two cases might happen. Either the graph is melonic, 
and  there are no enhanced faces {\it i.e.} $\rhot(\cG)=0$,
or the graph is non-melonic and the vertex might still contribute
or not to $\rhot(\cG)$, thus $\rhot(\cG) \le 1 $. 
Then we have shown that a, b, d, are true for any 1PI graphs. 
For 1PR graph, we simply observe that the presence of 
bridge at the external legs will not affect the counting
of internal faces visiting the vertex counting in $V_{(4)}$. 
Thus a,b and d are valid in this case.

A 1PI graph, with $V_{(4)}>1$,  has at most 2 external legs
per vertex. Consider a vertex having  exactly 1 external leg: then 
this vertex will contribute at most  1 to $\rhot$. 
If a vertex has 2 external legs, then 2 cases may occur: either the 2 legs are
on the same external face which cannot contribute to $\rhot$ or the legs are incident to partner
vertices. In the latter case, there are 2 external faces of that vertex
which cannot contribute  to $\rhot$. Hence, the upper bound
for $\rhot(\cG)$ is $2V_{(4)} - N_{\ext}/2$.

For a 1PR graph $\cG$, we cut all bridges  to obtain 1PI subgraphs of the graph $\tilde\cG$. 
On each component $\tilde\cG_{j}$, we use the 1PI general bound 
$\rhot(\tilde\cG_j) \le 2 V_{(4)}(\tilde\cG_j)
- {1 \over 2}N_{\ext}(\tilde\cG_j) $. 
As we perform in the proof of Lemma \ref{lem:rhob}, we can 
show that the sum over the components brings
$\rhot(\cG)=\rhot(\tilde\cG) \le 2 V_{(4)}- \frac{N_{\ext}}{2}- Br$.  

For a melonic graph, the above bounds must be refined.
According to the same discussion in the proof of Lemma \ref{lem:rhob}, we know that for a 1PI melonic graph, each vertex having external legs must have $N_{\ext} = 2$. (If $N_{\ext} =4$, then the vertex gets disconnected and this is the case with $V_{(4)}=1$.)
These two external legs must be on partner vertices $v$ and $\tilde v$. 
Hence the enhanced faces on this vertex are necessarily external and cannot contribute to $\rhot(\cG)$.
Repeating the argument for all vertices with external legs, we get $\rhot(\cG) \le  2 (V_{(4)} - {N_{\ext} \over 2}) = 2 V_{(4)}- N_{\ext}$.

Consider a melonic 1PR graph $\cG$. Using again the same strategy, we cut all the bridges
in $\tilde\cG$, and apply the relation $\rhot(\tilde\cG_j) \le 2 V_{(4)}(\tilde\cG_j)- N_{\ext}(\tilde\cG_j) $ for each 1PI component, we get
\be
\rhot(\tilde\cG^{\rm melon}) =
\sum_{j}\rhot(\tilde\cG_j) 
\le \sum_{j}[2 V_{(4)}(\tilde\cG_j)- N_{\ext}(\tilde\cG_j)  ]
 = 2 V_{(4)} - N_{\ext} - 2 Br\,,
\ee
which together with $\rhot(\cG^{\rm melon}) = \rhot(\tilde\cG^{\rm melon}) $ is the second relation for melonic graphs for $V_{(4)}>1$. 

\qed 

The bounds of Lemma \ref{lem:rhobx} should be chosen wisely 
when bounding the degree of divergence of the graph. 
Furthermore, again the generic case of $V_{(4)}>1$ will be the important
one that we will concentrate on.

\

\noindent{\bf Potentially renormalizable models.} 
We study only primitively divergent graphs and fix $Br=0$. 
Combining \eqref{degx} and \eqref{eq:face} and using $2 L = n \cdot V - N_{\ext}$, we obtain the bound for the degree of divergence in this model, at $3a\leq 2b$,
\bea
\omt(\cG)
 &=&  - {2 D \over (d^-)!} ( \omega(\cG_{\rm color}) - \omega (\partial \cG))  - D (C_{\partial \cG} - 1) 
-{1\over 2} \left[ ( D \, d^- - 2 b) N_{\ext} - 2 D \, d^- \right] 
\crcr
&&+ {1 \over 2} \left[ -2 D\, d^- + (D \, d^- - 2 b) n \right] \cdot V  + 2 a \rhot
+\sum_{\xi=a,2a,b}2\xi\rho_{2;\xi} \,.
\eea
Using the Lemma \ref{lem:rhobx}, and 
further inserting that $\omega(\cG_{\rm color})=0$ and $\omega (\partial \cG) = 0$ for melonic graphs, and $\omega(\cG_{\rm color}) - \omega (\partial \cG)  
\ge 
{1 \over 2} (d^- - 1) d^- !$ for non-melonic graphs, the following 
bound is true: 
\bea
&&
\omt(\cG^{\rm melon}) 
\le
- D (C_{\partial \cG} - 1) 
- {1 \over 2} \left[ (D  d^- - 2b + 4 a)N_{\rm ext} - 2D d^-\right] 
\cr
&&
- 2 b  V_2     - 2\sum_{\xi=a,2a}(b-\xi)V_{2;\xi}
+ ( D  d^- - 4 b + 4a) V_{(4)}
- 2 a \triangle^{\rm melon}_{\times}
\,,
\label{eq:1omegamelx0}
\\ \cr
&&
\omt(\cG^{\rm non-melon}) 
\le 
-  D (d^- - 1) - 
D (C_{\partial \cG} - 1) - {1 \over 2} \left[ (
D  d^- - 2b + 2a )N_{\rm ext} - 2 \, 
D d^-\right] \cr
&&
- 2b V_2     - 2\sum_{\xi=a,2a}(b-\xi)V_{2;\xi}
+ ( D d^- - 4b + 4a) V_{(4)}
- 2a\triangle^{\rm non-melon}_{\times}
\,, 
\label{eq:1omeganonmelx0}
\eea
where we define
\bea
\triangle^{\rm melon}_{\times} &=&  2 V_{(4)} 
- N_{\ext} - \rhot (\cG^{\rm melon}) \ge 0 \,,
\crcr
\triangle^{\rm non-melon}_{\times} &=& 2 V_{(4)}  - {N_{\ext} \over 2} 
- \rhot(\cG^{\rm non-melon}) \ge 0 \,,
\eea
and get the inequalities from   Lemma \ref{lem:rhobx}. 
Thus, we obtain 
\bea
&&
\omt(\cG^{\rm melon}) 
\le 
- 
D (C_{\partial \cG} - 1) 
- {1 \over 2} \left[ (
D  d^- - 2 b + 4 a) N_{\rm ext} - 2D d^-\right] \cr
&&
- 2 \, b \, V_2     - 2\sum_{\xi=a,2a}(b-\xi)V_{2;\xi}
+ ( D d^- - 4 b + 4  a ) V_{(4)}
\,,
\label{eq:1omegamelx}
\\
\cr
&&
\omt(\cG^{\rm non-melon}) 
\le 
- D (d^- - 1) 
- D (C_{\partial \cG} - 1) - {1 \over 2} \left[ (
D  d^- - 2\, b + 2\, a )N_{\rm ext} - 2 D d^-\right] \cr
&&
- 2b V_2      - 2\sum_{\xi=a,2a}(b-\xi)V_{2;\xi}
+ (D d^- - 4b + 4a) V_{(4)}
\,.
\label{eq:1omeganonmelx}
\eea
Seeking renormalizable models, we require 
\beq 
D \, d^- - 4 \, b + 4 \, a \leq  0 \,, \qquad 2a\leq b \,, 
\eeq
where the second condition,  more stringent than $3a \leq 2b$, will be kept. 
This gives for $a$,
\beq
a 
\leq   b - {1 \over 4} Dd^-  \,,  \qquad 
a \leq \frac{b}{2}
\,.
\label{eq:validity2x}
\eeq
To achieve just-renormalizability, we use $a=b - {1 \over 4} Dd^-$ (which implies 
$b\leq \frac{Dd^-}{2}$), \eqref{eq:validity2x}, in \eqref{eq:1omegamelx} and \eqref{eq:1omeganonmelx} and require that, for a number of external legs  higher
than 4, we have convergence:  
\bea
&&
 \omt(\cG^{\rm melon})|_{ N_{\ext} \ge 6 } < 0  \,,
\crcr
&&
  \omt(\cG^{\rm non-melon})|_{ N_{\ext} \ge 6 } < 0\,. 
\label{eq:rangex}
\eea
From \eqref{eq:1omegamelx} and \eqref{eq:1omeganonmelx}, we have:
\bea
&&
\omt(\cG^{\rm melon}) |_{N_{\rm ext} \ge 6} 
\le\cr
&&
\Big[
- 
D (C_{\partial \cG} - 1) 
+ D d^-
-b  N_{\rm ext} 
- 2 b V_2   -\frac12 Dd^-  V_{2;a}   -( Dd^- -2b)V_{2;2a}
\Big]
\Big|_{N_{\rm ext} \ge 6} 
\,,
 \cr\cr
&&
\label{eq:1omegamelxjust}
\\
&&
\omt(\cG^{\rm non-melon}) |_{N_{\rm ext} \ge 6} 
\le \cr
&& 
\Big[
D
- D (C_{\partial \cG} - 1) 
-{1 \over 4} D d^- N_{\ext}
- 2 b  V_2  -\frac12 Dd^-  V_{2;a}  -( Dd^- -2b)V_{2;2a}
\Big]\Big|_{ N_{\ext} \ge 6} 
\,.\cr\cr
&& 
\label{eq:1omeganonmelxjust}
\eea
The maximum value for $\omt(\cG)$ is reached at $N_{\ext}=6$, so we can always write an upper bound and further require convergence: 
\bea
&&
\omt(\cG^{\rm melon}) |_{N_{\ext} = 6}
\le 
 D \,d^-
-6 b 
< 0,
\label{eq:1omegamelxjust6}
\\
&&
\omt(\cG^{\rm non-melon}) |_{N_{\ext}=6}
\le 
- D ({3 \over 2} d^- - 1)
< 0 
\,.
\label{eq:1omeganonmelxjust6}
\eea
We note here that $d^->{2\over3}$ \eqref{eq:1omeganonmelxjust6} is trivially satisfied in our study in which we only consider tensors with rank $d \ge 3$.
Hence, for just renormalizability, we impose  
\be
{ D d^- \over 6} < b \leq \frac{Dd^-}{2} \,, \qquad 
a=b - {1 \over 4} D \, d^- \,. 
\label{eq:1bnonmelxjust6}
\ee
However \eqref{eq:1bnonmelxjust6} also entails 
$ a > -{D d^- \over 12}$. Restricting  to $a > 0$,  
the bound of $b$ given can be improved. 
For just-renormalizability ({\it i.e.,} the equality in \eqref{eq:validity2x}, and \eqref{eq:rangex} together with $a>0$),
we impose
\be
{\boxed
{
{ D d^- \over 4} < b \leq \frac{Dd^-}{2}  \,, \qquad 
a=b - {1 \over 4} D \, d^-  > 0
}
}
\label{eq:1bnonmelxjust6final}
\ee
whose values for given positive integer values of $D$ and $d$ are given  in Table \ref{table:table45}.

\begin{table}[H]
\setlength{\extrarowheight}{0.2cm}
\centering
\begin{tabular}{lccccccccccc
|}  \hline\hline
&$d^-=2$&$d^-=3$&$d^-=4$&$d^-=5$ \\
\hline\hline
$D = 1$ 
&\pbox{2.5cm} {$ 0 < a  \le {1 \over 2} $ \\ ${1 \over 2} <b  \le 1$ } 
& \pbox{2.5cm}{$0 < a \le {3 \over 4} $ \\ ${3 \over 4} <b \le {3 \over 2}$} 
&\pbox{2.5cm}{$0 < a  \le {1} $ \\ ${1} <b  \le { 2}$}
&\pbox{2.5cm}{$0 < a  \le {5 \over 4} $ \\ ${5 \over 4}<b  \le {5 \over 2} $}
\\ \hline
$D = 2$ 
&\pbox{2.5cm}{$0 < a \le{1} $ \\ ${ 1} <b \le {2}$} 
&\pbox{2.5cm}{$0 < a \le {3 \over 2}$ \\ ${3 \over 2}<b  \le { 3} $} 
&\pbox{2.5cm}{$0 < a  \le {2}$ \\ ${2} <b \le {4} $} 
&\pbox{2.5cm}{$0 < a  \le {5 \over 2} $ \\ ${ 5 \over 2} <b  \le{5}$} 
\\ \hline
$D = 3$ 
&\pbox{2.5cm}{$0 < a   \le {3 \over 2} $ \\ ${ 3 \over 2}<b \le {3} $} 
&\pbox{2.5cm}{$0 < a  \le { 9 \over 4} $ \\ ${ 9 \over 4} <b \le {9 \over 2} $}
&\pbox{2.5cm}{$0 < a \le { 3}  $ \\ ${3} <b \le{6} $}
&\pbox{2.5cm}{$0 < a  \le { 15 \over 4} $ \\ ${ 15 \over 4}<b  \le {15 \over 2}  $}
\\ \hline
$D = 4$ 
&\pbox{2.5cm}{$0 < a  \le 2 $ \\ $2 <b  \le 4$} 
&\pbox{2.5cm}{$0 < a   \le 3$ \\ $3 <b  \le 6$}
&\pbox{2.5cm}{$0 < a  \le 4$ \\ $4 <b \le 8 $}
&\pbox{2.5cm}{$0 < a   \le 5 $ \\ $5 <b \le 10 $}
\\ \hline\hline
\end{tabular}
\caption{Allowed region of the values of $a$ and $b$ for potentially just-renormalizable 
 models $\times$ with $d^- \le 5$ and $D \le 4$.
}
\label{table:table45} 
\end{table}

Let us understand what is entailed by the just-renormalizability condition $D \, d^- - 4 \, b + 4 \, a =0$,
at $N_{\ext}=4$. We have 
\bea\label{nmel4x}
\omt(\cG^{\rm non-melon}) |_{N_{\ext}= 4}
&\le& 
- \, 
D (d^- - 1) - \, 
D (C_{\partial \cG} - 1) - {1 \over 2} 
\left[ {D \, d^- \over 2} \cdot  4 - 2 \, D \,d^-\right] 
\cr
&&
- 2 b  V_2   -\frac12 Dd^-  V_{2;a} -( Dd^- -2b)V_{2;2a}
\crcr
&\le& 
- D (d^- - 1)   <  0 \,,
\eea
since we only consider tensors of rank $d \ge 3$. 
Thus,  non-melonic graphs with $N_{\ext} =4$ are found all convergent.
Similarly for melonic graphs, requiring just-renormalizability means 
$D \, d^- - 4 \, b + 4 \, a =0$, leading to 
\bea\label{mel4x}
&&
\omt(\cG^{\rm melon}) |_{N_{\ext} =4}
\le - 
D (C_{\partial \cG} - 1) 
- {1 \over 2} \left[ (2 a + {D d^- \over 2} )4 - 2 \, D \,d^-\right]  \\
&& 
- 2 b  V_2  -\frac12 Dd^-  V_{2;a} -( Dd^- -2b)V_{2;2a} 
\le 
- 4 a < 0 \,.
\nonumber 
\eea
Therefore,  all melonic graphs with $N_{\ext} =4$ are also convergent.

Further, we analyze graphs of $N_{\ext} = 2$ under the same condition
and find  
\bea
&&
\omt(\cG^{\rm melon})  |_{N_{\ext} =2}
\le 
- 
D (C_{\partial \cG} - 1) 
- {1 \over 2} \left[ 
2b \cdot 2 - 2 
Dd^-\right] \\
&&
- 2 b  V_2 -\frac12 Dd^-  V_{2;a} -( Dd^- -2b)V_{2;2a}
\le
-2 b + D d^-
<
{D d^- \over 2}  
\cr\cr
&&
\omt(\cG^{\rm non-melon}) |_{N_{\ext} =2}
\le
- D (d^- - 1) -
D (C_{\partial \cG} - 1) - {1 \over 2} \left[ {D d^- \over 2} \cdot 2 - 2 D d^-\right] \crcr
&&
- 2 b  V_2  -\frac12 Dd^-  V_{2;a} -( Dd^- -2b)V_{2;2a}
 \le 
 {1 \over 2} D(2 - d^- )
\leq 0 \,, 
\eea
where we used  \eqref{eq:1bnonmelxjust6final}, and $d^-\ge 2$. 
In summary, at $N_{\ext}=2$, both melonic and 
 non-melonic graphs might be divergent. 
A closer look shows that non-melonic graphs 
 can be at most logarithmically divergent at rank $d\le 3$. 
Furthermore, as observe above, if we increase $D$ or $d^-$, 
we  see that melons could be again the dominant amplitudes. 
 
We conclude that, for potentially just-renormalizable  models $\times$, {\it i.e.}, under  \eqref{eq:validity2x} and \eqref{eq:rangex}, only  graphs with $N_{\ext} = 2$
might be divergent. 

Let us emphasize that the model $\times$ appears as a new type of renormalizable
theory.  Indeed, the coupling constants $\lambda$ and $\rhop$ do not get any corrections, {\it i.e.,} do not get renormalized, but degree-2 vertices will do. 
In ordinary QFT and invariant tensor field theory, when a model acquires this property it becomes super-renormalizable, that is, there  is a finite number of graphs which contribute to the flow of the mass. That is for example the case, of the scalar 
 $P(\phi)_2$-model and even 
non-local super renormalizable tensor field theories \cite{Geloun:2013saa,Carrozza:2012uv}. However, in the present case, as we will see in the following, the model $\times$ at $d=3$ will have an infinite number of graphs which will contribute to the mass
renormalization. We attribute this property to the presence of 
 enhanced interactions in the model.

As a concrete study in section \ref{sect:renmox}, we will focus on  $a = {1 \over 2}$, $b = 1$ for $D = 1$ and $d^- = 2$ as satisfied in Table \ref{table:table45}.

\section{Rank $d$ just-renormalizable models $+$} 
\label{sect:renmo+}

In this section, we analyze a class of model $+$ which will be proved
renormalizable for arbitrary $d$ and $D$.
We  provide the list of their primitively divergent  graphs and 
proceed to the expansion of those around their local and diverging part. 
Our goal is to show that the divergent parts in this expansion recasts as a coupling
and so a subtracting scheme can be performed. 
Dealing exclusively with 
graphs with external legs, we have $C_{\bG}\geq 1$. Note also that 
the theory has bipartite graphs such that $N_{\ext}$ is an even number.

\subsection{List of divergent graphs}
\label{subsect:list+}

Consider an arbitrary model in the class of models $+$ at fix $(d,D)$, with $a = D(d^--1) /2, b = D(d^- -\frac12)/ 2$. 
Using \eqref{eq:1omegameldelta0} and \eqref{eq:1omeganonmeldelta0}, in the same notations and conditions
introduced above, the superficial degree of divergence is given by:
\beq
\omp (\cG^{\rm melon}) \le  - D\Big[(C_{\partial \cG} - 1) + {1 \over 2} ( (d^- -  {1 \over 2}) N_{\ext} - 2d^- ) +(d^- - {1 \over 2})\, V_2  
+\frac12 V_{2;a} 
+  (d^--1) \triangle^{\rm melon}_{+}\Big] 
\,,
\label{eq:1omegamel1}
\eeq
\bea
&& 
\omp  (\cG^{\rm non-melon}) \le - D\Big[ (d^-- 1) + (C_{\partial \cG} - 1) 
+ {1 \over 2} ( {1 \over 2} N_{\ext} - 2d^- ) \cr\cr
&& \qquad \qquad   + (d^--{1 \over 2}) V_2 
+ \frac12 V_{2;a} + (d^--1)V_{4} 
+ (d^--1) \triangle_{+}
\Big] 
\,.
\label{eq:1omeganonmel1}
\eea
We have already shown that for any graph such that $N_{\ext}\ge 6$, the amplitude
is convergent. At $N_{\ext}=4$, non-melonic graphs have maximal degree of divergence  0 (logarithmic divergence) and  melonic graphs converge. 

The following cases occur 
\begin{itemize}
\item[(i)]  $N_{\rm ext} = 4$,  
\bea
&&
\omp (\cG^{\rm melon})   \le - D(d^--1) <0 \,, 
\\
&&
\omp (\cG^{\rm non-melon})
\le - D\Big[  (C_{\partial \cG} - 1) 
+ (d^--{1 \over 2}) V_2 
+ \frac12 V_{2;a} + (d^--1)V_{4} \cr\cr
&&
+  (d^--1) \triangle_{+}\
\Big] \le  0 \,.
\nonumber
\eea
For non-melonic graphs, the upper bound saturates only if 
$C_{\partial \cG} = 1$, $V_4=V_2=V_{2;a}=0$, and  $\triangle_{+} = 0$, 
{\it  i.e.} $\rhop(\cG^{\rm non-melon})=V_{+;4}$.  

\item[(ii)] 
$N_{\rm ext} = 2$:  we can combine $V_{(4)}>1$ and $V_{(4)}=1$ at
$N_{\ext}=2$ from Lemma \ref{lem:rhob}. Thus we can write a single bound,
$V_{(4)}\ge 1$ as 
\be
\omp (\cG^{\rm melon}) \le   - D\Big[(C_{\partial \cG} - 1)     -  {1 \over 2}  +(d^- - {1 \over 2})\, V_2  
+\frac12 V_{2;a} 
+  (d^--1) \triangle^{\rm melon}_{+}\Big] 
\le  {D \over 2} \,.
\ee
Whenever $C_{\bG} -1>0$, or $V_2>0$, $V_{2;a}>1$, or $\triangle^{\rm melon}_{+}>0$,
  the graph becomes convergent. The only way to achieve
a divergence with $\omega_d (\cG^{\rm melon}) =\frac{D}{2}$ is to set the 
above  quantities to  0. 
Note that, for a graph with $N_{\ext} = 2$, $\triangle^{\rm melon}_{+}=0$ means  $\rhop(\cG^{\rm melon}) = V_{(4)} - 1$ from Lemma \ref{lem:rhob}.
But we also have the bound $\rhop(\cG^{\rm melon})  \le V_{+;4}$, therefore 
writing  $\rhop(\cG^{\rm melon}) = V_{+;4} - p$, $p\ge 0$, implies
that $V_{4} = 1-p \geq 0$. Thus $p=0$, yields $(\rhop(\cG^{\rm melon}) = V_{+;4}, V_4 =1)$ or $p=1$ and then $(\rhop(\cG^{\rm melon}) = V_{+;4} -1, V_4=0)$. 

The case $\omega_d (\cG^{\rm melon}) =0$
might occur for  $C_{\bG} -1=0$, $V_2=0$, $\triangle^{\rm melon}_{+}=0$, 
and $V_{2;a}=1$. Then $\triangle^{\rm melon}_{+}=0$ means that one of
the following two cases occurs, $(\rhop(\cG^{\rm melon}) = V_{+;4}, V_4 =1)$ or ($\rhop(\cG^{\rm melon}) = V_{+;4} -1, V_4=0)$ can be produced.

For a non-melonic graph, we have, $V_{(4)}\ge 0$,
\bea
&& 
\omp (\cG^{\rm non-melon}) \le - D\Big[  (C_{\partial \cG} - 1) 
- {1 \over 2} \cr\cr
&&   + (d^--{1 \over 2}) V_2 
+ \frac12 V_{2;a} + (d^--1)V_{4} 
+ (d^--1) \triangle_{+}
\Big]\le \frac{D}{2}  \,, 
\eea
and, the only way to achieve divergence is to set 
 $C_{\bG} = 1$, $V_2 = 0$, $V_4=0$, and 
$\triangle_{+}=0$. The last
condition translates as $\rhop(\cG^{\rm melon})=V_{+;4} $. 
 Likewise, we can have  $\omp (\cG^{\rm non-melon})=\frac D2 $ for $V_{2;a}=0$, 
or  $\omp (\cG^{\rm non-melon})=0 $ for $V_{2;a}=1$. 
\end{itemize}

We have thus completed the proof of the following statement: 

\begin{proposition}[List of primitively divergent graphs for model $+$]\label{prop:list+}
The $p^{2a}\phi^4$-model $+$ with parameters $a=D(d^--1)/2, b=D(d^--\frac12)/2$
for two integers $d>2$ and $D>0$, has primitively  divergent graphs  with $(\Omega(\cG)=\omega(\cexG) - \omega(\bG))$:

\begin{table}[H]
\centering
\begin{tabular}{lcccccccccccccccc}
\hline\hline
$\cG$ && $N_{\ext}$ && $V_{2}$ &&  $V_{2;a}$  && $V_{4}$ && $\rhop$   && $C_{\bG}-1$ && $\Omega(\cG)$ && $\omega_d(\cG)$  \\
\hline\hline
 && 4 && 0 && 0 && 0 && $V_{+;4}$ && 0 && 1&& 0\\
I && 2 && 0  && 0  && 0 && $V_{+;4}$ && 0 && 1 && ${D \over 2}$  \\
II && 2 && 0  && 0 && 0 && $V_{+;4}-1$  && 0 && 0 && ${D \over 2}$ \\
III && 2 && 0  && 0 && 1 && $V_{+;4}$  && 0 && 0 && ${D \over 2}$ \\
IV && 2 && 0  && 1  && 0 && $V_{+;4}$ && 0 && 1 && $0$  \\
V && 2 && 0  && 1 && 0 && $V_{+;4}-1$  && 0 && 0 && $0$ \\
VI && 2 && 0  && 1 && 1 && $V_{+;4}$  && 0 && 0 && $0$ \\
\hline\hline
\end{tabular}
\caption{List of primitively divergent graphs of the $p^{2a}\phi^4$-model $+$.} 
\label{tab:listprim1}
\end{table}

\end{proposition}

Some divergent 2-point graphs are illustrated in Figures \ref{fig:V1_1new} and \ref{fig:V2melon_1new}
in appendix \ref{app:mod+} specializing to $d=3$ and $D=1$. They will contribute to the  mass renormalization for this model. 
Secondly, consider the 4-point amplitudes associated with the graphs of  Figure \ref {fig:V2nonmelon_1new} 
in appendix \ref{app:mod+}. These will contribute to the renormalization of couplings $\eta_+$ or $\lambda$ depending
on the external momentum data of the correlators. We can construct an infinite
family of divergent 4-point graphs in this model.

At the end of this section, the proof of the next theorem will be completed: 
\begin{theorem}\label{theoren+}
The $p^{2a}\phi^4$ model $+$  with parameters $a=D(d^--1)/2, b=D(d^--\frac12)/2$ 
for arbitrary rank $d\ge 3$ and dimension $D>0$ with action defined by \eqref{eq:actiond} is just-renormalizable at all orders of perturbation
theory. 
\end{theorem}

\subsection{Renormalization}
\label{subsect:rentensr}

The subsequent part of the renormalization program consists 
in the proof that the divergent and local part of all divergent amplitudes can be
recast as terms which are present in the Lagrangian of the model  $+$ of section \ref{subsect:list+} with fixed parameter $a=D(d^--1)/2$ and $b=D(d^--1/2)/2$.
 For that purpose, we perform a Taylor expansion of the amplitudes of graphs listed in  Table \ref{tab:listprim1} and show
that the divergent terms in that expansion are associated with either the mass, 
counter-terms $CT_{2;\xi}$, $\xi=a,b$,  or interaction terms plus convergent remainders.

\medskip 

\noindent {\bf Renormalization of marginal 4-point functions.}
Marginal 4-point functions are given by the first
line of Table \ref{tab:listprim1}. Given a connected and bipartite 
boundary graph of a 4-point graph, 
it is simple to realize that the pattern of its external momenta should follow 
either the pattern of 
${\bf V}_{4;s}$
or of 
${\bf V}_{+;4;s}$
\eqref{vertexkernel} (see Figure \ref{fig:4vertex}). 
The locality principle of the present model tells us to 
consider a graph issued from the expansion of correlators
of the form \eqref{phi4} or \eqref{pPhi4+} which translate as
\bea
&&
\langle  \phi_{12\dots d} \,\bar\phi_{1'2\dots d}\,\phi_{1'2'\dots d'} \, \bar\phi_{12'\dots d'}\rangle \,, \label{ps0}\\\cr
&&
\langle |p_{1}^{\ext}|^{2a}\, \bar\phi_{1'2\dots d}\,\phi_{1'2'\dots d'} \, \bar\phi_{12'\dots d'}\rangle \,, 
\label{ps1}
\eea
with $|p_{1}^{\ext}|^{2a}$ an external momentum with color $s=1$. 
In the following, we will concentrate on an expansion of a graph 
with external data of the form of the operator 
${\bf V}_{+;4;s=1}$. 
In other words, we will focus on  $s=1$ 
and a graph coming from the expansion of the correlator \eqref{ps1}. However, 
as it will be clear, our   analysis is without loss of generality 
 since the method can be extended to 
${\bf V}_{4;1}$
and then 
to 
${\bf V}_{+;4;s}$, 
for any color $s$.  

Consider a 4-point graph with 4 external propagators attached to it with external momenta governed by the pattern of  \eqref{ps1}.  This 4-point graph  carries $2d$ momentum labels; these are associated with $2d$ external faces, which we denote by 
$$f\in \cF_{\ext}=\{f_{[1]},f_{2}, \dots, f_{d}, f_{1'},f_{2'}, \dots,f_{d'}\},$$ where we emphasize the
face which is enhanced by a square bracket (say [1]). 
Let  $A_{4}(\{p^{\ext}_f\})$ be the amplitude of  such a graph.  Two types of scale indices have to be considered in this amplitude: 
the external scales $j_l$ associated with external fields and 
which correspond to external propagators with labels $l$ and the (internal) scale $i$ 
of the $G^i_k$ graph. In short, a quasi-local graph $G^i_k$ implies 
that  $j_l \ll i$.

We have from \eqref{amplf} 
\bea\label{apl+}
&&
A_{4}
(\{p_{f_s}^{\ext}\})
=
 \kappa(\pet)
\sum_{p_{f_s}} 
\int \Big[\prod_{l\in \cL} d\alpha_l\, e^{-\alpha_l \mu} \Big]
\Big[\prod_{f_s\in \cF_{\ext}} e^{-(\sum_{l\in f_s} \alpha_l) \sum_\xi|p^{\ext}_{f_s}|^{2 \xi}}\Big]
\crcr
&&\times \Big[\prod_{f_s\in \cF_{\inter}} e^{-(\sum_{l\in f_s} \alpha_l)\sum_\xi |p_{f_s}|^{2\xi}} \Big]
 |p^{\ext}_{f_{[1]}}|^{2a}
\Big[\prod_{s=1}^{d}\prod_{v_s \in \cV_{+;4;s}} (\epsilon\,\tilde{p}^{\,2a})_{v_s} \Big]\,, \cr\cr\cr
&&
(\epsilon\,\tilde{p}^{\,2a})_{v_s} := 
\sum_{f_{s'}} \epsilon_{v_s,f_{s'}}(\tilde{p}_{f_{s'}})^{2a}\,, 
\eea
where $\kappa(\pet)$ includes symmetry factors
and coupling constants. 
We recall that $p^{\ext}_{f_s}$ 
are external momenta, and the last line shows  
$\tilde{p}_{f_s}$ which refers to an internal or an external momentum. 
Let us concentrate on the range of the parameters $\alpha$: 
for an internal line $l$, that we will now denote $\ell$, 
$\alpha_\ell \in [M^{- {(2 b)} \, i_\ell}, M^{- {(2 b)}\,  (i_\ell-1)}]$;
for an external line $l$, now denoted $\lex$,  $\alpha_\lex \in [M^{- {(2 b)} \, j_{\lex}}, M^{-{(2 b)} \,(j_{\lex}-1)}]$.
We are interested in a regime when $j_\lex \ll i \leq i_\ell$.

A Taylor expansion over an external face amplitude gives 
\bea
e^{-(\sum_{l \in f}\alpha_l)\sum_\xi |p_{f}^{\ext}|^{2 \xi}} 
&=& e^{-(\alpha_\lex+\alpha_{\lex'})\sum_\xi |p_{f}^{\ext}|^{2 \xi}} [1- R_f] 
\crcr
R_f &=&  \big(\sum_{\ell \in f }\alpha_\ell\big)\big(\sum_\xi |p_{f}^{\ext}|^{2 \xi}\big)
\int_0^1 e^{-t(\sum_{\ell \in f }\alpha_\ell) \sum_\xi|p_{f}^{\ext}|^{2 \xi}} dt \,,
\label{tayface}
\eea
where  $\sum_{\ell \in f}\alpha_\ell$ is  small  
($ \alpha_{\ell} \sim \mathcal{O} ({1 \over {|p_{f_s}|}^{2\xi}})\sim M^{-(2\xi)i_\ell}$). 
We insert that expansion for each external face in \eqref{apl+} and obtain: 
\bea
&&
A_{4}(\{p^{\ext}_f\})
=
\kappa( \pet)
\sum_{p_f} \int [\prod_{l\in \cL}d\alpha_l e^{-\alpha_l \mu
}] |p^{\ext}_{f_{[1]}}|^{2a}\Big[
\prod_{f\in \cF_{\ext}}
e^{-(\alpha_{\lex}+\alpha_{\lex'})\sum_\xi  |p_{f}^{\ext}|^{2 \xi}} \Big]\label{a6full}\\
&& \times\Big[1- \sum_{f \in \cF_{\ext}} R_f  + \sum_{f,f' \in \cF_{\ext}} R_f R_{f'} + \dots \Big] \Big[
\prod_{f \in \cF_{\inter}}
e^{-(\sum_{\ell \in f}\alpha_\ell)\sum_\xi  |p_{f}|^{2 \xi}}  \Big]
\prod_{s=1}^{d}\prod_{v_s \in \cV_{+;4;s}} [(\epsilon\,\tilde{p}^{\,2a})_{v_s} ].
\nonumber 
\eea
The dots are higher order products in the $R_f$'s. 
From Table \ref{tab:listprim1}, $\rhop $ \eqref{rho}
must be equal to $V_{+;4}$.  Hence, in each vertex kernel, we must collect
and integrate one momentum of a closed face. A divergent 4-point graph
satisfying the first row of Table \ref{tab:listprim1} 
 must be such that no external momenta can be found
within $\prod_{s=1}^{d}\prod_{v_s \in \cV_{+;4;s}} [(\epsilon\,\tilde{p}^{\,2a})_{v_s} ]$. 

We write the 0th order in expansion in $R_f$ as:
\bea
A_{4}(\{p^{\ext}_f\};0)
&=&
\kappa(\pet)
 \sum_{p_f} \int [\prod_{l}d\alpha_l e^{-\alpha_l \mu
}] |p^{\ext}_{f_{[1]}}|^{2a}\prod_{f\in \cF_{\ext}}\Big[
e^{-(\alpha_{\lex}+\alpha_{\lex'}) \sum_\xi |p_{f}^{\ext}|^{2 \xi}} \Big]
\crcr
&\times&
\Big[ \prod_{f \in \cF_{\inter}}
e^{-(\sum_{\ell \in f}\alpha_\ell)\sum_\xi |p_{f}|^{2 \xi}}  \Big] \Big[
\prod_{s=1}^{d}\prod_{v_s \in \cV_{+;4;s}} (\epsilon\,\tilde{p}^{\,2a})_{v_s} \Big]\,,
\cr\cr
&=& 
\kappa(\pet)
\Big[\int [\prod_{\lex}d\alpha_{\lex} 
e^{-\alpha_\lex \mu
}] |p^{\ext}_{f_{[1]}}|^{2a}
\prod_{f\in \cF_{\ext}}
e^{-(\alpha_\lex+\alpha_{\lex'})\sum_\xi |p_{f}^{\ext}|^{2 \xi}} 
\Big]
\label{a4}\\
&\times&
\Big[
\sum_{p_f} \int [\prod_{\ell}d\alpha_\ell e^{-\alpha_\ell \mu
}]
\prod_{f \in \cF_{\inter}}\Big[
e^{-(\sum_{\ell }\alpha_\ell) \sum_\xi |p_{f}|^{2 \xi}}  
\Big]
\Big[
\prod_{s=1}^{d}\prod_{v_s \in \cV_{+;4;s}} (\epsilon\tilde{p}^{\,2a})_{v_s} \Big]
\Big].
\nonumber
\eea
The factor involving external lines can be re-expressed 
as 
\bea
&&
\int [\prod_{\lex}d\alpha_{\lex} e^{-\alpha_{\lex} \mu
}] |p^{\ext}_{f_{[1]}}|^{2a}
\prod_{f\in \cF_{\ext}}
e^{-(\alpha_\lex+\alpha_{\lex'}) \sum_\xi |p_{f}^{\ext}|^{2 \xi}}
= \cr\cr
&& 
\int [\prod_{\lex}d\alpha_{\lex} ]  
|p^{\ext}_{f_{[1]}}|^{2a} \cr\cr
&& \times 
e^{-\alpha_{\lex_1}[\sum_\xi (|p_{f_{[1]}}^{\ext}|^{2\xi} + |p_{f_2}^{\ext}|^{2\xi}+\dots + |p_{f_d}^{\ext}|^{2\xi})+\mu]}
e^{-\alpha_{\lex_2}[\sum_\xi (|p_{f_{1'}}^{\ext}|^{2\xi} +| p_{f_2}^{\ext}|^{2\xi}+\dots + |p_{f_d}^{\ext}|^{2\xi})+\mu]}\cr\cr
&&  \times 
e^{-\alpha_{\lex_3}[\sum_\xi (|p_{f_{1'}}^{\ext} |^{2\xi}+ |p_{f_{2'}}^{\ext}|^{2\xi}+ 
\dots +|p_{f_{d'}}^{\ext}|^{2\xi})+\mu]}
e^{-\alpha_{\lex_4}[\sum_\xi (|p_{f_{[1]}}^{\ext} |^{2\xi}+| p_{f_{2'}}^{\ext}|^{2\xi}+\dots+| p_{f_{d'}}^{\ext}|^{2\xi})+\mu]}\,. 
\crcr
&&
\label{eq:finite}
\eea
Observe that the above expression describes 4 propagators glued in a way to produce the pattern of a vertex of type ${\bf V}_{+;4;1}$.
The term associated with the sum over internal momenta  is log-divergent. 
Therefore, the amplitude $A_{4}(\{p^{\ext}_f\};0)$ will renormalize the coupling $\pet$. 

We now prove that the remainders appearing in \eqref{a6full}
lead to convergence by improving the power counting.
 The first remainder calling a single  term $R_f$ is of the form: 
\bea
R_{4}  
&=&\kappa(\pet)
\sum_{p_f} \int [\prod_{l}d\alpha_l e^{-\alpha_l \mu
}]
|p^{\ext}_{f_{[1]}}|^{2a} 
\Big[
\prod_{f\in \cF_{\ext}}
e^{-(\alpha_{\lex}+\alpha_{\lex'})\sum_\xi |p_{f}^{\ext}|^{2 \xi}} \Big] \cr\cr
 &\times&
\Big[- \sum_{f \in \cF_{\ext}}\big(\sum_{\ell \in f}\alpha_\ell\big) \big(\sum_\xi |p_{f}^{\ext}|^{2 \xi}\big)
\int_0^1 e^{-t(\sum_{\ell \in f}\alpha_\ell) \sum_\xi |p_{f}^{\ext}|^{2 \xi}} dt  \Big] 
\cr\cr
&\times &
\Big[\prod_{f \in \cF_{\inter}}
e^{-(\sum_{\ell \in f}\alpha_\ell) \sum_\xi |p_{f}|^{2 \xi}}  \Big]
\prod_{s=1}^{d}\prod_{v_s \in \cV_{+;4;s}} [ (\epsilon \tilde{p}^{\,2a})_{v_s} ]\,.
 \label{remain4}
\eea
Using $i(G^i_k)=\inf_{\ell\in G^i_k} i_\ell $ and $e(G^i_k)=\sup_{l \in G^i_k}j_l$,
and recalling
$\alpha_\ell \in [M^{- {(2 b)} \, i_\ell}, M^{- {(2 b)}\,  (i_\ell-1)}]$ and
$\alpha_{\lex} \in [M^{- {(2 b)} \, j_\lex}, M^{-{(2 b)} \,(j_\lex-1)}]$, 
$\big(\sum_{\ell \in f}\alpha_\ell\big) |p_{f}^{\ext}|^{ 2 b} \le k_0 M^{-
({2 b}) \, (i(G^i_k)- e(G^i_k))}$,
then $R_4$ is bounded as, 
\be
|R_{4}| 
\leq 
K 
\prod_{(i,k)} M^{-
({2 b}) \, (i(G^i_k)- e(G^i_k))}
M^{\omp(G^i_k)}\,, 
\label{bound4}
\ee
for some constant $K$ (which includes $\kappa(\pet)$ and $k_0$, a constant
depending on the graph and a constant which bounds the integral in $t$). 
The factor   $M^{-({2 b}) \, (i(G^i_k)- e(G^i_k))}$ improves
the power counting (which is already logarithmic) and will be the source of decay 
to show that the sum over scale attributions is convergent in
the way  established in \cite{Rivasseau:1991ub}.  
Inspecting the higher order products in $R_f$'s, one realizes that they are obviously more convergent and so do not need further discussion. 

 If we remove $|p^{\ext}_{f_{[1]}}|^{2a} $ from the amplitude \eqref{apl+}, we will be in presence of an amplitude coming from \eqref{ps0}. The boundary data of the resulting amplitude will be of the form ${\bf V}_{4;1}$. Repeating step by step the previous analysis, we obtain at 0-th order  of the expansion a renormalization of the coupling $\lambda$
 and all remainders will lead exactly to the same convergence with power counting
 given by \eqref{bound4}. It is direct to see that the argument extends to any 
 color $s$. 

As a result of this analysis, we can comment that, although the renormalized coupling $\lambda$ does not receive any melonic corrections, it receives contributions from the coupling $\pet$. 
This is a new property of the perturbative Renormalization Group equations of this model.

\medskip

\noindent {\bf Renormalization of divergent 2-point  functions.}
We study  2-point functions that obey $\omp(\cG)=0,{D \over 2}$
and  characterized by the rows I through VI of Table \ref{tab:listprim1}. 
First, we will focus on the  row I and point out the differences with 
II and III at particular steps of the discussion.  We will sketch the analysis for the 
rows IV, V and VI.

A 2-point graph has a unique boundary which is given 
by the invariant $\Tr_2(\phi^2)$ of Figure \ref{fig:4vertex}. 
Because the vertices are enhanced, it may happen that
the external data of a two-point graph is not of the form of a mass
term, but rather a form of $CT_{2;a}$.
A careful discussion will be made about this in the text. 

Consider  an amplitude $A_{2}(\{p^{\ext}_f\})$ associated with a
2-point graph obeying the row I of Table \ref{tab:listprim1}. 
This graph has $d$ external faces that we label by 
$$f\in \cF_{\ext}=\{f_{1},f_{2}, \dots, f_{d}\}. $$
Keeping the same  notations as above (see paragraph
after \eqref{apl+}),  $\lex$ labels external propagators with scale index $j_{\lex}$,
$\ell$ labels internal lines with scale index $i_\ell$.   

A Taylor expansion of the external face factors out
in the same form as \eqref{tayface} and leads us to the following
expansion of the 2-point amplitude: 
 \bea
 && 
A_{2}(\{p^{\ext}_f\})
=
\kappa( \pet)
\sum_{p_f} 
\int [\prod_{l}d\alpha_l e^{-\alpha_l \mu
}]\Big[
\prod_{f\in \cF_{\ext}}
e^{-(\alpha_{\lex}+\alpha_{\lex'})\sum_\xi |p_{f}^{\ext}|^{2 \xi}} \Big] 
\label{rfrf2}\\
&& \times \Big[1- \sum_{f \in \cF_{\ext}} R_f  + \sum_{f,f' \in \cF_{\ext}} R_f R_{f'} + \dots \Big]\Big[
\prod_{f \in \cF_{\inter}}
e^{-(\sum_{\ell \in f}\alpha_\ell) \sum_\xi |p_{f}|^{2 \xi}}  \Big]
\Big[
\prod_{s=1}^{d}\prod_{v_s \in \cV_{+;4;s}} [ (\epsilon\,\tilde{p}^{\,2a})_{v_s} \Big]\,.
\nonumber 
\eea
The 0th order term in the expansion in $R_f$,
\bea
&&
A_{2}(\{p^{\ext}_f\};0)=
\kappa( \pet)
\sum_{p_f} 
\int [\prod_{l}d\alpha_l e^{-\alpha_l \mu
}]\cr\cr
&&\times 
\Big[
\prod_{f\in \cF_{\ext}}
e^{-(\alpha_{\lex}+\alpha_{\lex'})\sum_\xi |p_{f}^{\ext}|^{2 \xi}} \Big] 
\prod_{f \in \cF_{\inter}}\Big[
e^{-(\sum_{\ell \in f}\alpha_\ell) \sum_\xi |p_{f}|^{2 \xi}}  \Big]
\prod_{s=1}^{d}\prod_{v_s \in \cV_{+;4;s}} [ (\epsilon\,\tilde{p}^{\,2a})_{v_s} ] 
\cr\cr
&&=
 \kappa(\pet)
\Big[\int [\prod_{\lex}d\alpha_{\lex} ]  
e^{-\alpha_{\lex_1}[\sum_\xi (\sum_s |p_{f_{s}}^{\ext}|^{2\xi})  +\mu]}
e^{-\alpha_{\lex_2}[\sum_\xi (\sum_s |p_{f_{s}}^{\ext}|^{2\xi}) +\mu]}
\Big]
\crcr
&&
\times
\Big[
\sum_{p_f} \int [\prod_{\ell}d\alpha_\ell e^{-\alpha_\ell \mu
}]
\prod_{f \in \cF_{\inter}}\Big[
e^{-(\sum_{\ell }\alpha_\ell) \sum_\xi |p_{f}|^{2 \xi}}  
\Big]
\Big[
\prod_{s=1}^{d}\prod_{v_s \in \cV_{+;4;s}} (\epsilon\tilde{p}^{\,2a})_{v_s} \Big]
\Big].
\cr\cr
&&
\label{massren+}
\eea
Let us discuss the vertex density. A graph fulfilling the requirements
of the row I of Table \ref{tab:listprim1}, must be such that $\rhop = V_{+;4}$.
The same argument as in the case of 4-point functions applies: 
there are no external momenta in the product of vertex kernels and
all momenta will be summed.  The first factor involving external momenta
represents two propagators glued together by a degree-2 vertex;
the second factor gives by our power counting  a degree
of divergence $\frac D2$.  Hence the term \eqref{massren+} 
 renormalizes the mass term. 

Concerning a graph satisfying the rows II and III, we know that $\rhop=V_{(4)}-1$
and that the graph is melonic. Lemma \ref{lem:rhob} explained in its
proof that for 1PI melonic graphs, external legs must be hooked on
partner vertices. Hence a 2-point primitively divergent (1PI) melonic graph has at least a vertex $v_{0;s}$
which will not contribute to $\rhop$. Two cases might occur:

(II) The vertex $v_{0;s}$ must belong to $\cV_{+;4;s}$, then an extra factor of $|p^{\ext}_{f_{s}}|^{2a}$ must be added to the boundary data. This makes that graph of the form of $\cV_{2;a;s}$. 
Now, adding all colored symmetric contribution with respect to $s$ 
of this graph term renormalizes the coupling  $Z_{a}$;

(III)  The vertex $v_{0;s}$ must belong to $\cV_{4}$, then  the boundary data is that
of a mass  and so \eqref{massren+} again renormalizes the mass term.

The remainders of \eqref{rfrf2} are now treated for the 
row I of our table. The first order remainder involving  the sum $\sum_{f}R_f$ can be bounded as  follows
\bea
&&
R_{2} = - \kappa({\pet}) \Big[ 
\int [\prod_{\lex }d\alpha_\lex e^{-\alpha_\lex \mu
}]
\prod_{f\in \cF_{\ext}}
e^{-(\alpha_\lex+\alpha_{\lex'})\sum_\xi |p_{f}^{\ext}|^{2 \xi}} \Big]
\crcr 
&& \times \sum_{f \in \cF_{\ext}}\Big[
\int [\prod_{\ell }d\alpha_\ell e^{-\alpha_\ell \mu
}] 
\big(\sum_{\ell \in f}\alpha_\ell\big) \big(\sum_\xi |p_{f}^{\ext}|^{2 \xi}\big) \int_0^1 e^{-t(\sum_{\ell \in f}\alpha_\ell) \sum_\xi |p_{f}^{\ext}|^{2 \xi}} dt  \Big] 
 \cr\cr
&& \times
\sum_{p_f} \prod_{f \in \cF_{\inter}}\Big[
e^{-(\sum_{\ell \in f}\alpha_\ell) \sum_\xi |p_{f}|^{2 \xi}}  \Big]
\Bigg]
\prod_{s=1}^{d}\prod_{v_s \in \cV_{+;4;s}} [(\epsilon\,\tilde{p}^{\,2a})_{v_s} ]
\cr\cr
&&|R_2|
\leq 
K \prod_{(i,k)}M^{-2b(i(G^i_k)-e(G^i_k))} M^{\omp(G^i_k)={D \over 2}} \,,
\label{eq:rem2+}
\eea
where we used the same scheme leading to \eqref{bound4}, 
for some constant $K$. For a constant $C\geq 1$,
 $- D (d^--\frac12) C +\frac{D}{2} \le -  D(d^--1) \le -D<0$ ensures the convergence of the remainder.
 All higher order remainders can be proved more convergent.
  From this point, the summation over scale attributions can be again 
 performed after subtractions.  
The case of an amplitude of the rows II and III of the table can be addressed
in similar way.

Let us now discuss the rows IV, V, and VI. In these cases,
$\rho_{2;a}=1=V_{2;a}$ which means that the enhanced  momenta $|p_{f_s}|^{2a}$ associated with  the vertex which is counted in $V_{2;a}=1$ is necessarily integrated.
Hence the analysis of  the rows IV, V, and VI are completely similar to what we have done for the  rows I, II, and III, respectively.  
As a last remark, we stress that there are no corrections to the wave function $Z_b$ because all amplitudes at first order are already convergent. 
Hence we can put the coupling $Z_b$ to 0.

\

In conclusion, 

- the expansion of marginal 4-point functions around their local part gives a log--divergent terms which renormalize the coupling constants $\pet$ or $\lambda$.

- the expansion of  ${D \over 2}$--divergent or
log-divergent 2-point graphs around their  local
parts yield a ${D \over 2}$--divergent  
or log-divergent term renormalizing either the mass
or $Z_a$; 

- all remainders are convergent and will bring enough decay  for ensuring the final summability over scale attributions. From this point, the procedure for performing this last sum over attributions is standard  and will secure the renormalization at all orders of perturbation theory according to techniques developed in \cite{Rivasseau:1991ub}. 
Thus, Theorem \ref{theoren+} holds.

\section{A rank $d=3$ renormalizable model $\times$}
\label{sect:renmox}

We adopt the same strategy as in the previous section,
to prove the renormalizability of a model $\times$. 
After listing its primitively divergent connected graphs,  we proceed with 
the renormalization procedure. The same  conditions $C_{\bG}\ge 1$ and $N_{\ext}$ even
must be true.

\subsection{List of divergent graphs}
\label{subsect:listx}

We are interested in a rank $d=3$ model $\times$ with $a = {1 \over 2}$, $ b =1$
and $D=1$. This choice $b=1$ gives a Laplacian in the kinetic term, 
and an integer power of the interaction
$|p|\phi^4$, thus this model seems natural (that we can think as a single derivative coupling). 
The interactions are of the same form as given in Figure \ref{fig:phi4EnhancedMelonModel} with the sole difference that
we enhance by a product of $|p|$ both edges in the melonic
interaction.  

We start from \eqref{eq:1omegamelx} and \eqref{eq:1omeganonmelx},  
the superficial degree of divergence for generic graphs: 
\bea
\omt (\cG^{\rm melon}) \!\!\! &\le& \!\!\! - (C_{\partial \cG} - 1) - {1 \over 2} (2 N_{\rm ext} - 4 ) - 2 V_2 
 - V_{2;a} 
-\triangle^{\rm melon}_{\times}
\,,
\label{eq:1omegamel12}
\\
\omt  (\cG^{\rm non-melon})  \!\!\! &\le& \!\!\! - 1- (C_{\partial \cG} - 1) - {1 \over 2} (  N_{\rm ext} - 4 ) - 2  V_2 
- V_{2;a} 
- \triangle^{\rm non-melon}_{\times}.
\label{eq:1omeganonmel12}
\eea
We observe that both counter-terms $CT_{2;2a}$ and $CT_{2;b}$ 
disappear from the power counting. As degree-2 vertices, they are neutral for 
the power counting.

Our previous analysis shows that, for any $N_{\ext} \ge  4$,  the amplitude is convergent. 
We concentrate on the remaining case $N_{\ext} = 2$. 
From Lemma \ref{lem:rhobx}, we can still combine the 
analysis of $V_{(4)}=1$ and $V_{(4)}>1$ at $N_{\ext}=2$ and  have
 \bea\label{111}
&&\omt (\cG^{\rm melon}) \le - (C_{\partial \cG} -1) -2 V_2  - V_{2;a} 
-\triangle^{\rm melon}_{\times}
\le 0
\,,
\cr\cr
&&
\omt (\cG^{\rm non-melon}) \le  - (C_{\partial \cG} -1) - 2 V_2  - V_{2;a} 
-\triangle^{\rm non-melon}_{\times}
\le 0  .
\eea
The only way to achieve logarithmic divergence
is to have exactly:  $C_{\partial \cG} =1$, $V_2=0=V_{2;a}$. 
For a non-melonic graph, we further impose
$\triangle^{\rm non-melon}_{\times}=0$ from which 
we infer $\rhot=2V_{(4)}-1$. Knowing that $\rhot  \le 2V_{\times;4}$, 
this leads us to  $(\rhot =2V_{\times; 4}-1; V_{4}=0)$. 
 The case of a melonic graph yields 
 $\triangle^{\rm melon}_{\times}=0$ which gives $\rhot=2V_{(4)}-2$, 
which  together with $\rhot  \le 2V_{\times;4}$ yields two possibilities:
either $(\rhot =2V_{\times; 4}; V_{4}=1)$ or $(\rhot =2V_{\times; 4}-2; V_{4}=0)$.

We have the following proposition. 

\begin{proposition}[List of primitively divergent graphs for  model $\times$]
The $p^{2a}\phi^4$-model $\times$ with parameters $D = 1, d=3,a={1 \over2}, b=1$, 
has the following primitively divergent  graphs which  obey
$(\Omega(\cG)= \omega(\cexG) - \omega(\bG))$
\begin{table}[H]
\begin{center}
\begin{tabular}{lcccccccccccccccc}
\hline\hline
$\cG$ && $N_{\ext}$ && $V_{2}$  && $V_{2;a}$ && $ V_4$ && $\rhot$   && $C_{\bG}-1$ && $\Omega(\cG)$ && $\omega_d(\cG)$  \\
\hline\hline
I && 2 && 0  && 0 && 0 &&  $2 V_{\times;4} -1$ && 0 && 1 && ${0}$  \\
II && 2 && 0  && 0 && 0 && $2 V_{\times;4}-2$  && 0 && 0 && ${0}$ \\
III && 2 && 0  && 0 && 1 && $2 V_{\times;4}$  && 0 && 0 && ${0}$ \\
\hline\hline
\end{tabular}
{\caption{List of primitively divergent graphs of the $p^{2a}\phi^4$-model  $\times$. \label{tab:listprim2}}}
\end{center}
\end{table}
\end{proposition}

In appendix \ref{app:modx}, we have illustrated an infinite family of 2-point graphs with log-divergent amplitudes, see Figures \ref{fig:V1_1x}, \ref{fig:V2melon_1x}
and  \ref{fig:V2nonmelon_mixx}. Thus, this theory is not super-renormalizable
in the usual sense because it possesses an infinite family of corrections
to the mass, $Z_{a}$ and $Z_{2a}$ couplings. It  does not also fit the definition of a just-renormalizable theory because all corrections of the coupling $\lambda$ and $\eta_\times$ are  finite.  
Again this is a specific feature brought by the enhancement of non-local tensor interactions. 

With the above analysis, we can now prove that  the following
theorem holds.  
\begin{theorem}\label{theoremx}
The $p^{2a}\phi^4$ model $\times$  with parameters $D = 1, d=3,a={1 \over2}, b=1$,   with action defined by \eqref{eq:actiond}  is renormalizable at all orders of perturbation.
\label{theorenx}
\end{theorem}

\subsection{Renormalization}
\label{subsect:renx}

We follow a similar scheme as developed in section \ref{subsect:rentensr}.
We sketch  the expansion of amplitudes of the graphs listed in Table \ref{tab:listprim2} and check that their local parts indeed take the form of the terms in the Lagrangian of the model  $\times$ of section \ref{subsect:listx}.
Doing a Taylor expansion the amplitudes, we will also show that the subleading orders are convergent.

\

\noindent {\bf Renormalization of divergent 2-point  functions.}
In the model $\times$, we only have 2-point log-divergent graphs as listed in Table \ref{tab:listprim2}. 
For type I and II graphs,
note that $\rhot<2 V_{\times; 4}$, therefore one  or two edges of their boundary graph are touched by external faces which are enhanced. 
This entails that the boundaries of these graphs are equipped with $|p^{\ext}_{f[1]}|^{2 a}$  and $|p^{\ext}_{f[1]}|^{4 a}$ and therefore of the form of 
$CT_{2; a}$, $CT_{2; 2a}$, respectively.
On the other hand, for a log-divergent graph of the type III, we have $\rhot= 2 V_{\times;4}$ and its boundary graph does not have any enhanced edges
and so takes the form of the mass term.

In the following, we only address the 2-point graphs of type I and II and will give the main 
points leading to the treatment of type III graphs. 

Let us consider the amplitude $A_{2}(\{p^{\ext}_f\})$ of 
2-point non-melonic and melonic graphs obeying, respectively, 
the rows I and II of Table \ref{tab:listprim2}.  These graphs have 3 external faces labeled by 
$f\in \cF_{\ext}=\{f_{[1]},f_{2}, f_{3}\}$, with an enhanced color $1$ strand. 
By an argument of symmetry, our following study will give the same result 
for a graph with another enhanced color. 

We perform a Taylor expansion of  external face factors
 as given in \eqref{tayface} and the amplitude $A_{2}(\{p^{\ext}_f\})$
takes a similar form as \eqref{rfrf2}; we replace 
  $ \kappa(\pet)$ with $\kappa(\tet)$ and have extra term $|p_{f_{[1]}}^{\ext}|^{2 \xi}$ present, where $\xi = a$ for the type I  and $\xi =2a$ for type II graph.
Then the 0th order term in the expansion in $R_f$ expresses as 
\bea
&&
A_{2}(\{p^{\ext}_f\};0)
=
 \kappa(\tet)
\Big[\int [\prod_{\lex}d\alpha_{\lex} ]  
|p_{f_{[1]}}^{\ext}|^{2 \xi}
\crcr
&&\times
e^{-\alpha_{\lex_1}\sum_\xi (|p_{f_{[1]}}^{\ext}|^{2\xi} + |p_{f_2}^{\ext}|^{2\xi}+ |p_{f_3}^{\ext}|^{2\xi}+\mu)}
e^{-\alpha_{\lex_2}\sum_\xi (|p_{f_{[1]}}^{\ext}|^{2\xi} +| p_{f_2}^{\ext}|^{2\xi}+ |p_{f_3}^{\ext}|^{2\xi}+\mu)}
\Big]
\crcr
&&\times
\Big[
\sum_{p_f} \int [\prod_{\ell}d\alpha_\ell e^{-\alpha_\ell \mu
}]
\prod_{f \in \cF_{\inter}}\Big[
e^{-(\sum_{\ell }\alpha_\ell)\sum_\xi  |p_{f}|^{2 \xi}}  
\Big]
\Big[
\prod_{s=1}^{d}\prod_{v_s \in \cV_{\times;4;s}} (\epsilon\tilde{p}^{\,2a})_{v_s} \Big]
\Big].
\label{massrenx}
\eea
It is explicit here from the pattern of the external data, that this amplitude 
takes the form of the $CT_{2;\xi}$ term. 
We identify then the factor associated with the internal data 
as having a degree of divergence $\omt =0$, given by our power counting analysis in section \ref{subsect:listx}.
Adding all colored symmetric contribution with respect to $s$ of this graph, the
sum of these amplitudes renormalizes the coupling  $Z_{\xi}$.

We now treat the higher orders in the Taylor expansion in the form  $\sum_{f}R_f$
of $A_{2}(\{p^{\ext}_f\})$. 
The first order remainder involving  the sum $\sum_{f}R_f$ can be bounded in  the same
vein as  \eqref{eq:rem2+} and we find 
\be\label{rem22}
|R_{2}|
\leq 
K \prod_{(i,k)}M^{-2b(i(G^i_k)-e(G^i_k))} M^{\omp(G^i_k)=0}\,,
\ee
for some constant $K$. 
We are guaranteed of the convergence of this remainder,
and that all higher order remainders can be shown more convergent.
We are ensured about the summability over scale attributions in this case. 

For the log--divergent graphs III, the analysis is similar except that we do not have extra external momenta contributions $|p_{f_{[1]}}^{\ext}|^{2\xi}$ appearing, as discussed earlier. 

\

We conclude that,

- the expansion of  log--divergent 2-point graphs around their  local
parts yield  log-divergent terms renormalizing $Z_a$, $Z_{2a}$ and the mass.

- all remainders are convergent and bring a sufficient decay for ensuring the summability over scale attribution. This means that renormalization at all orders of perturbation theory according to \cite{Rivasseau:1991ub} can be achieved. 
Therefore, we conclude that Theorem \ref{theorenx} holds. 

- there is no wave function renormalization associated with 
$Z_b$ for the model $\times$  because all amplitudes at the leading order are already convergent. 

Without much work, we observe that several other models listed in Table 
\ref{table:table45} are renormalizable at $d=3$, just like the present model is (up to a change of  \eqref{eq:1omegamel12}, \eqref{eq:1omeganonmel12}, \eqref{111}, Table \ref{tab:listprim2} and \eqref{rem22}). 
More surprising perhaps, at fix $D$, we even suspect that it might have 
a continuum of renormalizable theories for a range of values of $b$.

\section{Conclusion}
\label{concl}

We have addressed the perturbative (at all orders) multi-scale renormalization analysis of the so-called enhanced quartic melonic tensor field theory at any rank $d$ of the tensor fields and for any group $(U(1)^D)^d$. Studied in the momentum space, 
the models are endowed with powers of momenta in the interaction terms which are roughly of the form $p^{2a}\phi^4$, $a>0$,
and which might be associated with derivative couplings. 
The case $a=0$ being well-studied  in the literature can be recovered 
at this limit. 
Through the enhancing procedure, amplitudes which were suppressed in models at $a=0$ now participate in the analysis at $a>0$.

In order to achieve renormalizability in enhanced models, the propagators need  
 a more general form than the usual Laplacian dynamics: we thus assume the propagators to be of  the form $(\sum_\xi |p|^{2\xi}+\mu)^{-1}$, $\xi=a,2a,b$ strictly positive.
Two types of models were introduced and studied in parallel in this work: 
an asymmetric model $+$ and a symmetric  model $\times$. 
From the multi-scale analyses of these models, we identify new combinatorial quantities which allow  us to write down a power counting theorem in terms of quasi-local subgraphs. 
At any rank $d>2$, group dimension $D\ge 1$, 
we have found intervals of values for $a$ and $b$ for which 
 both models are potentially renormalizable at all orders of perturbation 
theory. Let us give a summary of the particularities of each model. 

For arbitrary $d$ and $D$, which specify the rest of the parameters,
we find a two-dimensional grid ($(\mathbb{N}-\{0,1,2\})\times (\mathbb{N}-\{0\})$) of just-renormalizable models  $+$. 
As expected, the amplitudes of non-melonic diagrams start to contribute to the flow 
and dominate the melonic amplitudes in each model. 
In fact, all the 4-point melonic diagrams become convergent. 
We found that the enhanced coupling $\pet$ will contribute to the flow 
of the melonic coupling $\lambda$ whereas the opposite is not possible. 
Introducing enhanced interactions of 
$ \Tr_4 (p^{2 a}\phi^4)$ 
influences the study of the 2-point function as it requires to introduce a particular counter-term of the form  $\Tr_2(p^{2a}\phi^2)$.

The model $\times$ is also renormalizable for  a tuned set
of parameters. Again, melonic and non-melonic amplitudes  can be
of the same divergence degree, as expected with enhanced interactions. However, something puzzling happens in this model:
all $4$-point amplitudes prove to be finite and only 2-point amplitudes
may diverge. There are infinitely many 2-point divergent amplitudes. The detailed study of the 2-point
graphs imposes us to include two counter-terms for removing all divergences:
$\Tr_2(p^{2a}\phi^2)$ 
and 
$\Tr_2(p^{4a}\phi^2)$. 
Thus, the flow of this model
 is uniquely driven  by 2-point functions. Then, the model belongs
 neither to the class of just-renormalizable models nor 
 to the class of super-renormalizable models in the language of usual QFT.
 We conjecture that there are several other renomalizable models of this kind.

We recall that enhanced tensor models are introduced 
 from attempts to escape the branched polymer phase of  colored tensor models.
 If enhanced tensor models are turned into field theories, then the large $N$-limit becomes the UV-limit (large $p$).  As far as the present study  
 is concerned, the perturbative renormalizability of the enhanced models of the type presented
 here might not immediately tell us  anything about new limits or new phases 
  of these enhanced models. Having renormalizability rather ensures us that the field theory counterparts of these models are
 long-lived, defined through several layers of momentum scales, and might be UV-complete.  This is definitely an important and encouraging point to keep up with 
 their study. 

Another important aim that could be certainly reached from 
 our analysis is the computation of the perturbative $\beta$-functions
 for these models. There are several arguments putting forward that
 ordinary $\phi^4$ tensor field theories are perturbatively asymptotically free \cite{Rivasseau:2011hm, Geloun:2013saa}. The main ingredient  leading to  
 asymptotic freedom is the presence of a wave-function renormalization
 which  dominates the renormalized coupling constant. Nevertheless,
 the models addressed in this paper seem to belong to another class,
simply because of the presence of the several couplings $Z_\xi$, $\xi=a,2a,$
and the fact that $Z_b$ does not get any radiative corrections. 
For the model $+$, we realize that the RG equations might be more
involved than one might think because of the number of couplings in the theory. 
Thus, only careful computations of the $\beta$-functions
of this model could  help to understand the UV-behaviour of the model $+$. 

Beyond perturbation,  non-perturbative properties of these models can be sought in the future. In particular, the next steps of the program for enhanced models would be to find UV and IR fixed points which may exist and, from these,  perhaps complete trajectories from the UV to the IR.  The  proof of the perturbative renormalizability is again encouraging for this next level. 
The FRG approach has been applied to ordinary tensor field theories with interesting results. Extending the methods to the enhanced theory space
is again to be done. 
In particular,  if one shows the existence of stable IR fixed points in 
enhanced tensor field theories, it could give them a firmer underpinning
as interesting candidates undergoing a phase transition from discrete-like geometries
to some condensate-like geometry. 

\section*{Acknowledgements}
The research of RT is supported by the Netherlands Organisation for Scientific Research (NWO) within the Foundation for Fundamental Research on Matter (FOM) grant 13VP12.
RT thanks Max Planck Institute for Gravitational Physics, Potsdam-Golm (Albert Einstein Institute) for their hospitality while this work was in progress.
JBG thanks the Laboratoire de Physique Th\'eorique d'Orsay, 
Universit\'e Paris 11, for its hospitality.

\section*{Appendix}

\appendix

\renewcommand{\theequation}{\Alph{section}.\arabic{equation}}
\setcounter{equation}{0}

\section{Spectral sums}
\label{app:sums}

In this appendix, we perform spectral sums over internal momenta. 
We start from a basic sum and will go for more involved cases in dimension $D$. In the following, we use $B>0$
as a parameter, and  in the text, $B=M^{-i_l}$. Targeting upper bounds on $B^{\alpha}$,
we focus on the terms with $\alpha<0$. 

Noting that for constants  $B>0$, $a>0$ and $b>0$,  a single sum
over a $p_{s;l}$ behaves like 
\be\label{sum0}
\sum_{p=1}^{\infty} p^{a} e^{-B p^b} = k B^{-{(a+1) \over b}} (1+ O(B^{{(a+1) \over b}}))\,,
\ee
where $k$ is an  $a$-dependent constant. This relation has been proved for instance
in appendix A of  \cite{Geloun:2013saa}.  This sum can be generalized as 
\bea
&&
\sum_{p=1}^{\infty} p^{a} e^{-B (p^b + p^c)} 
 =\sum_{n=0}^{\infty}\frac{(-B)^n}{n!}\sum_{p=1}^{\infty} p^{a+cn} e^{-B p^b} \cr\cr
 && 
 =k\sum_{n=0}^{\infty}\frac{(-B)^n}{n!}    B^{-{(a+cn+1) \over b}} (1+ O(B^{{(a+cn+1) \over b}}))  
 \cr\cr
 && 
 =  k B^{-{(a+1) \over b}} e^{- B ^{1- \frac{c}{b}} } (1+ O(B^{{(a+1) \over b}})
 +  B^{(1-{c \over b})} O(B^{{(a+c+1) \over b}}) + 
   B^{2(1-{c \over b})} O(B^{{(a+2c+1) \over b}})) \cr\cr
   && 
 =  k B^{-{(a+1) \over b}} e^{- B ^{1- \frac{c}{b}} } (1+ O(B^{{(a+1) \over b}})) \,,
 \label{sum0abc}
\eea 
where we use \eqref{sum0} at an intermediate step. 
For the specific choice $c\leq b$, such that $1-\frac{c}{b}\geq 0$ the previous
result recasts as 
\be\label{sumfinabc}
\sum_{p=1}^{\infty} p^{a} e^{-B (p^b + p^c)} = 
k B^{-{(a+1) \over b}}  (  1 +  O ( B^{ 1 - { c \over b} } ) )\,. 
\ee
It is not difficult then to use the same routine and get, for $c+d\leq 2b$,
\bea
&& 
\sum_{p=1}^{\infty} p^{a} e^{-B (p^b + p^c + p^d)} 
 =\sum_{m,n=0}^{\infty}\frac{(-B)^{n+m}}{n!m!}\sum_{p=1}^{\infty} p^{a+cn+dm} e^{-B p^b} \cr\cr
 && 
 =k \sum_{m,n=0}^{\infty}\frac{(-B)^{n+m}}{n!m!}    B^{-{(a+cn+dm+1) \over b}} (1+ O(B^{{(a+cn+dm+1) \over b}}))  \cr\cr
 && 
 =   k B^{-{(a+1) \over b}} e^{-B^{2-\frac{(c+d)}{b} } }  (1+ O(B^{{(a+1) \over b}})) 
  =  k B^{-{(a+1) \over b}}   (1+ O(B^{2-{(c+d) \over b}}))  \,. 
  \label{sumfinabcd}
\eea 
Note that if the above calculations were approximated at $c\le b$,
using  $e^{-B(p^b + p^c)} \leq e^{-2B p^c}$ and, if  $c\le d \le b$, $e^{-B(p^b + p^c + p^d)} \leq  e^{-3B p^c}$, 
the sum behavior could change (using \eqref{sum0} with a modified $B$).
Hence using this bounds is not the optimal choice.

We will need a companion sum in dimension $D$:  
\bea
&&
\sum_{p_1, \dots, p_{D}=1}^{\infty} (\sum_{l=1}^D p_l^{a})^{n} 
e^{-B (\sum_{l=1}^Dp_l^b)} 
 =   \sum_{\sum_{i}n_i = n} 
\frac{n!}{\prod_{l=1}^D n_l!}
\sum_{p_1, \dots, p_{D}=1}^{\infty} \prod_{l=1}^D  p_l^{a n_l}  
e^{-B p_l^b}  
\cr\cr
&&
= c' \sum_{\sum_{i}n_i = n} 
\frac{n!}{\prod_{l=1}^D n_l!}
 B^{-{\sum_l (an_l+1) \over b}}\prod_{l=1}^D (1+ O(B^{{(an_l+1) \over b}})) 
 = c' B^{-\frac{(an+D)}{b}}  (1+ O(B^{\frac{1}{b}})),
\label{sums}
\eea
where $c' = c^D2^n$.
 We  extend this computation, in the 
case of multiple powers in the exponential, with $c\leq b$:
\bea
&&
\sum_{p_1, \dots, p_{D}=1}^{\infty} (\sum_{l=1}^D p_l^{a})^{n} 
e^{-B \sum_{l=1}^D(p_l^b+p_l^c)} 
= \cr\cr
&&
c' \sum_{\sum_{i}n_i = n} 
\frac{n!}{\prod_{l=1}^D n_l!}
 B^{-{\sum_l (an_l+1) \over b}} e^{-B^{1 -\frac{c}{b}}}\prod_{l=1}^D (1+ O(B^{{(an_l+1) \over b}}) \cr\cr
 &&
  = c' B^{-\frac{(an+D)}{b}} e^{-B^{1 -\frac{c}{b}}} (1+ O(B^{\frac{1}{b}}))
   = c' B^{-\frac{(an+D)}{b}} (1+ O(B^{\frac{1}{b}}) + O(B^{1 - \frac cb}))
\,,
\label{sumsabc2}
\eea
where \eqref{sum0abc} and \eqref{sumfinabc} have been used. 
Depending  on $b-c \leq 1$ or otherwise, the two big-O functions can be reduced
 into one. However,  being interested in the leading order, this relation 
is sufficient to proceed further. 
Using the same techniques, the last useful sum  evaluates as 
\bea
&&
\sum_{p_1, \dots, p_{D}=1}^{\infty} (\sum_{l=1}^D p_l^{a})^{n} 
e^{-B (\sum_{l=1}^D(p_l^b+p_l^c+p_l^d))} 
=
 c' B^{-\frac{(an+D)}{b}} e^{-B^{2 - \frac{(c+d)}{b}}}  (1+ O(B^{\frac{1}{b}}))
 \cr\cr
 &&
 = c' B^{-\frac{(an+D)}{b}}  (1+ O(B^{\frac{1}{b}})  + O(B^{2 - \frac{(c+d)}{b}}))
\,,
\label{sumsabc3}
\eea
where the last equality is obtained for $c+d\leq 2b$.

\section{Divergences in model $+$ $(d=3,D = 1, a = {1 \over 2},b = {3 \over 4})$}
\label{app:mod+}

\begin{figure}[H]\
\begin{center}
     \begin{minipage}{.7\textwidth}
     \centering
   \includegraphics[angle=0, width=5cm, height=1cm]{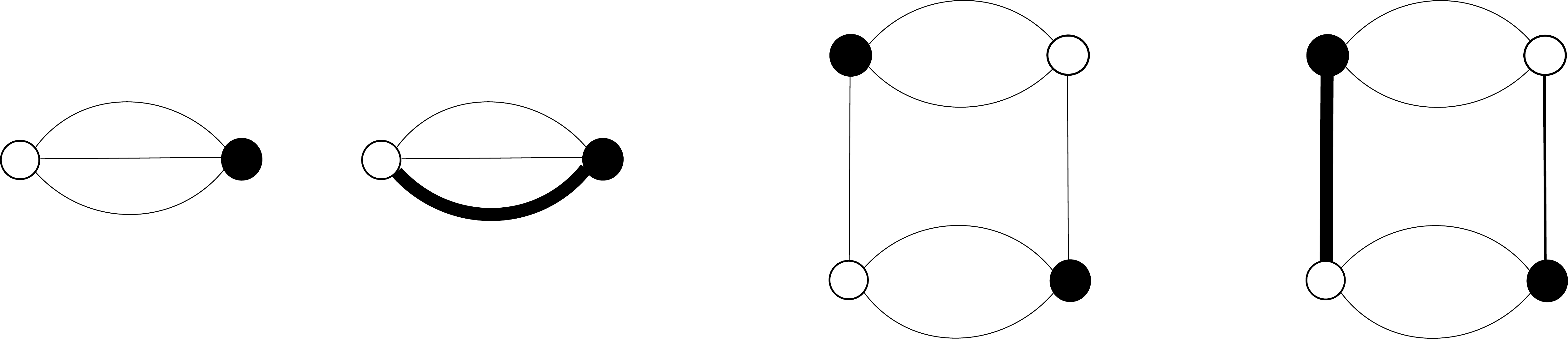}
\caption{ {\small Rank $d=3$, mass, enhanced $\phi^2$-interaction,
and (simple and enhanced) $\phi^4$-interactions. }}
\label{fig:phi4EnhancedMelonModel}
\end{minipage}
\end{center}
\end{figure}

We consider here a specific model $+$ with  parameter given as
$d=3, D=1,a=\frac12, b=\frac34$. 
The mass term and interactions are of the form given in Figure  \ref{fig:phi4EnhancedMelonModel}. 
In this appendix, we illustrate graphs which satisfy the power counting achieved in Table \ref{tab:listprim1} in  section \ref{subsect:list+}. 

The superficial degree of divergence of a graph $\cG$
is  given in \eqref{deg+} that we specialize for $D = 1$, $ a = {1 \over 2}$, and $b = {3 \over 4}$ as
\beq
\omp(\cG) = - {3 \over 2}  L +  F_{\rm int} +  \rhop  + \rho_{2;a} + {3 \over 2} \rho_{2;b} \,.
\eeq
We  use the bipartite colored graph representation of the Feynman
graphs of the model. Edges which are dashed are  propagators; edges 
in the interaction vertex can be in bold or not. If they are in bold that means that they receive
an enhancement factor of $p^{2a}$.  
  The figures illustrating graphs have red lines that facilitate
the identification of the face structure of the graphs. 
Given a colored graph, we emphasize a red cycle (made with alternating edges and dashed edges
with red color) that indicates a particular closed face. 
Naturally, this face will be the source
of an enhanced power counting if it contains a bold edge. 

We only list here some divergent graphs contributing  to the renormalization of the interactions.

(i)
We consider 2-point functions, $N_{\ext} = 2$.

\begin{figure}[H]
\centering
     \begin{minipage}{.7\textwidth}
\includegraphics[angle=0, width=12.5cm, height=2.2cm]{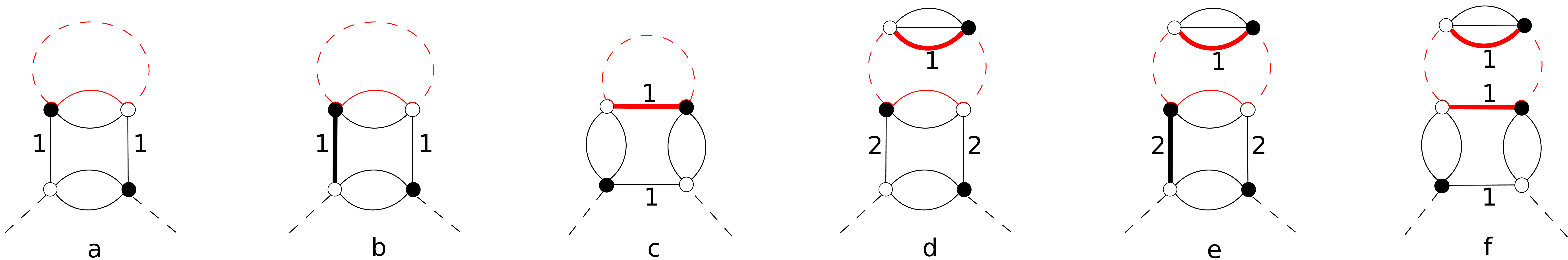}
\caption{ {\small Divergent graphs with $N_{\rm ext} = 2$ and $V_{(4)}=1$. }} 
\label{fig:V1_1new} 
\end{minipage}
\end{figure}

(i1) We consider $V_{(4)} = 1$ or divergent tadpoles given in Figure \ref{fig:V1_1new}. The graphs a and b are melonic  with $V_{(4)} = 1$,
$ \rhop  + \rho_{2;a} + {3 \over 2} \rho_{2;b}= 0$,  and $\omp = {1 \over 2}$,
whereas 
the graph c is non-melonic  with $V_{+;4} = 1$, $ \rhop =1,  \rho_{2;a} + {3 \over 2} \rho_{2;b}=0$ and 
$\omp = {1 \over 2}$. 
The graphs d and e with $V_{(4)} = 1$ and $V_{2;a} = 1$, $\rhop+  {3 \over 2} \rho_{2;b}=0,\rho_{2;a} =1 $ are melonic and log-divergent with
$\omp = 0$; 
the graph f is non-melonic  with $V_{+;4} = 1$, $V_{2;a} = 1$, 
$\rhop=1=\rho_{2;a}, \rho_{2;b}=0$, and also log-divergent with
$\omp = 0$.

\begin{figure}[H]
\begin{center}
     \begin{minipage}{.7\textwidth}
\begin{center}
\includegraphics[angle=0, width=12cm, height=6cm]{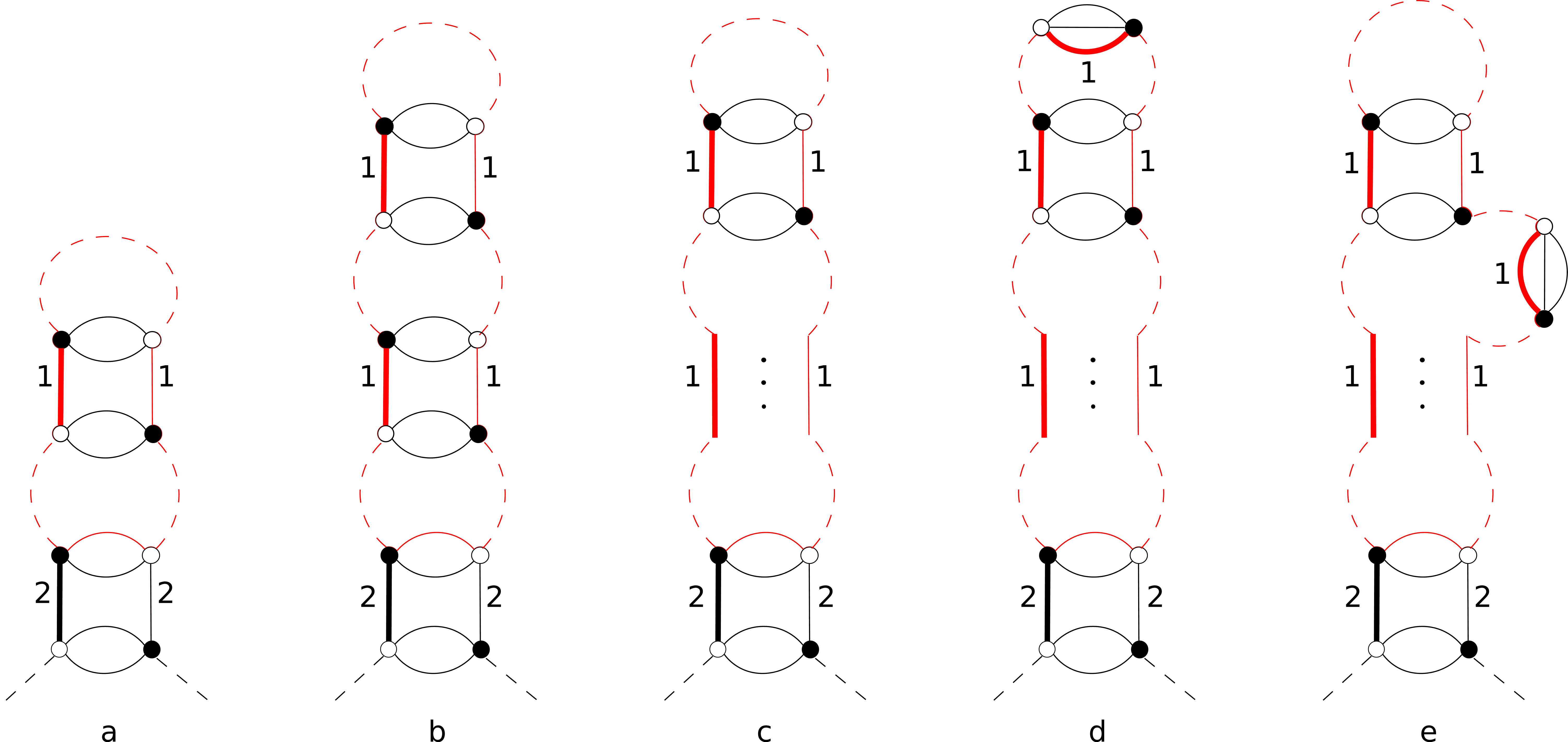}
\caption{ {\small 
A  family of melonic graphs (a,b and c)
with $N_{\rm ext} = 2$ and $\omp = {1 \over 2}$
and 
a family of log-divergent melonic graphs (d and f) with $N_{\rm ext} = 2$.}} 
\label{fig:V2melon_1new} 
\end{center}
\end{minipage}
\end{center}
\end{figure}

(i2) 
We consider $V_{(4)} = 2$ (in particular, $(V_4, \; V_{+;4}) = (1, \;1)$ or $(0, \;2)$) and its generalizations given in Figure  \ref{fig:V2melon_1new}.
Note that if we increase $V_{+;4}$ in the way of producing c
 ($V_{+;4}=3$ given by the graph b),  then we have for arbitrary $V_{+;4}$,
$L = 1 + 2(V_{(4)} -1) = 2 V_{(4)} -1$, $F_{\inter} = 2 V_{(4)} $, and $\rhop = V_{(4)}-1$, $\rho_{2;a} = 0$, and $\rho_{2;b} = 0$,
 therefore $\omp  = {1 \over 2}$, and is independent of $V_{(4)}$ (or $V_{+;4}$)
 which is expected. 
To these graphs, we can add the enhanced 2-point function $V_{2;a} =1$ to any of the internal lines as illustrated in the graphs d and e.
We have for arbitrary $V_{+;4}$ 
and $V_{2;a} = 1$, 
$L = 2 V_{(4)}  $, $F_{\inter} = 2 V_{(4)}  $, $\rhop = V_{(4)}-1 $, $\rho_{2;a} = 1$, and $\rho_{2;b} = 1$ 
therefore 
$\omp  = 0$ is 
 independent of
$V_{(4)}$ (or $V_{+;4}$).

\begin{figure}[H]
\begin{center}
     \begin{minipage}{.7\textwidth}
\begin{center}
\includegraphics[angle=0, width=12cm, height=4.5cm]{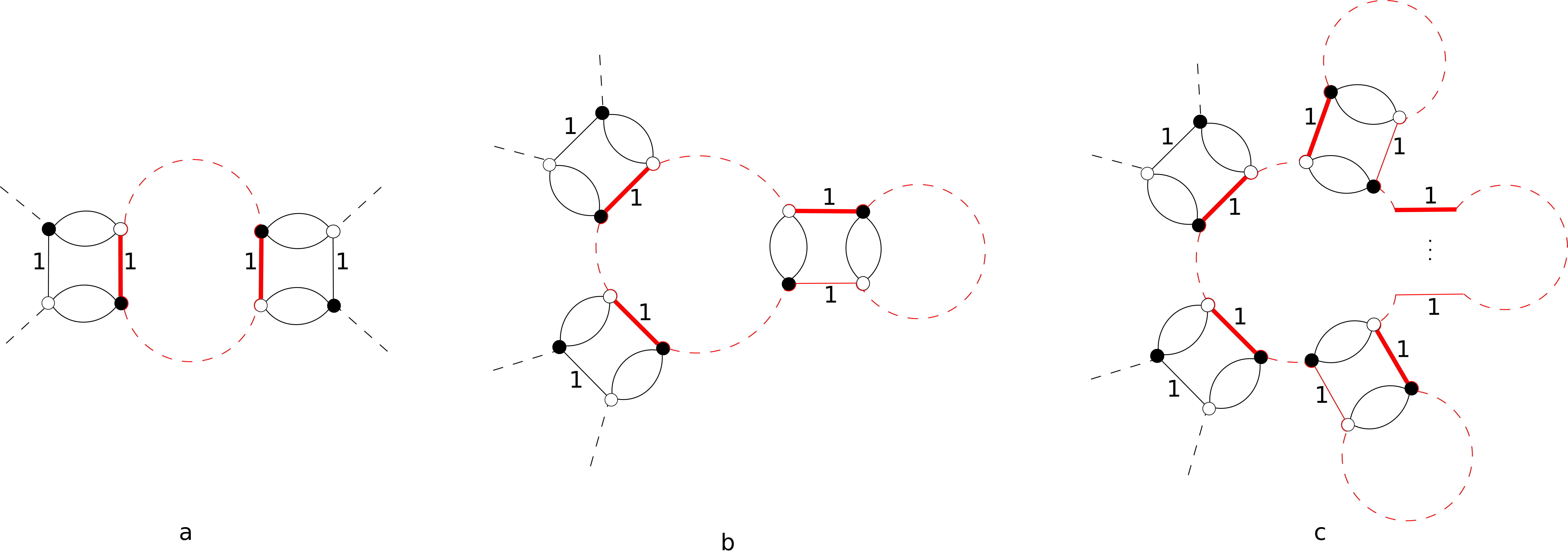}
\caption{ {\small 
A family of log-divergent  non-melonic graphs with $N_{\rm ext} = 4$ and $V_{+;4}\ge 2$.  
}} 
\label{fig:V2nonmelon_1new} 
\end{center}
\end{minipage}
\end{center}
\end{figure}

(ii)
We now consider 4-point functions, $N_{\rm ext} = 4$.
First, set $V_{+;4} = 2$. 
The non-melonic graph a of Figure \ref{fig:V2nonmelon_1new}  is logarithmic-divergent
since  $L=2$, $F_{\rm int} = 1$,  $\rhop =2$, $\rho_{2;a} = 0$, and $\rho_{2;b} = 0$.
This graph generalizes to b and then to c such that 
$F_{\rm int} = 1 + d^- (V_{+;4} - {N_{\rm ext} \over 2}) = -3 + 2 \, V_{+;4}$,  $ \rhop = V_{+;4}$,  $\rho_{2;a} = 0$ and $\rho_{2;b} = 0$.

\section{Divergences in model $\times$ $(d=3, D = 1,  a = {1 \over 2}, b =1)$}
\label{app:modx}

We illustrate some divergent amplitudes in the model $\times$ with parameters
given above.  We keep the same meaning of the graphical representation 
for graphs as in appendix \ref{app:mod+}. 

The superficial degree of divergence of a graph $\cG$ has been given in \eqref{degx} 
and that we evaluate at $D = 1$, $ a = {1 \over 2}$, and $b = {1}$ 
as 
\beq
\omt(\cG) = - 2  L +  F_{\rm int} +  \rhot  + \rho_{2;a} + 2 \rho_{2;b}\,.
\eeq

We consider 2-point functions, $N_{\rm ext} = 2$.

\begin{figure}[H]
\begin{center}
     \begin{minipage}{.7\textwidth}
\begin{center}
\includegraphics[angle=0, width=5cm, height=2cm]{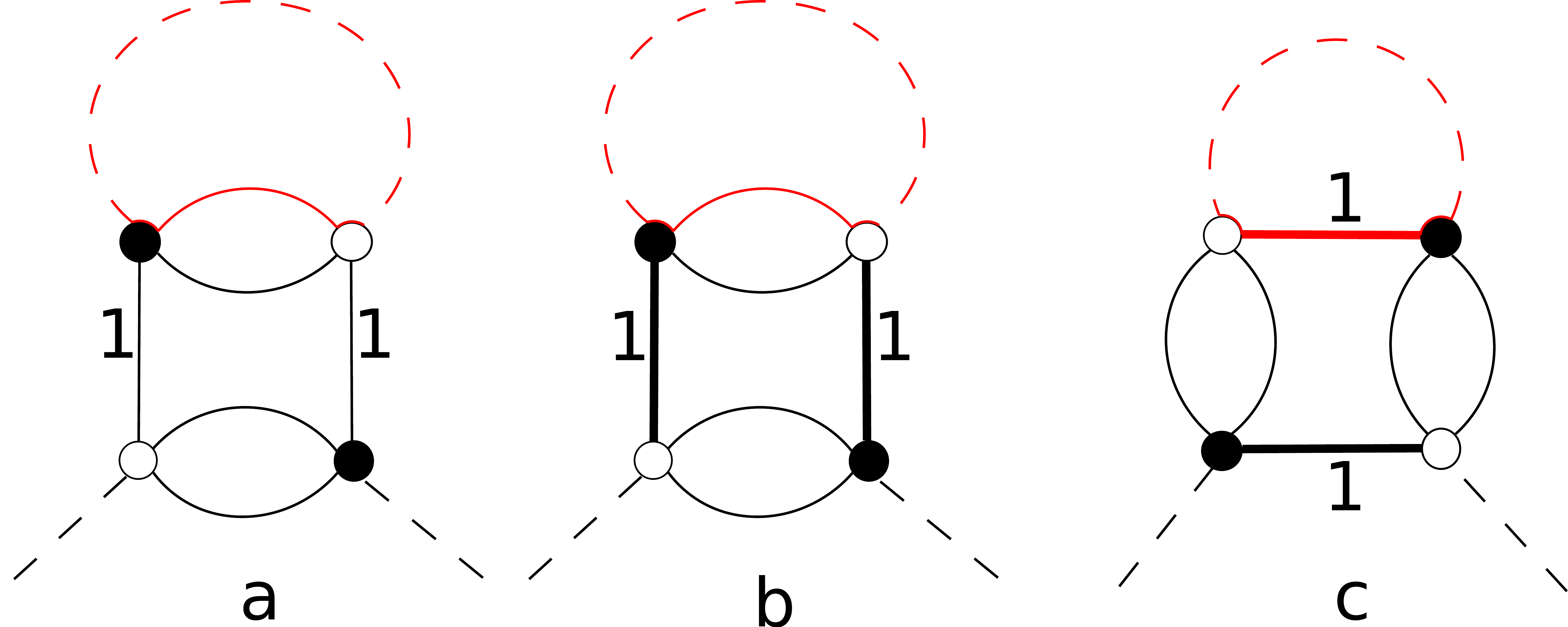}
\caption{ {\small 
Log-divergent graphs with $N_{\rm ext} = 2$ and $V_{(4)}=1$. 
}} 
 \label{fig:V1_1x} 
\end{center}
\end{minipage}
\end{center}
\end{figure}

(1) We consider tadpole graphs with $V_{(4)} = 1$ given in Figure  \ref{fig:V1_1x}
which are log-divergent. 
The graphs a and b are melonic 
whereas the graph c is non-melonic.

\begin{figure}[H]
\begin{center}
     \begin{minipage}{.7\textwidth}
\centering
\includegraphics[angle=0, width=12cm, height=6cm]{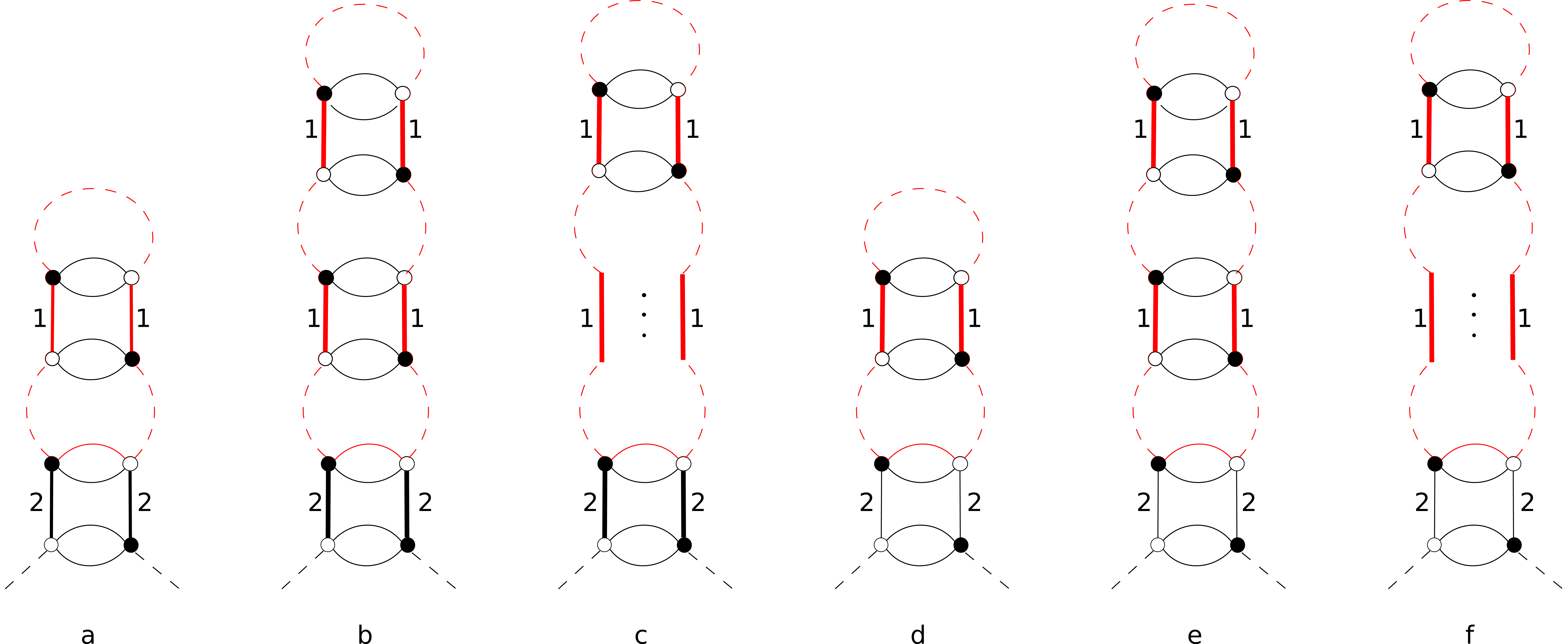}
\caption{ {\small 
Melonic graphs with $N_{\rm ext} = 2$ which give $\omt= 0$.}} 
\label{fig:V2melon_1x} 
\end{minipage}
\end{center}
\end{figure}

\begin{figure}[H]\
\begin{center}
     \begin{minipage}{.7\textwidth}
     \centering
\includegraphics[angle=0, width=6cm, height=5.8cm]{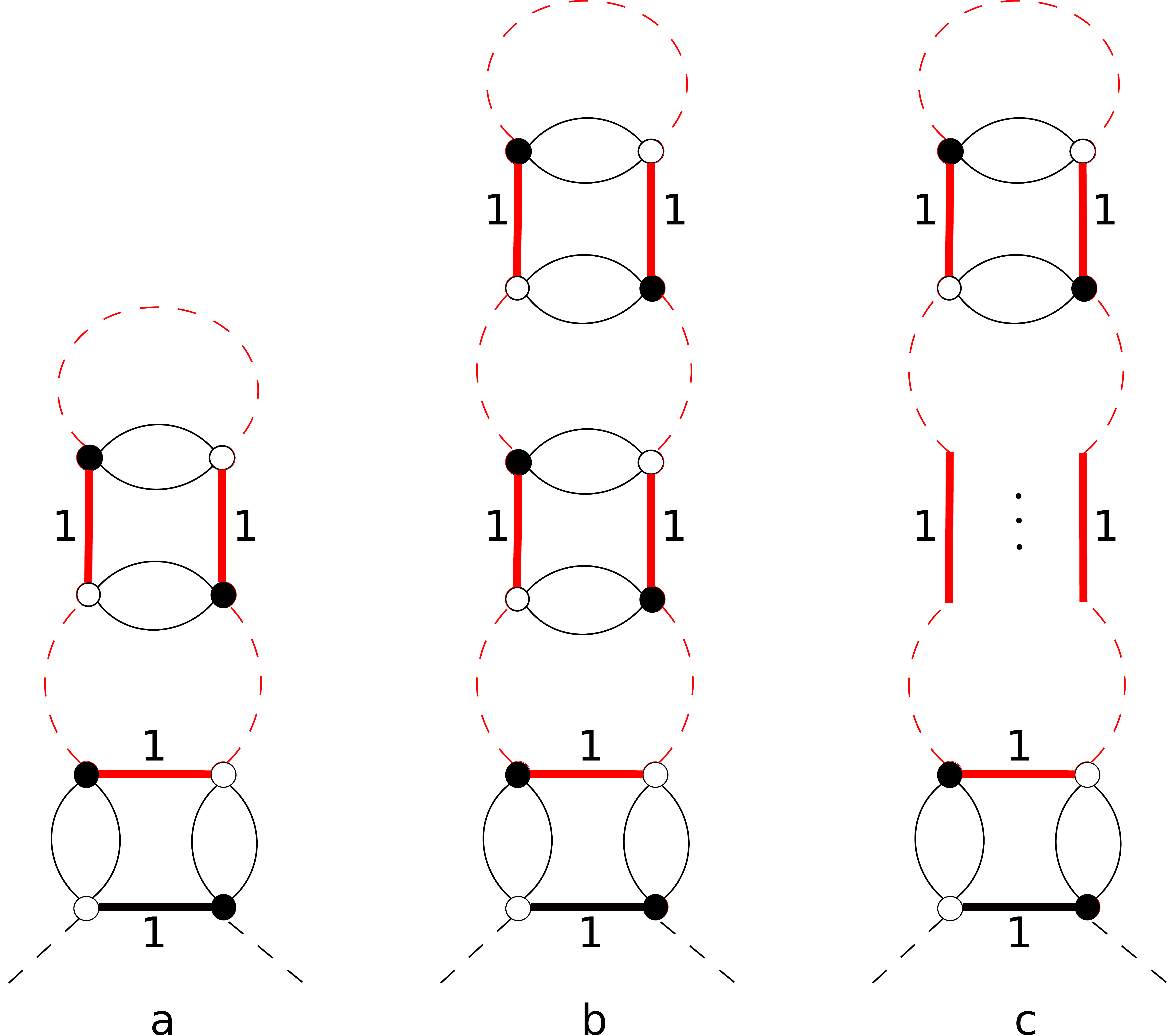}
\caption{ {\small Non-melonic graphs with $N_{\rm ext} = 2$ which give $\omt =  0$. }} 
\label{fig:V2nonmelon_mixx} 
\end{minipage}
\end{center}
\end{figure}

(2) Consider now $V_{(4)} \ge  2$.
Melonic graphs (resp. non-melonic graphs) with $V_{(4)}=2$ and  their generalizations for $V_{(4)} > 2$ are given in  Figure \ref{fig:V2melon_1x}  (resp. Figure \ref {fig:V2nonmelon_mixx}).

First, focus on the  melonic graphs of Figure \ref{fig:V2melon_1x}.
The melonic graphs, a and d with $N_{\rm ext} = 2$ and $V_{(4)}=2$ (but $V_{\times;4} \ge 1$) give 
$\omt=0$.
Increase $V_{\times;4}$ in  a way to have the graphs  c  and f (intermediate
steps are given by graphs b and e, respectively),  then we have for arbitrary $V_{\times;4}$,  $F_{\inter} = 2 V_{(4)}$, $\rhot = 2 (V_{(4)} -1)$, therefore $\omt  = 0$.
With these graphs, we confirm that this model has infinitely many 
log-divergent 2-point  graphs. 

Next, consider the non-melonic graphs of Figure \ref {fig:V2nonmelon_mixx}.
For arbitrary $V_{(4)} = V_{\times;4}$,   $F_{\inter} = 2 (V_{(4)} -1) + 1$, $\rhot = 1 + 2 (V_{(4)} -1)$, $\rho_{2;a} = 0$, and $\rho_{2;b} = 0$ therefore $\omt  = 0$. 
Again,  we see that this model has infinitely many graphs that are divergent.

\begin{figure}[H]\
\begin{center}
     \begin{minipage}{.7\textwidth}
     \centering
   \includegraphics[angle=0, width=4cm, height=2cm]{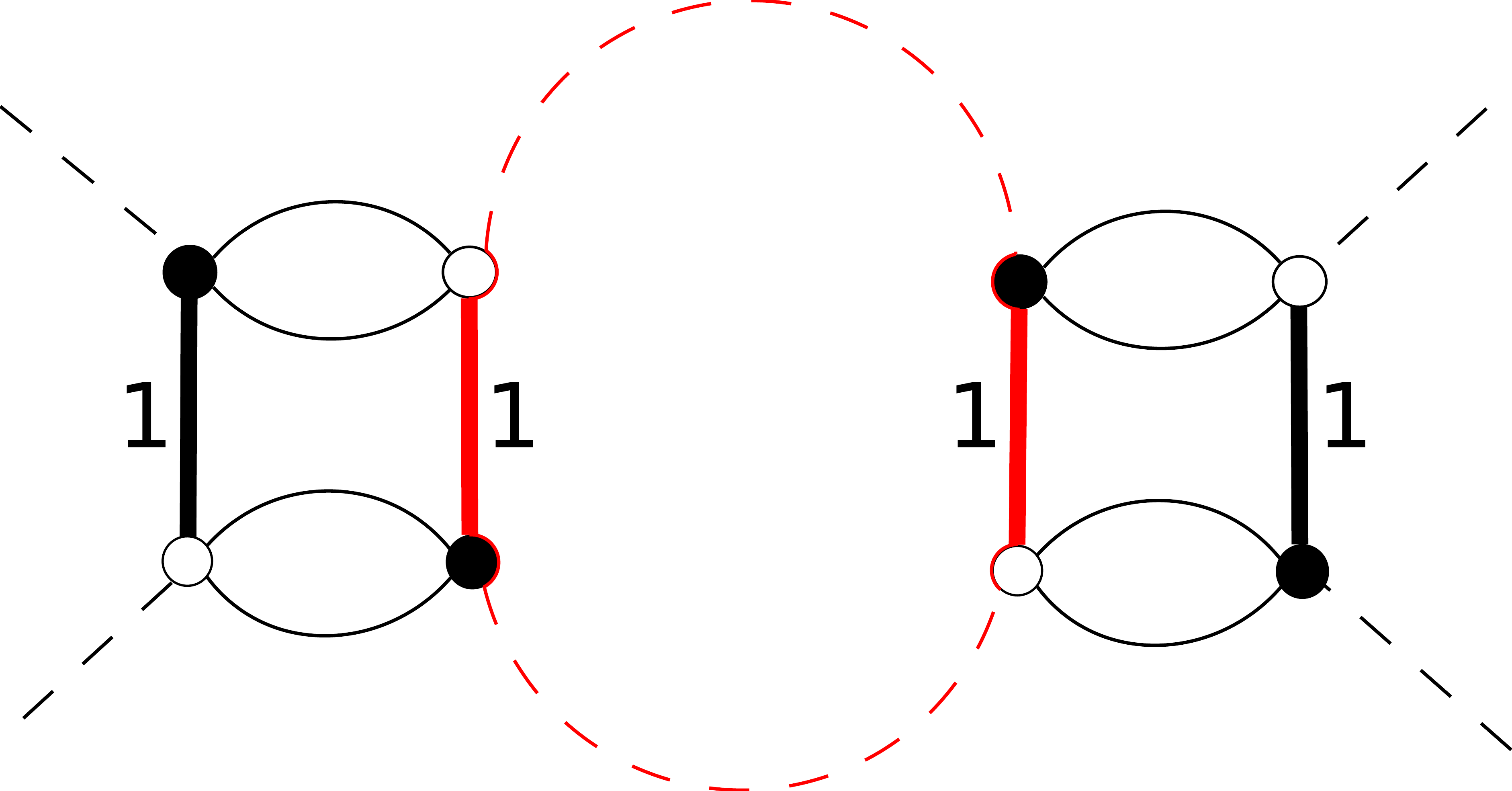}
\caption{ \small A convergent  non-melonic graph with $N_{\rm ext} = 4$ and $V_{(4)} = V_{\times; 4}=2$.  }
\label{fig:V2nonmelon_1x}
\end{minipage}
\end{center}
\end{figure}

(3)
A comparison with the previous model shows that the 4-point non-melonic
graph  with  $V_{(4)}=V_{\times;4} = 2$  of Figure \ref{fig:V2nonmelon_1x} is convergent 
$\omt= -1$, $F_{\rm int} = 1$,  $\rhot =2$, $\rho_{2;a} = 0$, and $\rho_{2;b} = 0$.

\end{document}